\documentclass[structabstract]{aa}  
\usepackage{graphicx}
\usepackage{natbib}
\bibpunct{(}{)}{;}{a}{}{,}
\usepackage{txfonts}
\usepackage{longtable}
\usepackage{multirow}

\begin{document}

\title{The ability of intermediate-band Str\"omgren photometry to
  correctly identify dwarf, subgiant, and giant stars and provide
  stellar metallicities and surface gravities\thanks{Appendices A and
    B are only available in electronic form at the {\tt
      http://www.aanda.org}. The table in Appendix B will be available
    through CDS.}}
 
\titlerunning{Metallicities and stellar classification from Str\"omgren photometry}

   \author{A. S. \'{A}rnad\'{o}ttir  \inst{1}
          \and
          S. Feltzing \inst{1}
          \and
          I. Lundstr\"{o}m \inst{1}
           } 

           \institute{Lund Observatory, Department of Astronomy and Theoretical Physics, Lund University, Box 43, SE-221 00 Lund, Sweden\\
             \email{anna, sofia, ingemar, @astro.lu.se} }

   \date{Received 25 October 2009: Accepted 10 Feb 2010}
 
  \abstract
  {Several large scale photometric and spectroscopic surveys
    are being undertaken to provide a more detailed picture of the Milky
    Way.  Given the necessity of generalisation in the determination
    of, e.g., stellar parameters when tens and hundred of thousands of
    stars are considered it remains important to provide independent,
    detailed studies to verify the methods used in the surveys.}
  {Our first aim is to critically evaluate available
    calibrations for deriving [M/H] from Str\"omgren
    photometry. Secondly, we develop the standard sequences for dwarf
    stars to reflect their inherent metallicity dependence. Finally,
    we test how well metallicities derived from $ugriz$ photometry
    reproduce metallicities derived from the well-tested system of
    Str\"omgren photometry.  }
  {We evaluate available metallicity calibrations based on Str\"omgren
    $uvby$ photometry for dwarf stars using a catalogue of stars with
    both $uvby$ photometry and spectroscopically determined iron
    abundances ([Fe/H]). The catalogue was created for this project.
    Using this catalogue we also evaluate available calibrations that
    determine $\log g$.  A larger catalogue, in which metallicity is
    determined directly from $uvby$ photometry, is used to trace
    metallicity-dependent standard sequences for dwarf stars. We also
    perform comparisons, for both dwarf and giant stars, of
    metallicities derived from $ugriz$ photometry with metallicities
    derived from Str\"omgren photometry.}
{We provide a homogenised catalogue of 451 dwarf stars with
  $0.3<(b-y)_0<1.0$. All stars in the catalogue have $uvby$ photometry
  and [Fe/H] determined from spectra with high resolution and high
  signal-to-noise ratios (S/N).  Using this catalogue, we test how
  well various photometric metallicity calibrations reproduce the
  spectroscopically determined [Fe/H]. Using the preferred metallicity
  calibration for dwarf stars, we derive new standard sequences in the
  $c_{1,0}$ versus $(b-y)_0$ plane and in the $c_{1,0}$ versus
  $(v-y)_0$ plane for dwarf stars with $0.40 < (b-y)_0 < 0.95$ and
  $1.10 < (v-y)_0 < 2.38$.  }
{We recommend the calibrations by {Ram{\'{\i}}rez} \&
  {Mel{\'e}ndez}\,(2005) for deriving metallicities from Str\"omgren
  photometry and find that intermediate band photometry, such as
  Str\"omgren photometry, more accurately than broad band photometry
  reproduces spectroscopically determined [Fe/H]. Str\"omgren
  photometry is also better at differentiating between dwarf and
  giant stars.  We conclude that additional investigations of the
  differences between metallicities derived from $ugriz$ photometry and
  intermediate-band photometry, such as Str\"omgren photometry, are
  required.}


\maketitle

\section{Introduction}

The photometric system introduced by Bengt Str\"omgren
\citep{1963QJRAS...4....8S,1964ApNr....9..333S} provides a means of
reliably estimating stellar parameters for stars with a wide range
of spectral classes.  For instance, metallicities can be determined
for many types of stars. In particular, the system can accurately
identify stars at different evolutionary stages \citep[see discussion
in, e.g.,][]{1963QJRAS...4....8S}.  This makes it possible to
determine the distances of stars with no parallax measurements. If
reddening is not known, the system must, however, be complemented with
$H\beta$ photometry.

The advent of CCD photometry has meant that larger and deeper areas of
sky can be scanned to determine the properties of stars in the field
and from them infer the properties of the stellar populations in the
Milky Way. For broad-band photometry, this approach has been very
successful, e.g., \citet{1983MNRAS.202.1025G}, who inferred the
existence of the thick disk and, e.g., \citet{2001Natur.412...49I} and
\citet{2005ApJ...622L.109F}, who studied the stellar structures in the
Andromeda galaxy. Arguably the most important large study of this kind
is the Sloan Digital Sky Survey (SDSS) \citep{2000AJ....120.1579Y},
which provides deep photometry of stars for roughly half the sky.

However, in contrast, the usage of narrow and medium band photometry
for Galactic studies was for a much longer time severely hampered by
the relative inefficiency of the CCDs, which required too long exposure
times to make these techniques competitive. This combined with
relatively small fields of view (mainly due to small filters on the
cameras equipped with suitable filters) meant that only very small
portions of the sky could be usefully studied.  Additionally, the size
of telescopes that have cameras with Str\"omgren filters and
relatively low efficiency in the blue also hampered observations in
the $u$ filter \citep[e.g.,][]{1992AJ....104.1765V}.  All of this
meant that systems, such as that designed by Str\"omgren, were mainly
applied to the study of globular and open clusters \citep[two fairly
recent examples are provided
by][]{2002A&A...395..481G,2003AJ....125.1383T} or to individual stars
\citep[e.g.,][]{1994A&AS..106..257O,1995A&A...295..710O,1989A&A...222...69S,
  2004A&A...422..527S,2006A&A...445..939S}. Recent attempts to use
Str\"omgren photometry to study the properties of the Milky Way
stellar disks away from the solar neighbourhood are few.  Interesting
examples being \citet{1993ApJ...407..115V} and
\citet{1995A&A...298..799J}.

Advancements in technology have meant that we now also have
access to larger CCD areas on telescopes equipped with large
$uvby$-filters, enabling an efficient study of  stellar properties
across larger areas of the sky.

We have published two studies based on Str\"omgren photometry of the
red giant branches of dwarf spheroidal galaxies in the Local Group
using the Wide Field Camera (WFC) on the Isaac Newton Telescope on La
Palma \citep{2007AA...465..357F,2009A&A...506.1147A}. This camera is
equipped with large filters that allow an, almost, unvignetted field
of view of half by half a degree. However, far more can be achieved
with this dataset.  It provides the largest database of Str\"omgren
photometry for Milky Way disk stars without any kinematic or colour
biases. The stars are situated at distances between 0.5 and 4\,kpc
away from the Sun and in the directions of the four dwarf spheroidal
galaxies Draco, Sextans, Hercules, and Ursa Major\,II. We intend to
apply this unique dataset to explore the properties of the
Milky Way disk(s) in some detail.

As part of a series of papers on the properties of the Milky Way disks
using Str\"omgren photometry, we have undertaken a critical evaluation
of the available calibrations for metallicity and $\log g$
determinations for dwarf and sub-giant stars.  We have also determined
new standard sequences \citep[compare, e.g.,][]{1984A&AS...57..443O}
to improve the identification of dwarf and giant stars in the distant
disk and halo.  We also provide a basic comparison of metallicities
derived using Str\"omgren photometry and metallicities derived for
dwarf and giant stars from SDSS $ugriz$ photometry using the
calibration in \citet{ivezic2008}.

The paper is organised as follows. Section\,\ref{SecStromgren} provides
a short introduction to the Str\"omgren photometric system and 
background to the work presented here, Sect.\,\ref{Sect:cats} details
the catalogues we compile to test the metallicity
calibrations available for dwarf stars, which is described in
Sect.\,\ref{Sect:met} where we also compare the Str\"omgren
metallicities with those derived by the SDSS project
\citep[DR7][]{2009ApJS..182..543A}. Section\,\ref{sect:seq} considers
the system's ability to distinguish between giant and dwarf stars of
similar colour. We also provide new, metallicity-dependent stellar
sequences for dwarf stars in this section. These new sequences are
compared to model predictions (e.g., isochrones) in
Sect.\,\ref{sect:model}.  Section\,\ref{sect:sum} summarises our
findings and provides a few suggestions for future work.

\section{A short  introduction to the Str\"{o}mgren
  photometric system}
\label{SecStromgren}

\begin{figure}   
\includegraphics[width=9cm]{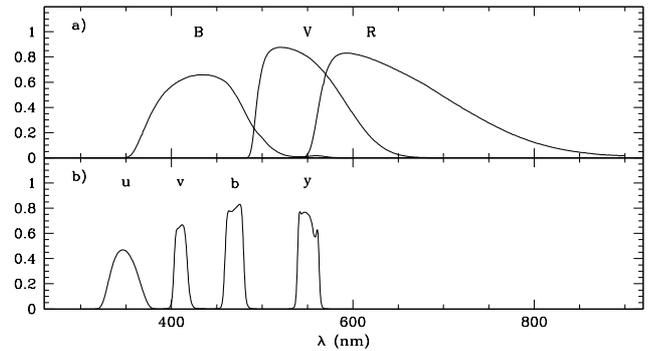}
\caption{Filter throughput curves for broad-band and Str\"omgren
  filters. Filter curves are from  the database of filters used with the
  wide-field camera on the Isaac Newton Telescope. The  database is available
      at {\tt http://catserver.ing.iac.es/filter/}.  {\bf a)} Harris $B$,
    $V$, and, $R$ filters, and {\bf b)} Str\"omgren $u$, $v$, $b$, and
    $y$ filters.}
\label{fig:filters}
\end{figure}

\begin{figure}   
{
\includegraphics[width=7cm,angle=-90]{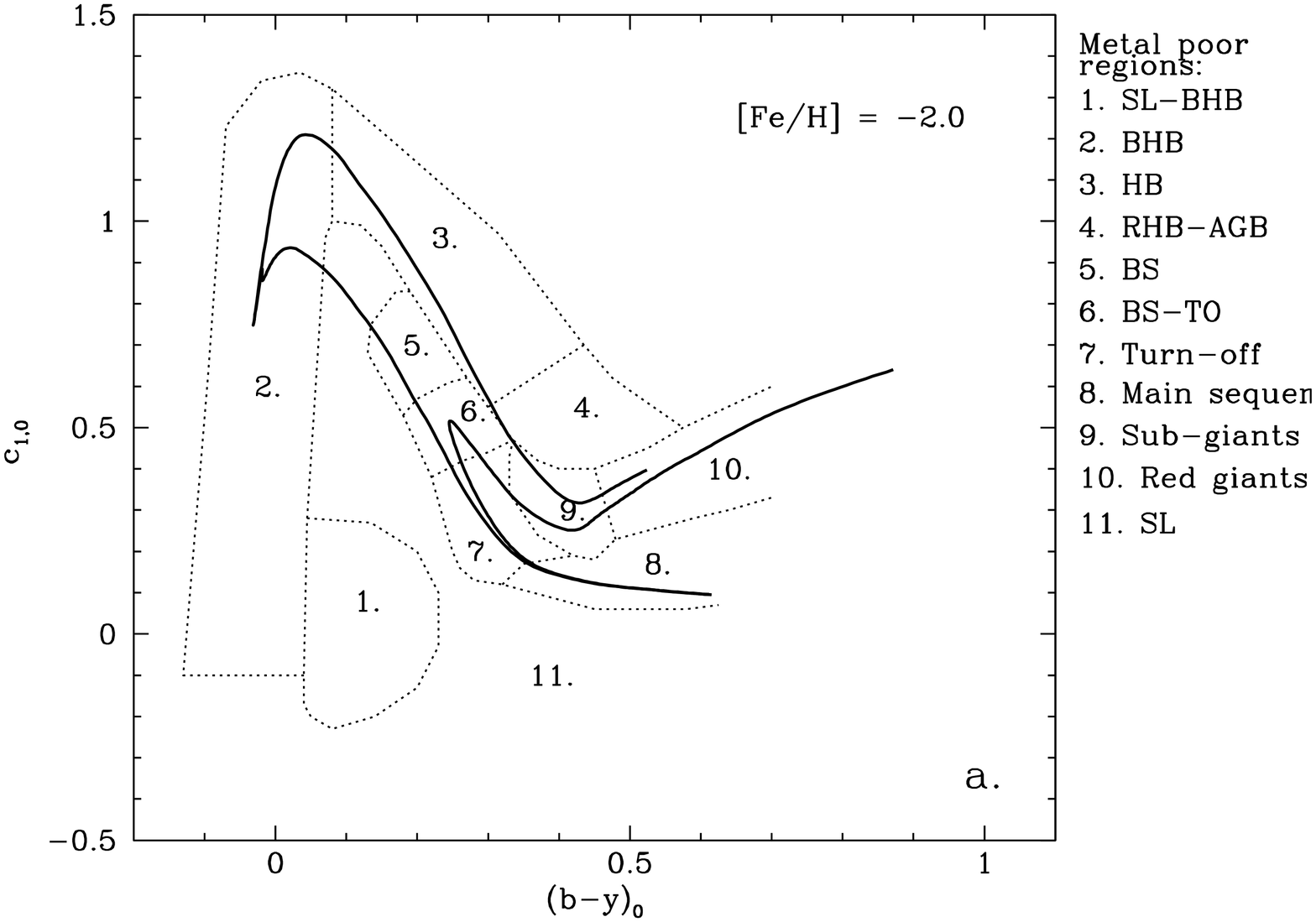}
\includegraphics[width=7cm,angle=-90]{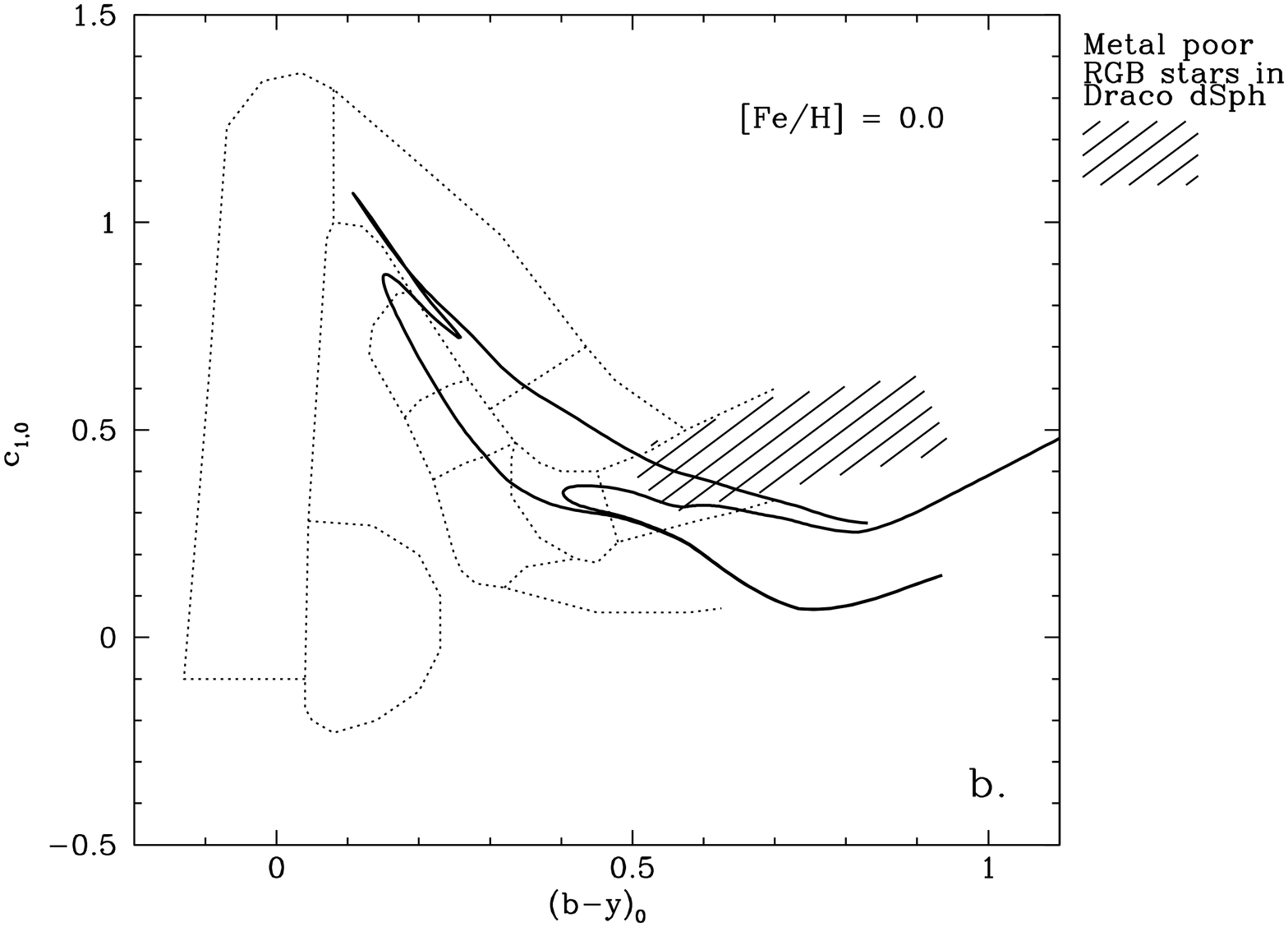}}
\caption{Illustration of the $uvby$ system's ability to identify stars
  at different evolutionary
  stages. The classification scheme by \citet{2004A&A...422..527S} is
  indicated by dotted lines. Evolutionary stages are identified in panel
  {a.} as: 1. SL-BHB: sub-luminous -- blue horizontal branch
  transition, 2. BHB: blue horizontal branch, 3. HB: horizontal
  branch, 4. RHB-AGB: red horizontal branch -- asymptotic giant branch
  transition, 5. BS: blue-straggler stars, 6. BS-TO: blue-straggler --
  turn-off transition, 7. Turn-off: turn-off stars, 8. main sequence,
  9. sub-giants, 10. red giants, and, 11. SL: sub-luminous stars. Two
  isochrones by \citet{2006ApJS..162..375V} using the
  temperature-colour transformation by \citet{2004AJ....127.1227C} are
  shown as full lines (age $=1$Gyr and $10$Gyrs). The metallicities of
  the isochrones are indicated in the panels. The region occupied by
  metal-poor red giants in the Draco dwarf spheroidal galaxy
  \citep{2007AA...465..357F} is indicated by a hashed area in panel
  {b}.}
\label{fig:scheme}
\end{figure}

The Str\"omgren system consists of the four medium-width filters $u$,
$v$, $b$, and $y$ (hereafter collectively denoted as $uvby$), where
the $y$ magnitude is calibrated to be the same as the $V$ magnitude in
the $UBV$ system \citep[e.g.,][see also
\citet{1984A&AS...57..443O} and
Fig.\,\ref{fig:filters}]{1953ApJ...117..313J}. The filters are centred
on 350, 410, 470, and 550\,nm and their half-widths are 38, 20, 10,
and 20\,nm, respectively \citep[e.g.,][page
180]{1974ASSL...41.....G}. In addition, the system relies on the three
colour indices (differences) that are constructed in the following way
\citep[compare, e.g.,][]{1963QJRAS...4....8S}

$$(b-y)$$
$$m_1 \equiv (v-b)-(b-y)$$
$$c_1 \equiv (u-v)-(v-b).$$

These indices are designed to measure important properties of the
stars and were first introduced by Bengt Str\"omgren in a series of
papers, including \citet{1963QJRAS...4....8S} and
\citet{1964ApNr....9..333S}. Work on the system continued by
establishing standard stars
\citep[e.g.,][]{1970AJ.....75..978C,1976A&AS...26..155G,1983A&AS...54...55O,1987PASP...99.1184P,1993A&AS..102...89O}. However,
as discussed in \citet{1997A&AS..122..559C}, the establishment of
standard fields akin to those available for $UVBRI$ photometry
\citep{1992AJ....104..340L} have only very recently been attempted. An
additional problem is that the primary standards are too bright for
most available combinations of cameras with $uvby$ filters and
telescopes. Although \citet{1997A&AS..122..559C},
\citet{1987SAAOC..11...93C}, and \citet{1988A&AS...73..225S} provide
secondary fainter standards, the situation for both standard fields
and secondary standards that can be used with large telescopes remains
unsatisfactory.

There are two main sets of established standard stars for the $uvby$
system, those of \citet{1980ApJS...44..517B} and
\citet{1993A&AS..102...89O}. There are some non-negligible differences
between the two sets and \citet{1995A&A...295..710O} provides a detailed
discussion of this subject. He concludes that the main difference
concerns the $c_1$ index and is caused mainly by the
$u$-filter. Hence, if we wish to compare results based on the two sets
of standards we need to apply corrections \citep[compare, e.g.,
  Fig.\,15 in][]{2007AA...465..357F}. We adopt
observations calibrated to the system established by
\citet{1993A&AS..102...89O}.

The system was originally designed to study earlier types of
stars \citep[A2 to G2,][]{1963QJRAS...4....8S}.  Later work
has, however, shown that the system and its properties can be
extended to later types of stars. Particularly
important extensions of the application of the system have
been presented by 
\citet{1970ApJS...22..117B} (for metal-poor giants),
\citet{1978bs...symp..145G} (for red horizontal branch stars),
\citet{1984A&AS...57..443O} (for G and K dwarf stars),
\citet{1989A&A...221...65S} (for metal-poor stars),
\citet{1994AJ....107.1577A} (for giants), and
\citet{2007AJ....134.1777T} (for G and K dwarf stars).  The theoretical
foundations of these extension can be found in, e.g.,
\citet{1978A&AS...34..229B} and \citet{1979A&A....74..313G}, and
more recently \citet{2009A&A...498..527O}.  Applications to yellow
super-giants have also been successful \citep[see,
e.g.,][]{1993AJ....106.2516A}.

The colour-index $(b-y)$ is relatively unaffected by blanketing
effects and can thus be used to measure the stellar temperature (if
the reddening is known). Recent examples of colour-temperature
calibrations are given for dwarf stars by \citet{1996A&AS..117..227A},
and for giant stars by
\citet{1999A&AS..140..261A}. \cite{2005ApJ...626..465R} provide
calibrations for both giant and dwarf stars.

In contrast, the $m_1$ index is designed specifically to measure the
amount of blanketing in a region around 410\,nm
\citep[e.g.,][]{1975AJ.....80..955C} or as originally stated by
\citet{1963QJRAS...4....8S} is ``a colour difference that is a measure
of the total intensity of the metal lines in the $v$-band''. It is
thus sensitive to the total amount of metals present in the stellar
atmosphere. However, it was soon recognised that these metallicity
lines in population\,I stars are strong enough to depend mainly on
microturbulence ($\xi_t$) and less on metallicity. It was later shown
that $\xi_t$ is not a free parameter and hence the dependence still
prevails \citep[see, e.g., discussion in][]{1972A&A....19..261G}.
Because of the properties of the $m_{1}$ index it can be used to
derive metallicities for a variety of late-type stars (e.g., F to K
and V to III).  Recent examples of metallicity calibrations include
for giants \citet{2000A&A...355..994H} and \cite{2007ApJ...670..400C},
and for dwarf stars \citet{1984A&AS...57..443O},
\citet{1989A&A...221...65S}, and \citet{2007A&A...475..519H} (see
Sect.\,\ref{Sect:met} for a more complete list).  The calibrations for
giant stars include only linear terms in the different indices and
none include $c_{\rm 1}$.  For dwarf stars, the relations are more
complex and less straightforward, including dependencies also, e.g.,
on the $c_{1}$ index and quadratic terms. The reliability of the
metallicity calibrations for dwarf stars is one of the main topics of
this paper.

Finally, the $c_1$ index is designed to measure the Balmer
discontinuity \citep{1963QJRAS...4....8S}. For early-type stars, B and
A, the $c_1$ index is a measure of the temperature but for later type
stars (F and G stars) it provides a measure of the surface gravity.
Hence, for stars with spectral class later than roughly A, already by
design this system is able to identify different types of stars in a
reliable way. This was, in fact, the main advantage of the system as
it was used in early applications. Note that the identification works
equally well if the reddening is known or if all stars can be assumed
to suffer from the same amount of reddening.  For stars with spectral
type later than A, it was possible, by measuring $(b-y)$ and $c_{1}$
and comparing to standard sequences, to determine an absolute
magnitude for the star once it had been classified
\citep[e.g.,][]{1963QJRAS...4....8S}. It thus became important to
develop standard sequences in the $c_{1}$ vs. $(b-y)$ diagram so that
stars could be reliably classified according to spectral class and
evolutionary stage.  We return to the issue of standard sequences
for late-type dwarf stars later in this paper.

The ability to classify stars at different evolutionary stages using
the $uvby$ system has been elaborated upon. For metal-poor stars,
\citet{2004A&A...422..527S} developed a finely tuned classification
scheme to identify main sequence, turn-off, blue stragglers, red
giant, horizontal branch and asymptotic branch stars (see
Fig.\,\ref{fig:scheme}).  \citet{2009A&A...506.1147A} used this
classification scheme to successfully trace the faint ($V\sim 21.1$)
horizontal branch of the Hercules dwarf spheroidal galaxy.

The scheme developed in \citet{2004A&A...422..527S} extends only to
about $(b-y)_0$\footnote{The subscript $0$ indicates that the
  photometry has been dereddened. In the following, we explicitly
  indicate which photometry has been dereddened and which has not. All
  metallicity and other calibrations are based on the star's ``true''
  colours, i.e., the dereddened photometry. However, the separation of
  dwarf and giant stars with the help of the $c_{\rm 1}$ index (see,
  e.g., Fig.\,\ref{fig:scheme}) is effective using photometry that has not
  been dereddened as long as both types of stars are represented and
  all stars are affected by the same amount of reddening. This is, for
  example, the case for the dwarf spheroidal galaxies.} of 0.4 for
dwarf stars and about 0.6 for giants. However, the ability of the
$uvby$ system to distinguish different evolutionary stages (for all
metallicities) improves as we move to redder
colours. A simple illustration of this is given in
Fig.\,\ref{fig:scheme}. In this figure, we reproduce the classification
scheme of \citet{2004A&A...422..527S} and overlay two sets of
isochrones by \citet{2006ApJS..162..375V}, which use the
temperature-colour transformation by \citet{2004AJ....127.1227C}
\citep[but see][for a critical discussion of the reliability of the
intermediate metallicity isochrones based on this temperature-colour
transformation]{2007AA...465..357F}.

Finally, $uvby$ photometry is often complemented with observations in
additional filters. In particular, many studies have been performed
using the $\beta$ index \citep[e.g.,][]{1996A&AS..117..317S}. For
late-type stars, this index provides a temperature estimate that is
essentially independent of reddening. However, the two filters
included in this index are both narrow or very narrow, hence for
large-scale studies of fainter stars observing times become
prohibitively long. Here we are therefore not concerned with the
$\beta$ index.

Other studies have also developed systems that use additionally
information, e.g., Ca\,II H and K photometry \citep[see,
e.g.,][]{1998AJ....116.1922A}. For the same reasons given for the
$\beta$ index, we do not address these systems but
rather consider only $uvby$, where, in terms of observing time,
 $u$ is by far the most expensive filter.

\section{Two catalogues}
\label{Sect:cats}

Before testing available metallicity and $\log g$ calibrations and
deriving new standard relations we will first detail how we selected
the stars used to perform these tasks.  Below we describe the
construction of two catalogues for dwarf stars, one with $uvby$
photometry only and one with both $uvby$ photometry and iron
abundances determined from high-resolution spectroscopy.

\subsection{Reddening}
\label{sect:reddening}

For both catalogues we  need to decide whether the photometry
  for the stars should be dereddened or not and which reddening map
  to use. We only consider stars that have parallaxes in
  the Hipparcos catalogue
  \citep{1997A&A...323L..49P,2007hnrr.book.....V} and use the same
  method to deredden the photometry in the two catalogues. In brief,
we assume that the dust in the Galactic disk can be modelled as a
thin exponential disk with a scale-height of 125\,pc \citep[following,
e.g.,][]{2000A&AS..145..473B,2002AJ....124..931B}. Since most of the
stars are nearby, they are inside this dust
disk. We reduce the extinction accordingly using

\begin{eqnarray}
\label{EqExtScaling}
  E(B-V)_{star} = [ 1 - \exp (- |d \sin b| / h) ] \cdot E(B-V)_{\rm LOS},
\end{eqnarray}

\noindent
where $E(B-V)_{\rm LOS}$ is the full colour-excess along the line of
sight (LOS) taken from the dust maps by \citet{1998ApJ...500..525S},
$d$ is the distance \citep[here we use the parallaxes from the new
reduction of the Hipparcos catalogue of][]{2007hnrr.book.....V}, $b$
is the galactic latitude, and $h$ is the scale-height of the thin dust
disk (taken to be 125\,pc, see above).

Following, for instance, \citet{2004A&A...418..989N} we assume that
stars with $E(B-V)$ below $0.02$ are un-reddened and do not apply any
dereddening to the photometry for these stars. We discuss the
implications of this in Sect.\,\ref{sect:dwarftest}.

Several studies have noted that the dust maps of
\citet{1998ApJ...500..525S} overpredict $E(B-V)$ when $E(B-V)>0.15$
\citep[see,
e.g.,][]{1999ApJ...512L.135A,2002AJ....124..931B,2007AJ....134..698Y}.
Our catalogues are dominated by nearby stars with low $E(B-V)$.  For
the spectroscopic catalogue discussed in Sect.\,\ref{Sect:speccat} and
used to test the metallicity calibration in
Sect.\,\ref{sect:dwarftest}, only two stars have $E(B-V)>0.15$. In the
photometric catalogue used to trace dwarf-star sequences in
Sect.\,\ref{sect:seq}, there are 38 of 3645 stars that have
$E(B-V)>0.15$.  Since so few stars are affected by a possible
overprediction of the reddening we chose not to apply any corrections
to the reddening values found from the map by
\citet{1998ApJ...500..525S}.

To deredden the $uvby$ photometry we use the relation for
$A_{\lambda}/E(B-V)$ from Table\,6 (Col. 8) in Appendix B of
\citet{1998ApJ...500..525S}. For individual magnitudes, this amount to
$x_0=x - E(B-V)\cdot k_x$, where $x$ is any of $uvby$ and $k_x=$
5.231, 4.552, 4.049, and 3.277 for $uvby$, respectively, and the
subscript $0$ corresponds to the dereddened photometry.

\subsection{The photometric catalogue}
\label{photcat}

\begin{figure}   
\centering
\includegraphics[width=9cm]{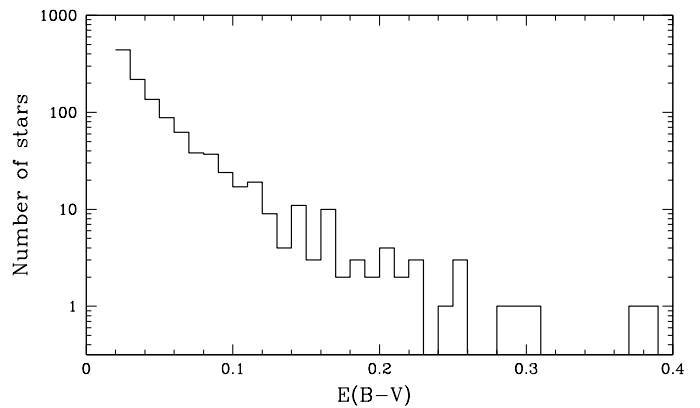}
\caption{Distribution of $E(B-V)$ for our photometric catalogue (see
  Sect.\,\ref{photcat}). There are 2502 stars with $E(B-V)<0.02$,
which are not shown.}
\label{fig:EBV}
\end{figure}

\begin{figure*}   
\centering
\includegraphics[width=9cm]{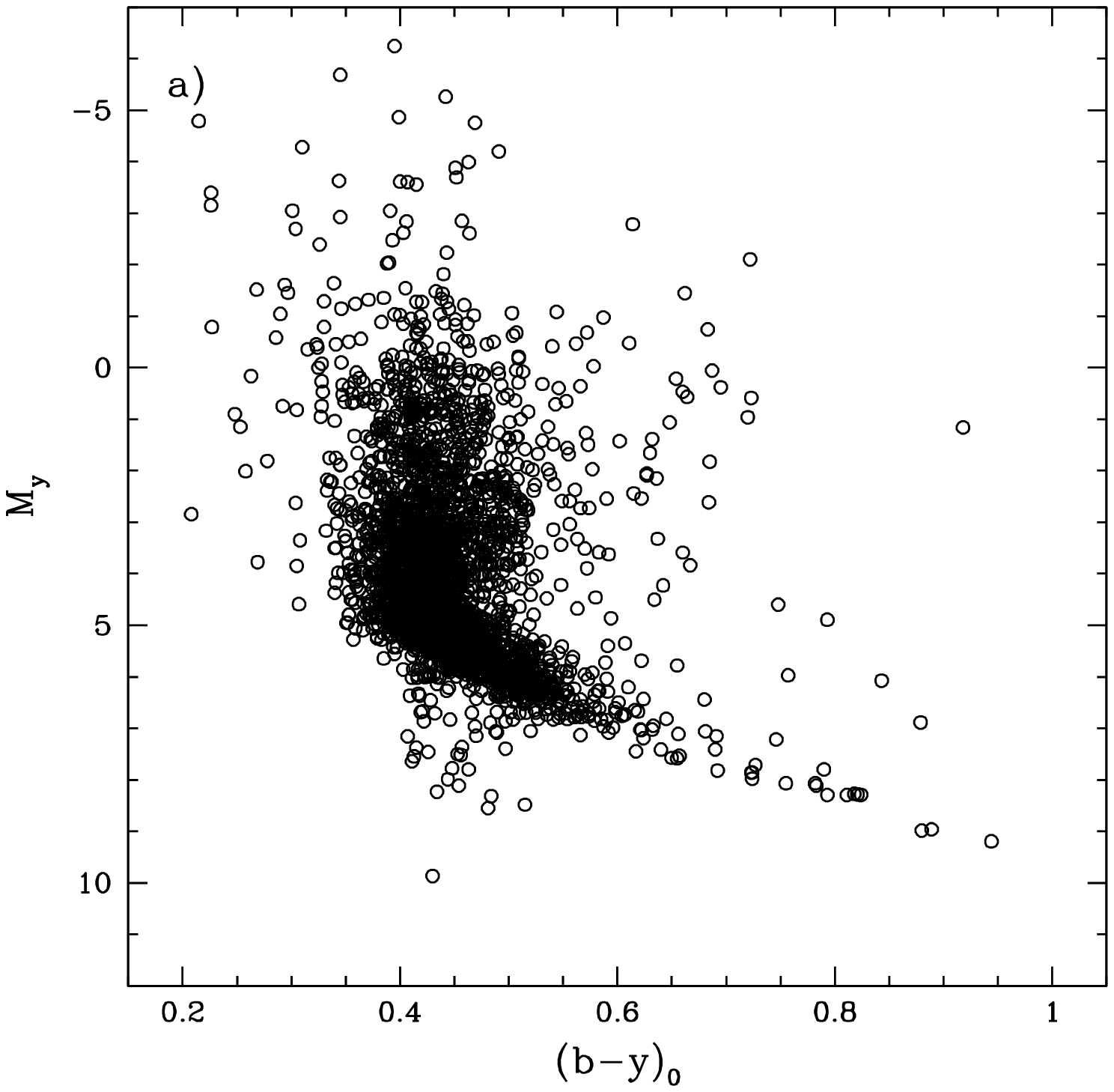}\includegraphics[width=9cm]{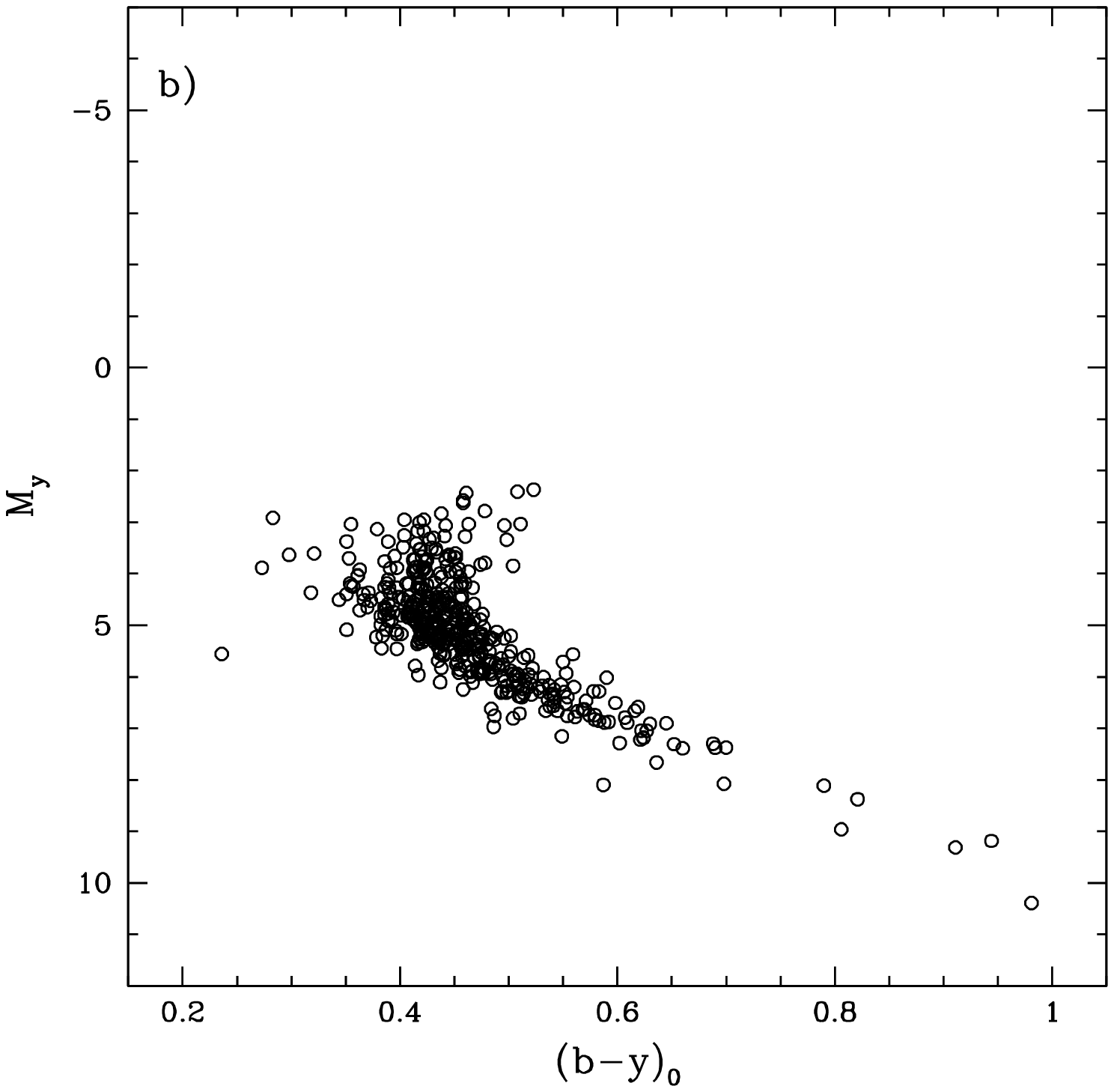}
\caption{{\bf a)} HR diagram for the photometric catalogue of dwarf
  stars (see Sect.\,\ref{photcat}). {\bf b)} HR-diagram for the dwarf
  stars in the spectroscopic catalogue (see
  Sect.\,\ref{Sect:speccat}). }
\label{fig:hr}
\end{figure*}

The three studies by \citet{1993A&AS..102...89O},
\citet{1994A&AS..104..429O}, and \citet{1994A&AS..106..257O} represent
one of the largest homogeneous catalogues of high quality $uvby$
photometry for nearby dwarf stars that also includes spectral
classification of the stars. The stars were classified into three main
groups: sub-giant stars (or the BAF group), giant stars (or the GKIII
group), and dwarf stars (or the GKV group). For our final catalogue,
we only include stars classified as dwarf stars by Olsen (the GKV
group).  Whenever a star has an entry in more than one of the three
studies we adopt the most recent set of measurements.

Dereddening was performed as described in Sect\,\ref{sect:reddening}.
The majority of the stars in \citet{1993A&AS..102...89O},
\citet{1994A&AS..104..429O}, and \citet{1994A&AS..106..257O} have
parallaxes from Hipparcos
\citep[][]{1997yCat.1239....0E,1997A&A...323L..49P,2007hnrr.book.....V}.
Stars that have no parallax from Hipparcos were simply discarded from
the photometric catalogue.  Known binary stars were excluded using the
SIMBAD database.  The resulting catalogue consists of 3645 dwarf
stars. Figure\,\ref{fig:hr} a shows the distribution of the stars in
the HR-diagram.

\subsection{The spectroscopic catalogue}
\label{Sect:speccat}

\begin{figure}   
\centering
\includegraphics[width=9cm]{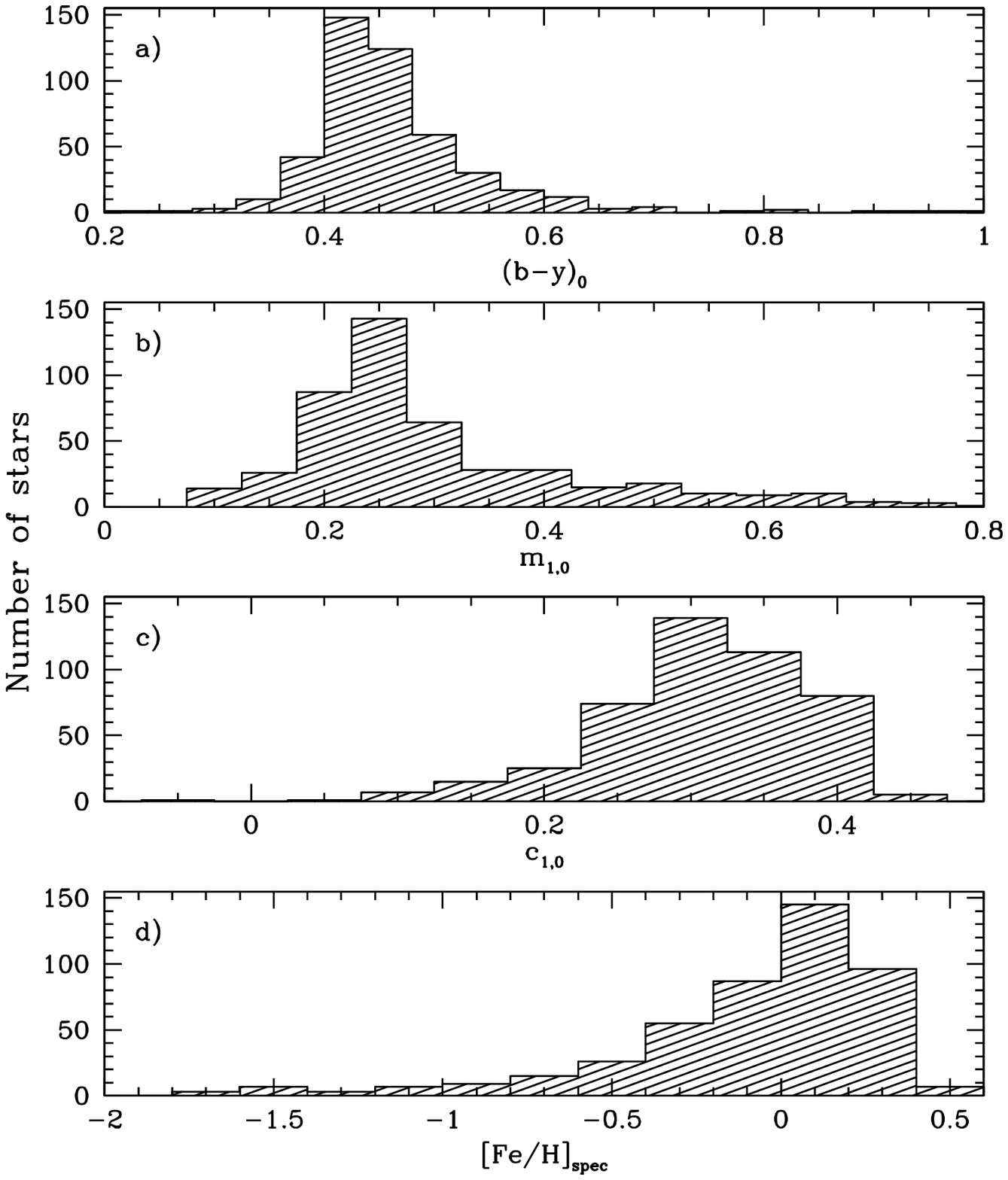}
\caption{Histograms showing the distribution of the photometric
  indices and [Fe/H] for the spectroscopic catalogue (Table\,B.\ref{FCC}
  and Sect.\,\ref{Sect:speccat}). {\bf a)} The number of stars as a
  function of $(b-y)_0$, {\bf b)} the number of stars as a function of
  $m_{1,0}$, {\bf c)} the number of stars as a function of $c_{1,0}$,
  and {\bf d)} the number of stars as a function of [Fe/H].}
\label{fig:speccat}
\end{figure}

\begin{table*}
  \caption{Coefficients for Eq.\,(\ref{twarogeqn}). 
   } 
\label{VFstrapping}
\center
\begin{tabular}{l r r c r r r r l}
\hline \hline
Study & Ref. & \# of stars & \# of stars with &   \multicolumn{1}{c}{$a$}  & \multicolumn{1}{c}{$b$} & \multicolumn{1}{c}{$c$} & \multicolumn{1}{c}{$d$} & \multicolumn{1}{c}{$\sigma$} \\
 & & & $(b-y)_{0}>0.6$\\
\hline
Favata et al. (1997)            & 2    &   46 & 1 & 1.0608 & --0.9662 & --0.7918 &   7.1781 & 0.07 \\
Feltzing \& Gustafsson (1998)   & 3    &   23 & 2 & 0.7903 & --0.8958 & --0.1252 &   3.9841 & 0.05 \\
Chen et al. (2000)              & 4    &   28 & 1 & 1.1759 & --2.0582 & --0.0767 &   8.2072 & 0.06 \\
Thor\'{e}n \& Feltzing (2000)   & 5    &   12 & 4 & 0.9918 &   0.0163 & --0.2020 &   0.8187 & 0.08 \\
Santos et al. (2001)            & 6    &   61 & 1 & 1.0405 & --0.9088 & --0.0586 &   3.6431 & 0.04 \\
Heiter \& Luck (2003)           & 7    &   75 & 0 & 0.8985 &   0.7027 & --0.1373 & --2.0303 & 0.05 \\
Yong \& Lambert (2003)          & 8    &    6 & 2 & 1.1258 &   2.0980 & --0.2534 & --6.3255 & 0.05 \\
Mishenina et al. (2004)         & 9    &   93 & 1 & 1.1434 & --1.8958 & --0.0888 &   7.5583 & 0.06 \\
Santos et al. (2004)            & 10   &  141 & 1 & 1.0098 & --1.2361 & --0.0838 &   5.0008 & 0.04 \\
Bonfils et al. (2005)           & 11   &   19 & 0 & 0.8761 & --0.5454 & --0.0422 &   2.2801 & 0.07 \\
Luck \& Heiter (2005)           & 12   &   65 & 6 & 0.9736 &   0.7346 & --0.0249 & --2.6251 & 0.06 \\
Santos et al. (2005)            & 13   &   64 & 7 & 1.0495 & --1.1400 & --0.0204 &   4.3450 & 0.04 \\
Woolf \& Wallerstein (2005)     & 14   &    8 & 6 & 1.0192 &   0.0846 & --0.4226 &   1.5298 & 0.04 \\
Sousa et al. (2006)             & 15   &   57 & 1 & 0.9360 & --0.9590 & --0.0805 &   3.9268 & 0.02 \\
\hline
\end{tabular}
\begin{list}{}{}
\item[] Column 1 lists the study that is being moved onto the
  \citet{2005ApJS..159..141V} system and Col. 2 the reference number
  used in Table\,B.\ref{FCC}.  Column 3 lists the number of stars in
  common with \citet{2005ApJS..159..141V}.  These are used to obtain
  the coefficients. Column 4 lists the number of stars redder than
  $(b-y)_0=0.6$.  Columns 5 to 8 list the coefficients used in
  Eq.\,(\ref{twarogeqn}), and Col. 9 lists the $\sigma$ for the
  difference between [Fe/H] in the study listed in Col. 1 and the
  [Fe/H] derived once the data have been put on to the
  \citet{2005ApJS..159..141V} system.  The difference is calculated in
  the sense [Fe/H]$_{\rm original}$ minus [Fe/H]$_{\rm corrected}$.
\end{list}
\end{table*}

To test the available metallicity and $\log g$ calibrations for dwarf
stars, we need a homogeneous catalogue of stars, which have both
$uvby$ photometry and spectroscopically determined [Fe/H]\footnote{We
  adopt the usual notation where [Fe/H] $ \equiv \log (N_{\rm Fe}
  /N_{\rm H})_{\star} - \log (N_{\rm Fe}/N_{\rm H})_{\sun}$ and use
  [Fe/H] exclusively for iron abundances determined from
  high-resolution spectroscopy. Metallicities determined from
  photometric calibrations will be either called just that or denoted
  [M/H].} and $\log g$. The [Fe/H] should preferably have been derived
using parallaxes, but ionisation equilibrium might also be acceptable
\citep[compare discussion in][]{2005A&A...433..185B}.

Because we place special emphasis on the redder dwarf stars, we
started our search by looking in the General Catalogue of Photometric
Data \citep{1997A&AS..124..349M} for stars with $(b-y)>0.6$., and found
such stars in four studies: \citet{1984A&AS...57..443O},
\citet{1988A&AS...73..225S}, \citet{1993A&AS..102...89O}, and
\citet{1994A&AS..104..429O}. 

As discussed above, in both \citet{1993A&AS..102...89O} and
\citet{1994A&AS..104..429O} the stars were classified according to
their evolutionary stages. In these two papers, we found 97 and 29
dwarfs stars, respectively, that are redder than
$(b-y)=0.6$. \citet{1984A&AS...57..443O} and
\citet{1988A&AS...73..225S} do not provide stellar classifications, we
therefore used the $c_{1}$ vs. $(b-y)$ diagram, compare
Fig.\,\ref{fig:scheme}, to exclude any obvious giant or early type
stars. We found 37 and 27 stars, respectively, in these two papers
which are likely dwarf stars with $(b-y)>0.6$.

In total, we found 190 probable dwarf stars with $(b-y)>0.6$. Upon
further inspection, it was found that 44 entries in this list were
duplications.  We decided to keep the most recent photometric
measurements when more than one set of measurements were available for
a given star.

Eleven additional stars were excluded (5 stars were marked as binaries
in one of the four papers and 6 stars had been observed to be
variables during those observing campaigns). Finally, we used the
SIMBAD database to identify any additional binaries, variables, or
unclassified stars. In total, 37 additional stars were excluded by
this check: 5 because they had no identification at all in SIMBAD,
being possible miss-identifications, 28 stars because they were
identified as variable, spectroscopic binaries, carbon stars, T Tauri
stars or peculiar; and 4 stars were giants.

For the remaining 98 dwarf stars with $(b-y)>0.6$, we searched the
literature for metallicity determinations using the SIMBAD and VizeR
databases \citep{2000A&AS..143...23O}. Fifty-seven of the stars had no
previous metallicity determinations at all. Thirteen stars had only
metallicities derived from photometry. We were thus left with 28 stars
with $(b-y)>0.6$ and [Fe/H] derived from high-resolution spectroscopy.

The 28 red dwarf stars were  found in 15 studies using high-resolution
spectroscopy to determine [Fe/H]:
\citet{2005ApJS..159..141V}, \citet{1997A&A...323..809F},
\citet{1998A&AS..129..237F}, \citet{2000A&AS..141..491C},
\citet{2000A&A...363..692T}, \citet{2001A&A...373.1019S},
\citet{2003AJ....126.2015H}, \citet{2003PASP..115...22Y},
\citet{2004A&A...415.1153S}, \citet{2004A&A...418..551M},
\citet{2005MNRAS.356..963W}, \citet{2005A&A...437.1127S},
\citet{2005AJ....129.1063L}, \citet{2005A&A...442..635B}, and
\citet{2006A&A...458..873S}. 

Several of these 15 studies also include large numbers of dwarf stars
bluer than $(b-y)=0.6$. This is especially true for
\citet{2005ApJS..159..141V}, which includes [Fe/H] for 1040 stars. Our
aim is to use this compilation to the test available calibrations for,
mainly, F- and G-type dwarf stars. We therefore decided that
\citet{2005ApJS..159..141V} should be the baseline for our
compilation.

Following \citet{2007AJ....134.1777T}, the [Fe/H] determined in the 15
spectroscopic studies (referred to as the 'original studies' below)
were moved onto the system of \citet{2005ApJS..159..141V} in the
following way. For each study, we took all stars (i.e., including
stars with $(b-y)<0.6$) in common between the study and
\citet{2005ApJS..159..141V} and performed a least squares fit to
determine the coefficients of the equation that transforms [Fe/H] onto
the metallicity-scale by \citet{2005ApJS..159..141V} given by

   \begin{eqnarray}
   \label{twarogeqn}
    {\rm [Fe/H]_{\rm VF05}} = a {\rm[Fe/H]} + b \log T_{\rm eff} + c \log g + d,
   \end{eqnarray}

\noindent
where [Fe/H] is the iron abundance, $T_{\rm eff}$ is the effective
temperature, and $\log g$ is the surface gravity derived in the
original study, that is being moved onto the metallicity-scale by
\citet{2005ApJS..159..141V}.  ${\rm [Fe/H]}_{\rm VF05}$ is the [Fe/H]
derived in \citet{2005ApJS..159..141V}.  The coefficients, $a$, $b$,
$c$, and $d$, together with the number of stars in common between
\citet{2005ApJS..159..141V} and the original study are listed in
Table\,\ref{VFstrapping}.

These transformations were then used to move all entries in the 15
studies onto the common metallicity scale. We then used the General
Catalogue of Photometric Data \citep{1997A&AS..124..349M} to find
$uvby$ photometry for these stars from the catalogues by Olsen and
Schuster and collaborators.  In total, 451 stars had [Fe/H] derived
from high-resolution spectroscopy and $uvby$ photometry. As
before, if a star had more than one set of $uvby$ measurements the
most recent was kept. The spectroscopic catalogue can be found in
Table\,B.\ref{FCC}.

Also for this catalogue we dereddened the photometry as described in
Sect.\,\ref{sect:reddening}. We recall that, the photometry for stars
with $E(B-V)<0.02$ were not corrected. The implications of this are
discussed in Sect.\,\ref{sect:dwarftest}. Fifty stars in the catalogue
have $E(B-V)>0.02$. The stellar distances are based on the reanalysed
Hipparcos parallaxes \citep{2007hnrr.book.....V}. Five stars
HD\,23261, HD\,69582, HD\,180890, HD\,192020, and PLX\,1219 do not
have Hipparcos parallaxes.  Their extinction was estimated using the
method of \citet{1983AJ.....88..623C} which is based on $VJK$
photometry.  These five stars do not have a Hipparcos number in
Table\,B.\ref{FCC}.

For two of the 15 studies, we note that no star redder than
$(b-y)_{0}>0.6$ remained after the dereddening (see
Table\,\ref{VFstrapping}). These studies were nevertheless kept in the
compilation as they provide valuable additional stars close to this
border.  Figure\,\ref{fig:hr} b. shows the distribution of the stars
in the HR-diagram and Fig.\,\ref{fig:speccat} shows the distributions
of both the Str\"omgren indices and [Fe/H] for the spectroscopic
catalogue.

\section{Metallicities from $uvby$ photometry - a critical evaluation}
\label{Sect:met}

\begin{table*}
\caption{Metallicity calibrations evaluated in Sect\,\ref{Sect:met}. }
\label{Cal-mean-sigma}
\center
\begin{tabular}{l l l l l r r r  l}
\noalign{\smallskip}
\hline \hline
\noalign{\smallskip}
Reference    & \multicolumn{2}{c}{$(b-y)_0$} &  \multicolumn{2}{c}{[Fe/H]} & \multirow{2}{*}{$<$[Fe/H]$-$[M/H]$>$}& \multirow{2}{*}{$\pm$}& \multirow{2}{*}{$\sigma$} & \multirow{2}{*}{Comment} \\
& min & max &  min & max & &  \\
\noalign{\smallskip}
\hline
\noalign{\smallskip}
Olsen (1984)       & 0.29 & 1.00      & --2.60 & 0.39  &  0.11 &$\pm$ &0.34 & Their Eq.\,(15)\\
                         & 0.514 & 1.000 & --2.60 & 0.39 &   0.04&$\pm$&0.39 & Their Eq.\,(15)\\
                        & 0.514 & 1.000 & --0.25 & 0.60 &   0.02&$\pm$&0.17 & Their Eq.\,(16) \\
Schuster \& Nissen (1989b)  & 0.22  & 0.38  & --3.5  & 0.2  &  \multirow{2}{*}{0.06}&\multirow{2}{*}{$\pm$}&\multirow{2}{*}{0.16} & F-type dwarfs \\
                                 & 0.37  & 0.59  & --2.6  & 0.4  &  && & G-type dwarfs \\
Haywood (2002)            &  0.22  & 0.59  & --2.0  & 0.5  &   0.00&$\pm$&0.18 & \\
Martell \& Laughlin (2002) & 0.288 & 0.591 & --2.0  & 0.5  &   0.05&$\pm$&0.13 &\\
Martell \& Smith (2004)    & 0.288 & 0.591 & --2.0  & 0.5  &   0.06&$\pm$&0.21 &\\
Nordstr\"{o}m et al. (2004)& 0.18  & 0.38  & --2.0  & 0.8  & --0.17&$\pm$&0.52 &\\
                                      & 0.44  & 0.59  & --2.0  & 0.8  &   0.13&$\pm$&0.08 & \\
Ram\'{i}rez \& Mel\'{e}ndez (2005a) & 0.19  & 0.35   & --3.5 & 0.4  & \multirow{2}{*}{0.04}&\multirow{2}{*}{$\pm$}&\multirow{2}{*}{0.14} & F-type dwarfs \\
                                      & 0.35  & 0.80   & --2.5 & 0.4  & && & G-type dwarfs \\
Holmberg et al. (2007)     & 0.24  & 0.63  & --1.00 & 0.37 &   0.08&$\pm$&0.16 &\\
\noalign{\smallskip}
\hline
\noalign{\smallskip}
\"{O}nehag et al. (2008)       & 0.22 & 0.59  &   --3.5 & 0.4  & 0.33&$\pm$&0.30 &\\
\noalign{\smallskip}
\hline
\end{tabular}
\begin{list}{}{}
\item[] Column 1 lists the reference for the calibration.  In Cols.\,2
  to 5 we quote the ranges, for for $(b-y)_0$ and [Fe/H], within which
  the calibrations is valid.  Column\,6 gives the mean difference
  between [Fe/H] and [M/H] and the associated $\sigma$. Column\,7
  provides additional comments.
\end{list}
\end{table*}

\begin{figure*}   
\center
\includegraphics[width=8.5cm]{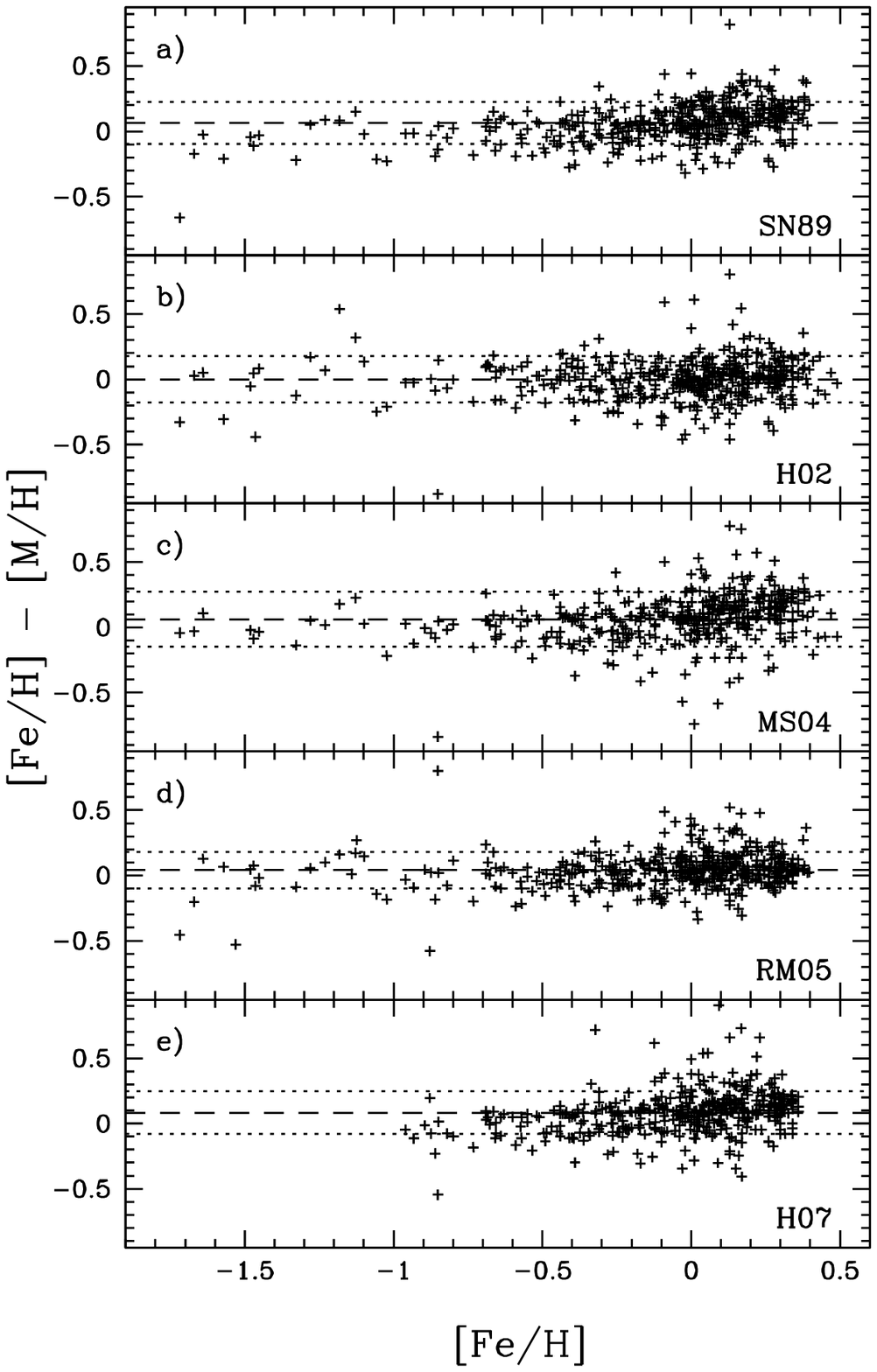}{\includegraphics[width=8.5cm]{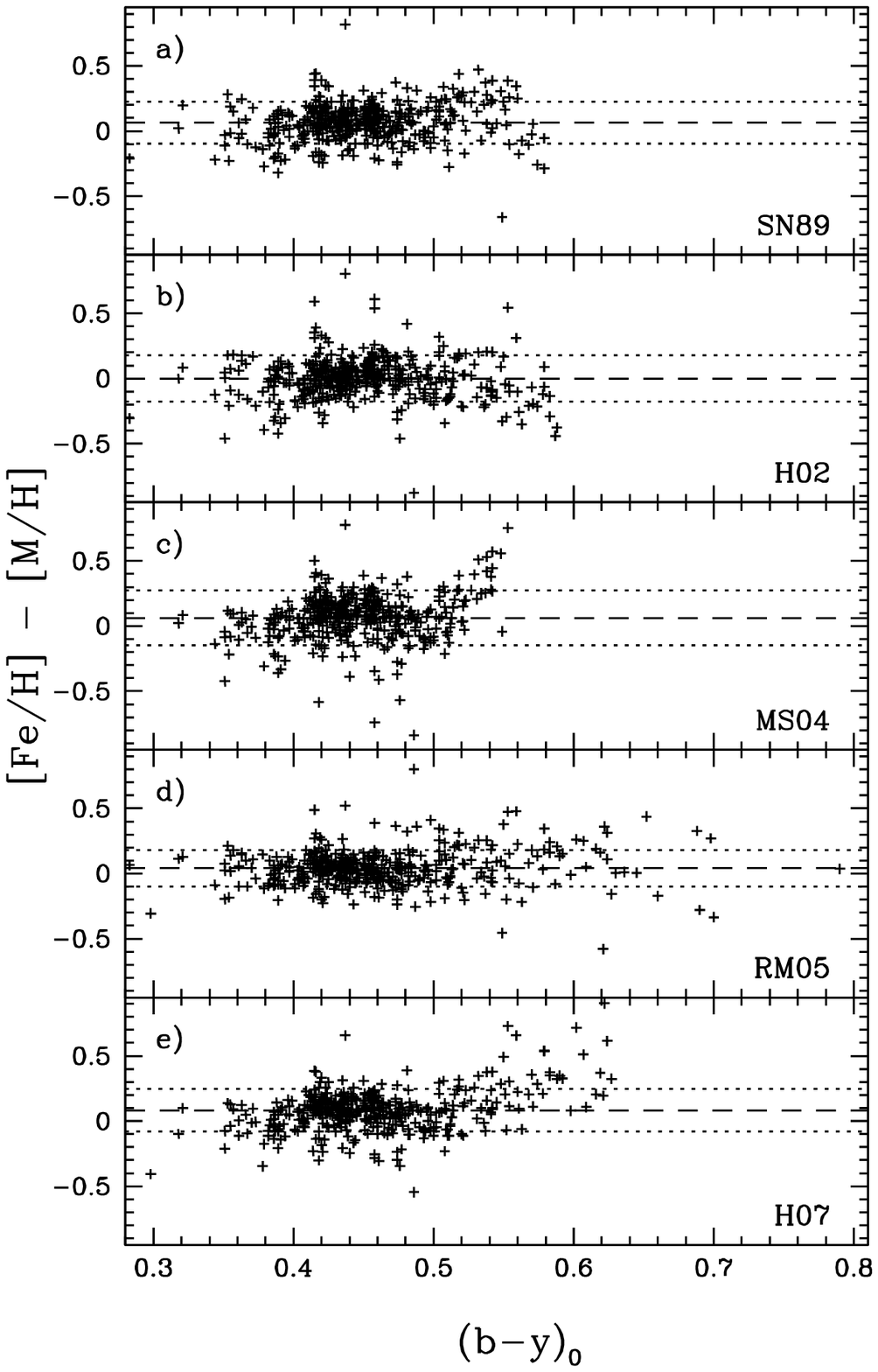}}
\caption{The difference between [Fe/H] and [M/H], derived from the
  calibrations listed in Table\,\ref{Cal-mean-sigma}, as a function of
  [Fe/H] (left hand panels) and $(b-y)_{0}$ (right hand panels).  The
  metallicity calibrations used are labelled as follows: SN89 for
  \citet{1989A&A...221...65S}, MS04 for \citet{2004PASP..116..920M},
  H07 for \citet{2007A&A...475..519H}, H02 for
  \citet{2002MNRAS.337..151H}, and RM05 for
  \citet{2005ApJ...626..446R}. The mean differences (dashed lines) and
  the $\sigma$ (dotted lines) are listed in Table\,\ref{Cal-mean-sigma}.}
\label{Fig:compfe}
\end{figure*}

\begin{figure*}   
\center
{\includegraphics[width=8.5cm]{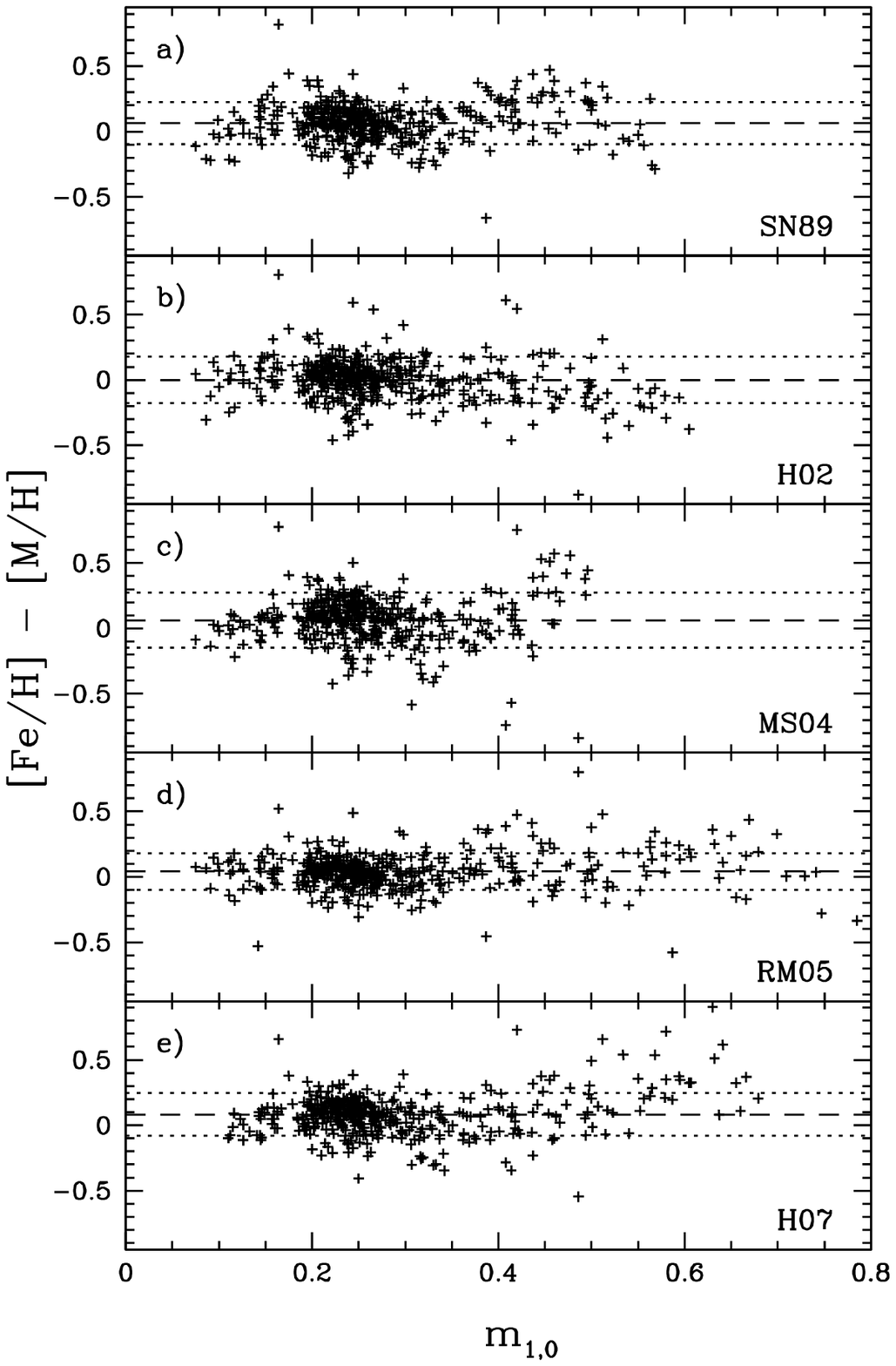}\includegraphics[width=8.5cm]{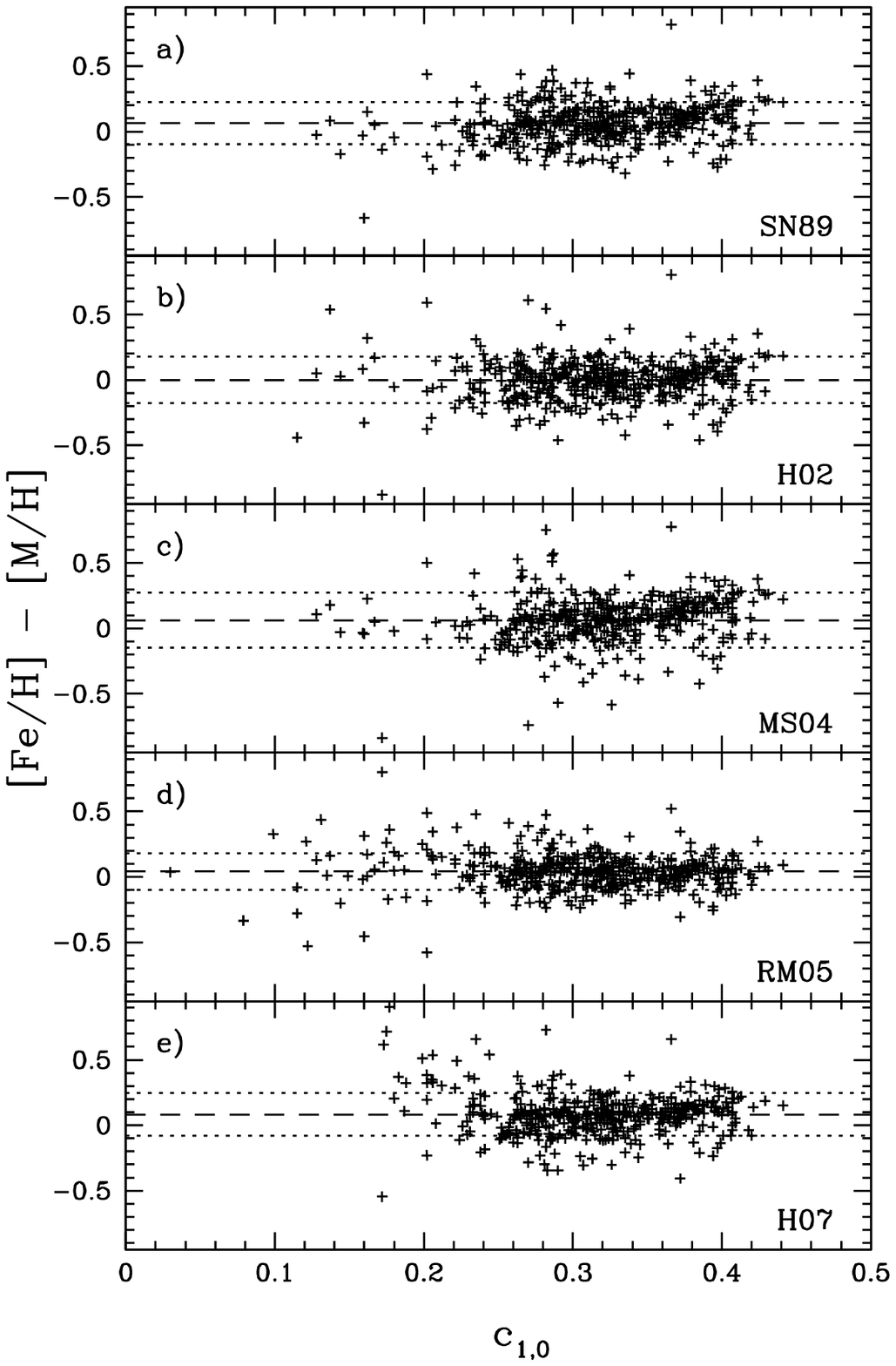}}
\caption{The difference between [Fe/H] and [M/H], derived from the
  calibrations listed in Table\,\ref{Cal-mean-sigma}, as a function of
  $m_{1,0}$ (left-hand panels) and $c_{1,0}$ (right-hand panels).  The
  metallicity calibrations used are labelled as follows: SN89 for
  \citet{1989A&A...221...65S}, MS04 for \citet{2004PASP..116..920M},
  H07 for \citet{2007A&A...475..519H}, H02 for
  \citet{2002MNRAS.337..151H}, and RM05 for
  \citet{2005ApJ...626..446R}. The mean differences (dashed lines) and
  the $\sigma$ (dotted lines) are listed in Table\,\ref{Cal-mean-sigma}.}
\label{Fig:comp}
\end{figure*}

\begin{figure}   
\centering
\includegraphics[width=9cm]{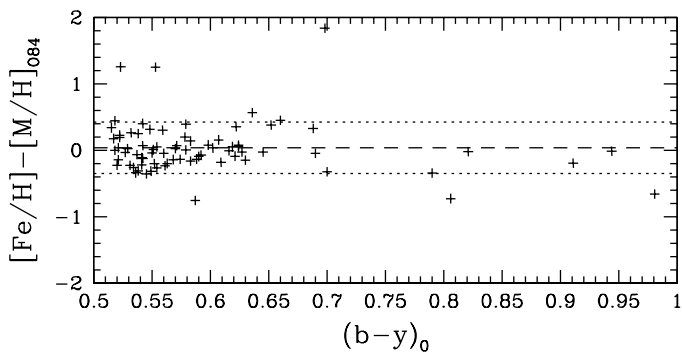}
\caption{The difference between [Fe/H] and [M/H] calculated using the
  calibration by \citet{1984A&AS...57..443O} ([M/H]$_{\rm O84}$). The
  comparison is made in the colour interval $0.514<(b-y)_0<1.000$.
  The mean difference is $0.03$\,dex (dashed line) with a $\sigma$ of
  $0.39$\,dex (dotted lines).}
\label{compO84}
\end{figure}

\begin{figure}   
\centering
\includegraphics[width=9cm]{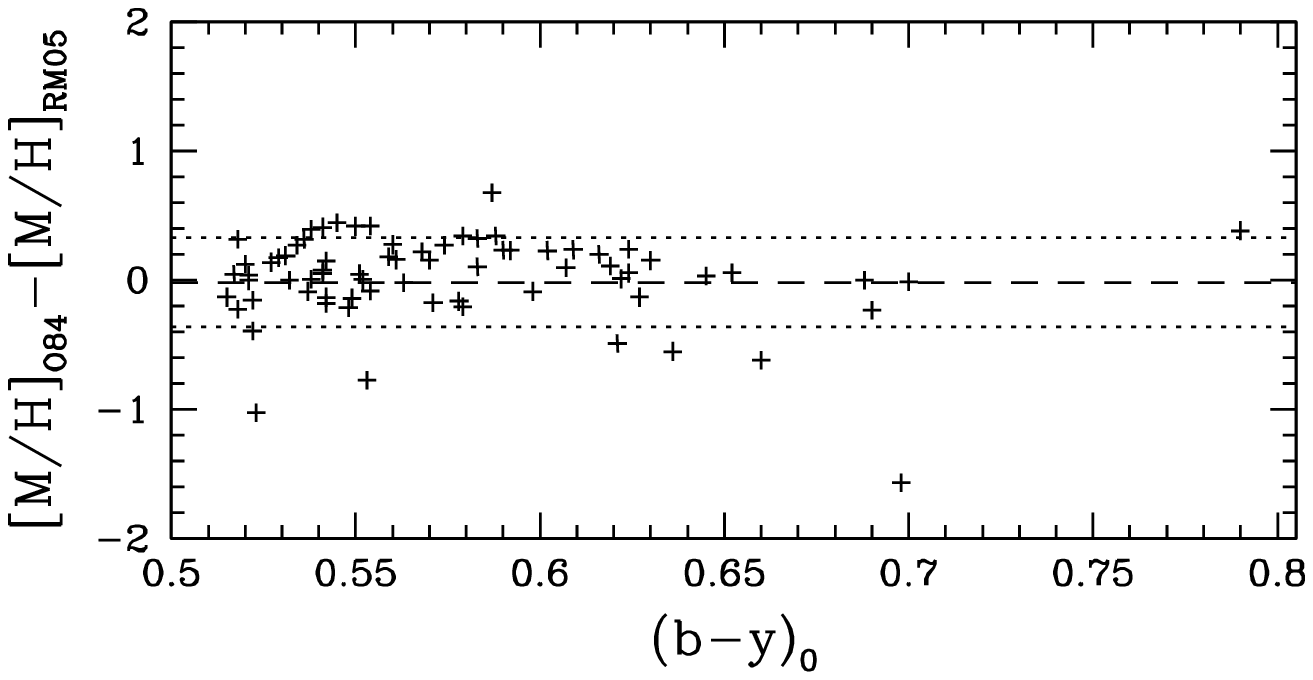}
\caption{A comparison of the [M/H] calculated using the calibration by
  \citet{1984A&AS...57..443O} ([M/H]$_{\rm O84}$) and the calibration
  by \citet{2005ApJ...626..446R} ([M/H]$_{\rm RM05}$). The comparison
  is made in the colour interval $0.514<(b-y)_0<0.800$.  The mean
  difference is $-0.02$\,dex (dashed line) with a $\sigma$ of
  $0.39$\,dex (dotted lines).}
\label{compR05O84}
\end{figure}

The literature contains many calibrations that make it possible to
derive metallicities from Str\"omgren photometry. Most of them are
empirical but theoretical investigations also exist \citep[see,
e.g.,][for a recent example]{2009A&A...498..527O}.  The early
metallicity calibrations
\citep{1964ApNr....9..333S,1975AJ.....80..955C,1984A&AS...57..443O}
were mostly based on how much the colour indices $m_{\rm 1}$ and
$c_{\rm 1}$ differed from a given standard relation, $\delta m_{\rm 1}
= m_{\rm 1,std}-m_{\rm 1,obs}$ and $\delta c_{\rm 1} =
c_{1,obs}-c_{1,std}$. 
The $m_{1,0} - (b-y)_0$ and $c_{1,0} - (b-y)_0$ relations used in
these calibrations are usually derived from observations of stars
belonging to the Hyades stellar cluster (for $m_{1,0} - (b-y)_0$) and
from field stars that are believed to be on the ZAMS (for $c_{1,0} -
(b-y)_0$).  \cite{1984A&AS...57..443O} provides an example of how the
preliminary standard sequences were derived.

More recent calibrations for dwarf stars have abandoned the use of
standard relations \citep[with the exception
of][]{2002MNRAS.337..151H} and derive [Fe/H] directly from the colour
indices $(b-y)_0$, $m_{1,0}$, and $c_{1,0}$
\citep{1989A&A...221...65S,1994A&AS..108..441M,
  2002MNRAS.337..151H,2002ApJ...577L..45M,2004PASP..116..920M,
  2004A&A...418..989N,2005ApJ...626..446R,2007A&A...475..519H}.  The
metallicity calibration by \citet{1984A&AS...57..443O} is the only
calibration that extends all the way to $(b-y)_{\rm 0} = 1.0$. No
calibration exists for dwarf stars redder than $(b-y)_{\rm 0}=1.0$.

In addition, some metallicity calibrations for dwarf stars require the
use of the $\beta$ index
\citep[e.g.,][]{1981A&A....97..145N,2007AJ....134.1777T} or additional
broadband photometry
\citep{2002MNRAS.336..879K,2005A&A...442..635B,1997MNRAS.286..617F}.
These will be not be investigated here.

Already \citet{1980ApJS...44..517B} found tentative evidence of a
metallicity dependence in the Str\"omgren indices for red giant stars
in the field, which was further investigated by
\citet{1994AJ....107.1577A}, who also derived metallicity dependent
standard sequences of red giants in the $c_{1,0}$ vs. $(b-y)_0$
diagram. Theoretical studies of the stellar colours of red giant stars
found that the colours show clear dependencies on both metallicity and
the amount of CNO in the atmospheres of the stars
\citep{1979A&A....74..313G}. \citet{2000A&A...355..994H} provided an
updated calibration based on both field stars and red giant branch
stars in globular clusters. However, the number of metallicity
calibrations derived directly for red giant stars is limited. The list
includes \citet{1980ApJS...44..517B}, \citet{1992A&A...253..359G},
\citet{1998AJ....116.1922A}, \citet{2000A&A...355..994H}, and
\citet{2007ApJ...670..400C}.

\subsection{A test of metallicity calibrations for dwarf stars}
\label{sect:dwarftest}

We now use our compilation of dwarf stars in Table\,B.\ref{FCC} to
evaluate how well various metallicity calibrations can reproduce
[Fe/H]. We investigate the  calibrations by
\citet{1984A&AS...57..443O}, \citet{1989A&A...221...65S},
\citet{2002MNRAS.337..151H}, \citet{2002ApJ...577L..45M},
\citet{2004PASP..116..920M}, \citet{2004A&A...418..989N},
\citet{2005ApJ...626..446R}, and \citet{2007A&A...475..519H}. The
common aspect of these calibrations is that they are relatively recent
and/or have been influential. In Sect.\,\ref{sect:ind}, we
discuss the ability of model atmospheres to reproduce the observed
Str\"omgren indices \citep{2009A&A...498..527O}.

We note that there are two metallicity calibrations in
\citet{1984A&AS...57..443O}.  Both calibrations depend on $\delta
m_{\rm 1}$, but while Eq.\,(16) in \citet{1984A&AS...57..443O} is a
linear equation in $\delta m_{\rm 1}$, Eq.\,(15) includes a quadratic
term in $\delta m_{\rm 1}$.  We investigate both calibrations.

Each calibration was applied only to stars with photometric indices in
the range where the calibration is valid (as indicated in the original
study).  In Table\,\ref{Cal-mean-sigma}, we list the mean difference
between [Fe/H] and [M/H]. As can be seen, the mean offset is, in most
of the cases, smaller than 0.1\,dex. Two calibrations yield larger offsets,
\citet{1984A&AS...57..443O} (full range of Eq.\,(15)) and
\citet{2004A&A...418..989N}. These calibrations also have some of the
largest scatters (compare Table\,\ref{Cal-mean-sigma} and
Figs.\,\ref{Fig:compfe} and \ref{Fig:comp}).

Figure\,\ref{Fig:compfe} compares the differences between [Fe/H] and
[M/H] as a function of [Fe/H].  There is a tendency for some of the
calibrations \citep[notably][]{1989A&A...221...65S,
  2004PASP..116..920M, 2007A&A...475..519H} to show a declining trend
towards lower [Fe/H]. The second and third of these calibrations also
show obvious trends with $(b-y)_{\rm 0}$ when $(b-y)_{\rm 0}\geq 0.5$.
Hence, even if these calibrations formally extend all the way to about
0.6, it is clear that there are shortcomings for the redder colours.

A comparison of the difference as a function of $m_{\rm 1,0}$
(Fig.\,\ref{Fig:comp}) indicates that two of the calibrations
\citep{2004PASP..116..920M,2007A&A...475..519H} fall short at the
redder end of the distribution. Finally, studying the difference as
function of $c_{\rm 1,0}$ we note that \citet{2007A&A...475..519H}
appears to show some real trend for the lower $c_{\rm 1,0}$ and that
\citet{2004PASP..116..920M}, and possibly \citet{2002MNRAS.337..151H},
show an overall trend such that the metallicity is underestimated at
low $c_{\rm 1,0}$ and overestimated at high $c_{\rm 1,0}$.

In summary, we find that both \citet{1989A&A...221...65S} and
\citet{2005ApJ...626..446R} perform very well in all four
comparisons. However, as \citet{2005ApJ...626..446R} covers a much
larger parameter space we would recommend it over
\citet{1989A&A...221...65S}, but again recall that in the regions
where the two calibrations overlap they perform equally well.

However, \citet{2005ApJ...626..446R} extends only to $(b-y)_{\rm
  0}=0.8$.  We therefore investigated the redder calibration of
\citet{1984A&AS...57..443O}. In Fig.\,\ref{compO84}, we compare the
[Fe/H] with the resulting [M/H] from that calibration, finding good
agreement.  In Fig.\,\ref{compR05O84}, we compare the results from
\citet{1984A&AS...57..443O} with the results from
\citet{2005ApJ...626..446R} as a function of $(b-y)_0$, and again find
close agreement. From these tests, we conclude that
\citet{1984A&AS...57..443O} provides an adequate extension of
\citet{2005ApJ...626..446R} for stars redder than $(b-y)_0=0.8$.

As discussed in Sect.\,\ref{sect:reddening}, if the reddening towards
a star is less than 0.02 we do not apply a reddening correction
(Table\,B.\ref{FCC}). The effect of this omission is small.  For
example, if we use the calibration of \citet{2005ApJ...626..446R} to
calculate [M/H] and assume that stars with $E(B-V)<0.02$, have an
$E(B-V)=0.02$ the mean difference between [Fe/H] and [M/H] changes
from 0.041$\pm$0.140 to --0.003$\pm$0.148. The trends with [Fe/H] and
the photometric indices change very little. To the eye, it appears
that, e.g., for redder $(b-y)_0$ the scatter increases.  Similar
trends are seen for the other indices.

\subsection{Metallicity calibrations for  red giant branch stars}
\label{sect:RGB}

\begin{figure*}   
\centering
\includegraphics[width=18cm]{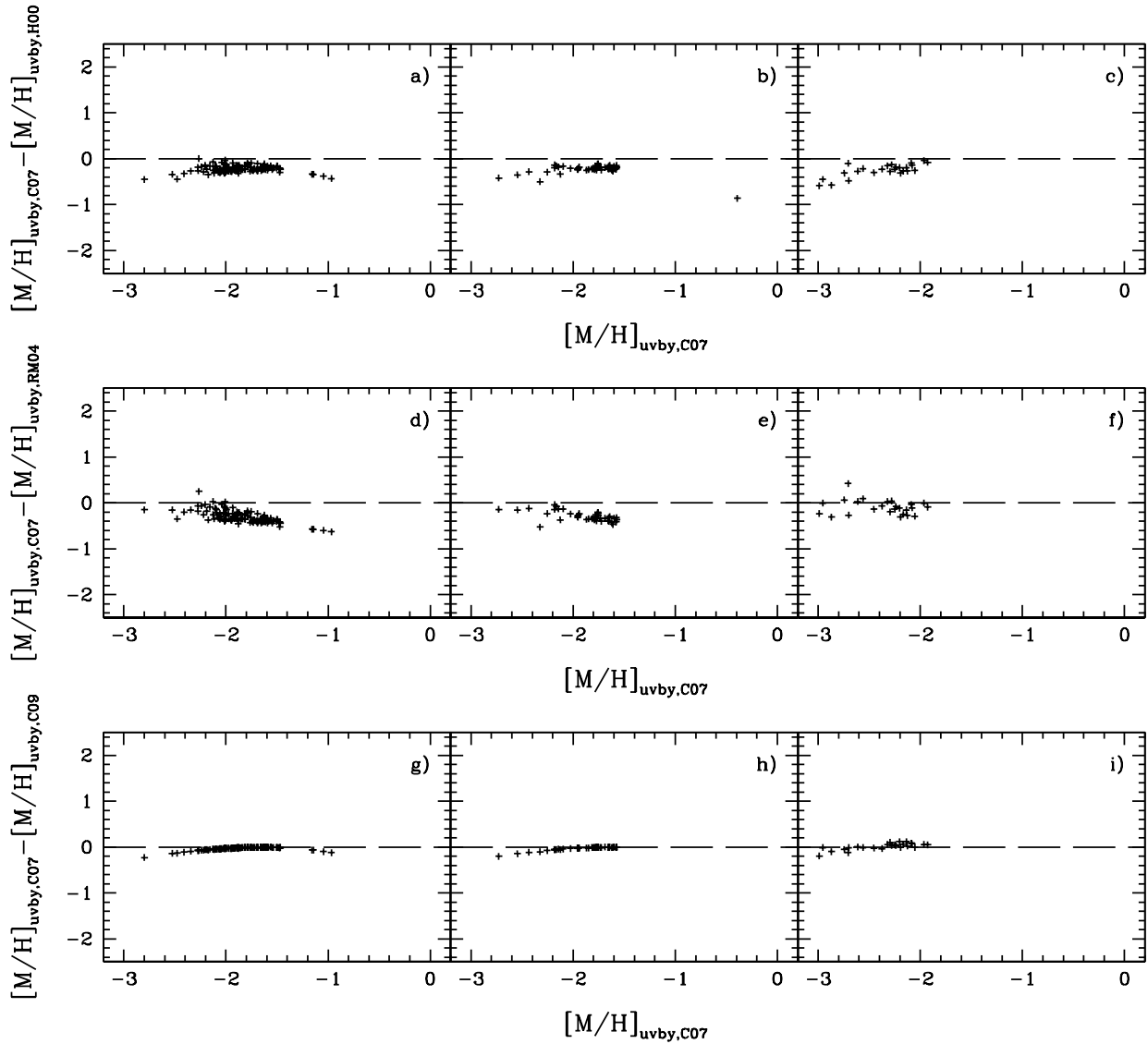}
\caption{A comparison of [M/H] derived for giant stars using the 
four most recent metallicity calibrations for $uvby$ photometry. We 
use the calibration of \citet{2007ApJ...670..400C} as the reference
for all comparisons. Panels a, d, and g shows the data for 
stars in the Draco dwarf spheroidal galaxy, panels b, e, and h
the data for stars in the Sextans dwarf spheroidal galaxy, and
panels c, f, and i data for giant stars in the Hercules dwarf
spheroidal galaxy.}
\label{fig:uvbycalibsRGB}
\end{figure*}

\citet{2007AA...465..357F} undertook a detailed investigation of the
calibrations then available and found that the calibration of
\citet{2000A&A...355..994H} was by far the most successful when
comparing with high-resolution spectroscopy. However,
\citet{2007AA...465..357F} only gives a limited comparison of
metal-poor, faint red giant stars in the Draco dwarf spheroidal
galaxy. \citet{2004A&A...417..301R} undertook a comparison with field
giants in the Milky Way ranging from solar all the way down to
--2.5\,dex. They found that the \citet{2000A&A...355..994H}
calibration underestimated the intermediate metallicities but
overestimated the lowest metallicities when compared to the
spectroscopically derived iron abundances. Solar metallicities were
well reproduced. \citet{2004A&A...417..301R} provide a correction
formula to place the calibrations of \citet{2000A&A...355..994H} onto
the spectroscopic scale.  Since then, \citet{2007ApJ...670..400C}
presented a new, and very comprehensive, study of metallicities of red
giant stars and their iron abundance. This study used giant stars in
globular clusters as a reference for their
calibration. \citet{2007ApJ...670..400C} used the more metallicity
sensitive index $(v-y)_0$, rather than $(b-y)_0$ used in
\citet{2000A&A...355..994H}. As discussed already by
\citet{1963QJRAS...4....8S}, the position of the $v$ filter provides a
measure of the total decrement due to the presence of metallicity
lines. We refer the reader to \citet{2007ApJ...670..400C} and
\citet{2009ApJ...706.1277C} \citep[which provides an update
to][]{2007ApJ...670..400C} for an extended discussion of the
derivation of their metallicity calibration for red giant stars.

Figure\,\ref{fig:uvbycalibsRGB}  compares  the different
calibrations applied to metal-poor red giant branch stars in
three nearby dwarf spheroidal galaxies (Draco, Sextans, and
Hercules). For this comparison, we use the calibration by
\citet{2007ApJ...670..400C} as reference.  Data for Draco and Sextans
are taken from Ad\'en et al. (in prep.) and data for Hercules from
\citet{2009A&A...506.1147A}. The data in Ad\'en et al. (in prep.) will
supersede those of \citet{2007AA...465..357F}.

The comparison between \citet{2007ApJ...670..400C} and
\citet{2000A&A...355..994H} shows the same banana shape noted by
\citet{2004A&A...417..301R}. This is most prominently seen for stars
in the Draco dwarf spheroidal galaxy. The difference between
\citet{2007ApJ...670..400C} and \citet{2009ApJ...706.1277C} is, as
expected, very small, the major difference being at the most
metal-poor end. Comparing \citet{2007ApJ...670..400C} and the
corrected \citet{2000A&A...355..994H} calibration by
\citet{2004A&A...417..301R} indicates that the calibration by
\citet{2004A&A...417..301R} would produce a more metal-poor as well as
more concentrated metallicity distribution function for the three
galaxies than if we used the calibration by
\citet{2007ApJ...670..400C}.  \citet{2009ApJ...706.1277C} use $(v-y)$
and $(u-v)$ for their calibrations; although these colours are more
sensitive to metallicity than $(b-y)$ they are also sensitive to CH
and CN. It appears, however, from the comparison carried out here,
that the choice of colours to use in the calibration might not be very
sensitive to the presence of molecules (at least for the giant stars
in the dwarf spheroidal galaxies). This should, however, be further
studied.

We note that all of these calibrations are poorly constrained at the
metal-poor end and more calibration data are required to improve the
calibrations.  Many studies currently target stars in the metal-poor
dwarf spheroidal galaxies and these data will thus become available
soon. We also note that to date only the calibration by
\citet{2004A&A...417..301R} extends to solar metallicity, which is an
important property for investigations where more metal-rich stars can
be expected.

Calibrations of $uvby$ photometry for red giant stars with
metallicities below --2\,dex have not been rigorously tested because
$uvby$ photometry and iron abundances based on high-resolution
spectroscopy for metal-poor field red giant have been largely
unavailable. However, a first look at data for Hercules (Ad\'en et
al. submitted) indicates that [Fe/H] based on high resolution
spectroscopy for about ten red giant branch stars infers lower iron
abundances than predicted from photometry using any of the metallicity
calibrations discussed here. In addition, preliminary comparisons with
data from \citet{2008ApJ...685L..43K} find the same result
\citep[][and Ad\'en et al., submitted]{2009A&A...506.1147A}.  This
conclusion is supported by a comparison with the new Draco data by
\citet{2009ApJ...701.1053C}, who obtained high-resolution spectroscopy
of eight of the brighter red giants in the Draco dwarf spheroidal. We
have Str\"omgren photometry for six of these stars. A comparison with
[M/H] derived using the calibration of \citet{2009ApJ...706.1277C}
gives a mean difference of --0.21\,dex and a $\sigma$ of 0.19\,dex. A
similar comparison but using the calibration by
\citet{2004A&A...417..301R} gives a mean difference of --0.25\,dex and
a $\sigma$ of 0.22\,dex. \citet{2009ApJ...701.1053C} noted a similar
difference when they compared their spectroscopic [Fe/H] with those
metallicities derived using the calibration of
\citet{2000A&A...355..994H}. We note that the most metal-poor stars in
the sample cause the largest deviations. Above about --2\,dex, the
comparison is very favourable.  As part of our ongoing work on
$uvby$ photometry for red giant stars in dwarf spheroidal galaxies, we
are evaluating the possibilities to extend current metallicity
calibrations for $uvby$ photometry to metallicities below --2\,dex.

We also compared the iron abundances of giant stars in the Draco dwarf
spheroidal galaxy determined in \citet{2009ApJ...701.1053C} with
metallicities derived from $ugriz$ photometry using the calibration of
\citet{ivezic2008}. The scatter is very large and some of the
metallicities are clearly incorrect. The differences are such that
even with a very large sample and considering, e.g., only the mean
metallicity of the sample the conclusions would be at best indicative
(see also Sect.\,\ref{sect:sdsscomp} below).

\subsection{A comparison with photometric metallicities from SDSS --
  both dwarf and giant stars}\label{sect:sdsscomp}

\begin{figure*}   
\centering
\includegraphics[width=9cm]{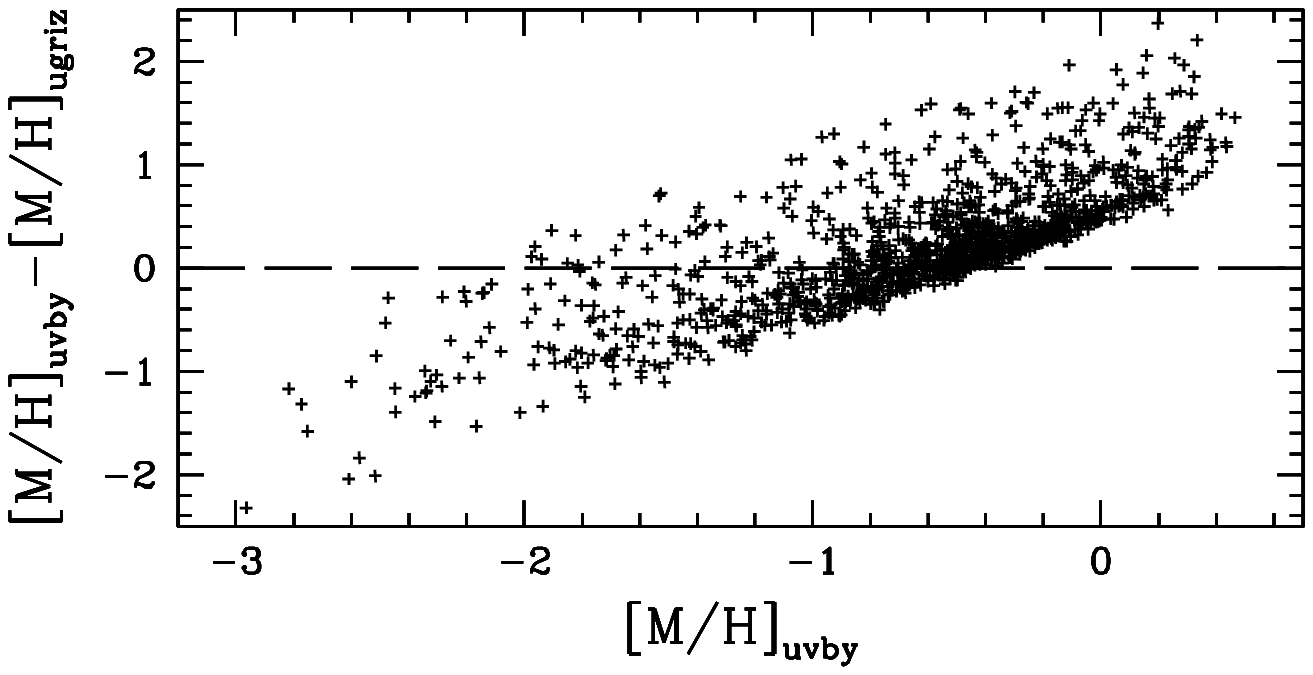}\includegraphics[width=9cm]{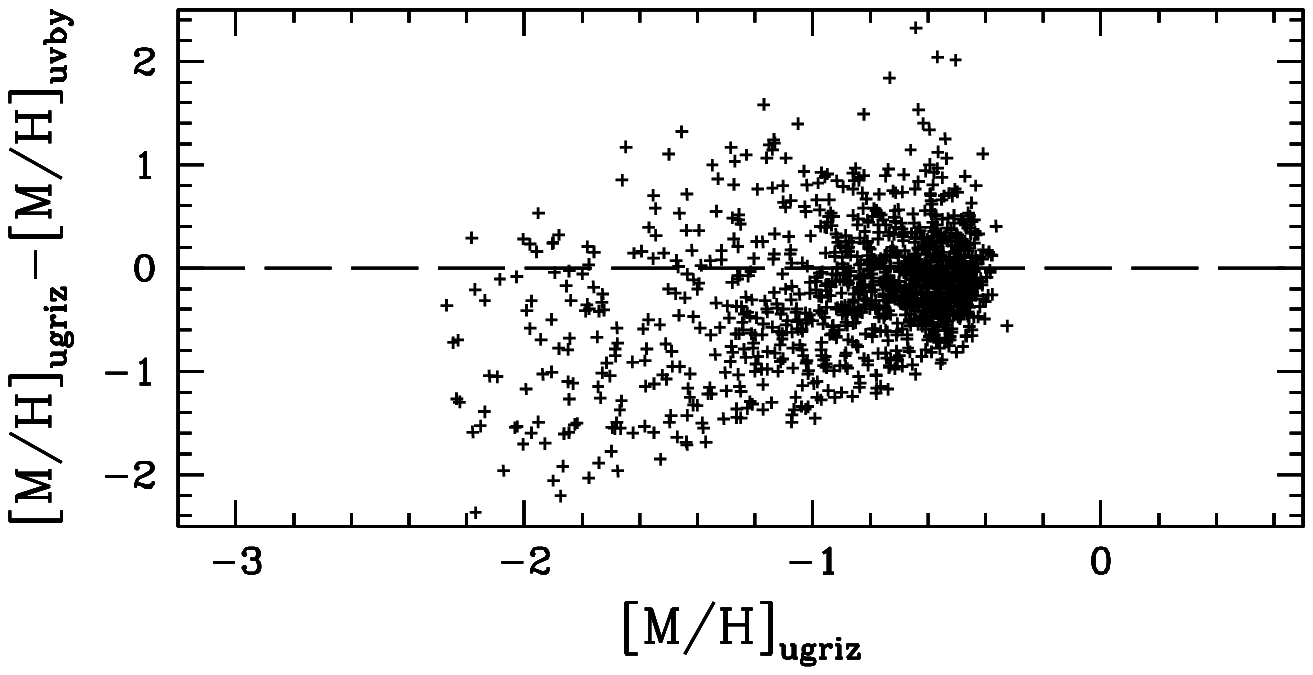}
\caption{A comparison of metallicities for dwarf stars derived from
  $uvby$ photometry ([M/H]$_{uvby}$) using the calibrations by
  \citet{2005ApJ...626..446R} and \citet{1984A&AS...57..443O} and
  metallicities derived from SDSS $ugriz$ photometry ([M/H]$_{ugriz}$)
  using the calibration of \citet{ivezic2008}.  The stars are along
  the lines-of-sight in the directions of the Hercules, Draco, and
  Sextans dwarf spheroidal galaxies. A full description of how these
  stars were selected will be provided in \'{A}rnad\'{o}ttir et
  al. (in preparation). All stars have $15<V_0<18.5$. The dashed line
  indicates a metallicity difference of zero.  On the abscissa the
  left-hand panel has [M/H]$_{uvby}$ and the right-hand panel has
  [M/H]$_{ugriz}$.}
\label{fig:sdssdwarfs}
\end{figure*}

\begin{figure*}   
\centering
\includegraphics[width=18cm]{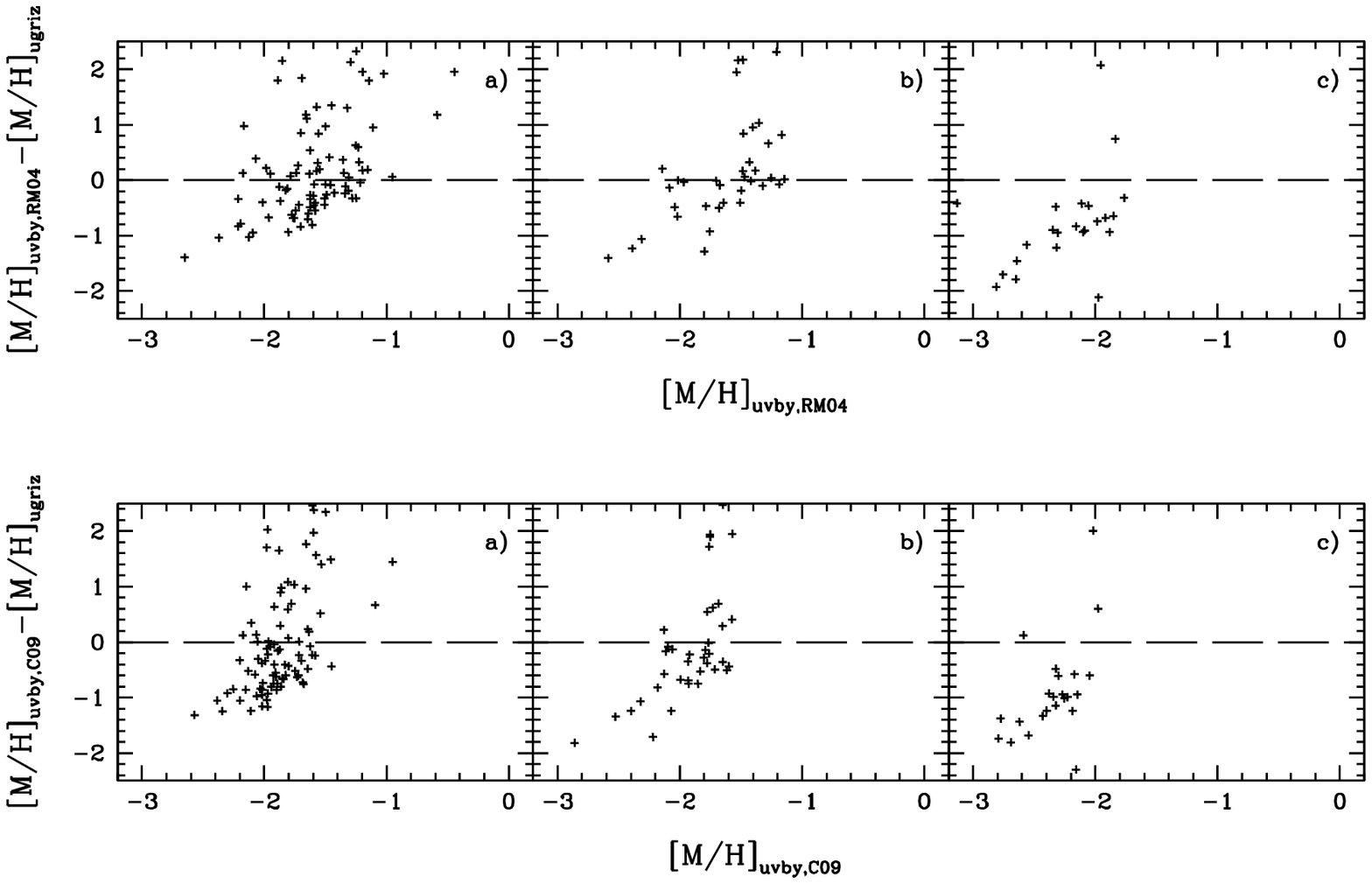}
\caption{A comparison of metallicities derived from $uvby$ photometry
  and $ugriz$ photometry \citep[using the calibration of][]{ivezic2008},
  respectively, for giant stars in dwarf spheroidal galaxies. The top
  panels uses the calibration by \citet{2004A&A...417..301R} and the
  bottom panels the calibration by \citet{2009ApJ...706.1277C} to
  obtain metallicities from $uvby$ photometry.  {\bf a.} Comparison
  for red giant branch stars in the Draco dwarf spheroidal galaxy
  ($uvby$ photometry: Ad\'en et al. in prep.). {\bf b.}  Comparison
  for red giant branch stars in the Sextans dwarf spheroidal galaxy
  \citep[$uvby$ photometry: Ad\'en et al. in prep. and
  ][]{Carina08}. {\bf c.}  Comparison for red giant branch stars in
  the Hercules dwarf spheroidal galaxy \citep[$uvby$
  photometry:][]{2009A&A...506.1147A}. }
\label{fig:sdssgiants}
\end{figure*}

The SDSS \citep{2000AJ....120.1579Y} is one of the most influential
studies covering a very large portion of the sky. The stellar part
contains not only $ugriz$ photometry but also spectra for a large
fraction of the objects. This and additional spectroscopic campaigns
provide [M/H] \citep[e.g.,][]{2008AJ....136.2050L}. It is of great
interest to attempt to derive calibrations to use the $ugriz$
photometry to provide stellar parameters and in particular [M/H]
\citep{ivezic2008}. If good calibrations can be obtained, much new
information about the thick disk and the halo can be obtained
\citep[see, e.g.,][]{2008Natur.451..216C}. Because of the potential
impact of SDSS, it remains important to test the calibrations against
independent metallicity measures. Our Str\"omgren photometry provides
an opportunity to do so for a large sample of fairly faint dwarf and
giant stars.

To perform these comparisons we use $uvby$ photometry of dwarf stars
from \'{A}rnad\'{o}ttir et al. (in preparation) and data for red giant
stars from \citet{fariathesis}, \citet{2007AA...465..357F},
\citet{Carina08}, \citet{2009A&A...506.1147A}, and Ad\'en et al. (in
prep.). The identification of dwarfs and giants is unambiguous for the
stars we use \citep[see
e.g.,][]{2007AA...465..357F,2009A&A...506.1147A}. The $ugriz$
photometry is from SDSS DR7 \citep{2009ApJS..182..543A}.

For the $uvby$ photometry, we use the calibrations of
\citet{2005ApJ...626..446R} and \citet{1984A&AS...57..443O} (for dwarf
stars) and \citet{2007ApJ...670..400C} (for giant stars) to calculate
[M/H]$_{uvby}$.  For the $ugriz$ photometry, we use the calibration of
\citet{ivezic2008} to calculate [M/H]$_{ugriz}$.  The comparisons
between [M/H]$_{uvby}$ and [M/H]$_{ugriz}$ are shown in
Figs.\,\ref{fig:sdssdwarfs} and \ref{fig:sdssgiants}.

We first note that for the dwarf stars in Fig.\,\ref{fig:sdssdwarfs}
there is good agreement at metallicities around --1\,dex, but that
agreement quickly deteriorates as we move to higher or lower
metallicities. There is some scatter but there is a distinctive linear
relation such that [M/H]$_{ugriz}$ is higher than [M/H]$_{uvby}$ at
low metallicities and the opposite is true for solar metallicities. At
solar metallicity, the offset is about 0.5\,dex and at [M/H]$_{uvby}$
= --2 the offset is about 1.5\,dex. Given the fairly extensive tests
that have been performed to compare [M/H]$_{uvby}$ to [Fe/H] derived
from stellar spectroscopy provided both in this study (see
Sect.\,\ref{Sect:met} and Figs.\,\ref{Fig:compfe} and \ref{Fig:comp})
and elsewhere,e these differences are a concern.

A comparison for metallicities for giants presented in
Fig.\,\ref{fig:sdssgiants} is perhaps even less encouraging. For $-3
<{\rm [M/H]}_{uvby} < -2$, there is a trend similar to that for the
dwarf stars, but at higher metallicities the relation appears to break
down completely.  We note that our datasets for the giant stars are
small but we believe that the more populated red giant branch of the
Draco dwarf spheroidal galaxy provides a fairly unambiguous result.
It is beyond the scope of this paper to explain these
differences. However, given the very large discrepancies in some cases
caution is required when using [M/H]$_{ugriz}$ to infer the properties
of the halo, where clearly many of the targets will be giants. Given
the overall scatter for giant stars of metallicity --2\,dex, a typical
halo metallicity, in Fig.\,\ref{fig:sdssgiants} these inferences must
be regarded as only indicative.

The comparison between [M/H]$_{uvby}$ and [Fe/H] from high resolution
spectroscopy indicates that [M/H]$_{uvby}$ is overestimated
(Sect.\,\ref{sect:RGB}). If [M/H]$_{uvby}$ were corrected to more
closely match [Fe/H], then the difference between [M/H]$_{ugriz}$ and
[M/H]$_{uvby}$ would be even greater.

\section{The $uvby$ system's ability to distinguish between dwarf,
  sub-giant, and giant stars -- New stellar sequences}
\label{sect:seq}

The Str\"omgren $uvby$ system has a proven ability to distinguish
between dwarf and giant stars for certain colour ranges. We have 
used this in two studies of dwarf spheroidal galaxies
 to remove the foreground contamination by Milky Way dwarf
stars \citep[][]{2007AA...465..357F,2009A&A...506.1147A}. In the most recent
paper, we showed that about 30\% of the stars that would otherwise be
assumed to be radial velocity members of the Hercules dwarf spheroidal
galaxy are instead foreground dwarf stars. This result has lead to a
re-evaluation of the minimum common mass for such galaxies \citep[compare,
e.g.,][]{2008Natur.454.1096S,2009ApJ...706L.150A}.

A significant drawback is that the stellar sequences merge around
$(b-y)_{0}=0.55$ in the $c_{1,0}$ vs. $(b-y)_0$ diagram.  For bluer
colours, the lower red giant branch almost meets the main sequence and
the subgiant branch and turn-off forms a loop (see
Fig. \ref{fig:scheme}).  \citet{2007AJ....134.1777T} investigated
whether a new index could be developed to distinguish between dwarf,
sub-giant, and giant stars at bluer colours. We also performed fairly
extensive tests with our datasets described in Sect.\,\ref{Sect:cats}
based on our studies of dwarf spheroidal galaxies
\citep{2007AA...465..357F,2009A&A...506.1147A}; we found that for
larger datasets the proposed new index does not appear to have the
desired ability to distinguish between the bluer dwarf, sub-giant, and
giant stars.

\subsection{Metallicity-dependent dwarf star sequences}
\label{Sect:dwarfseq}

\begin{figure*}   
\centering
\includegraphics[width=9cm]{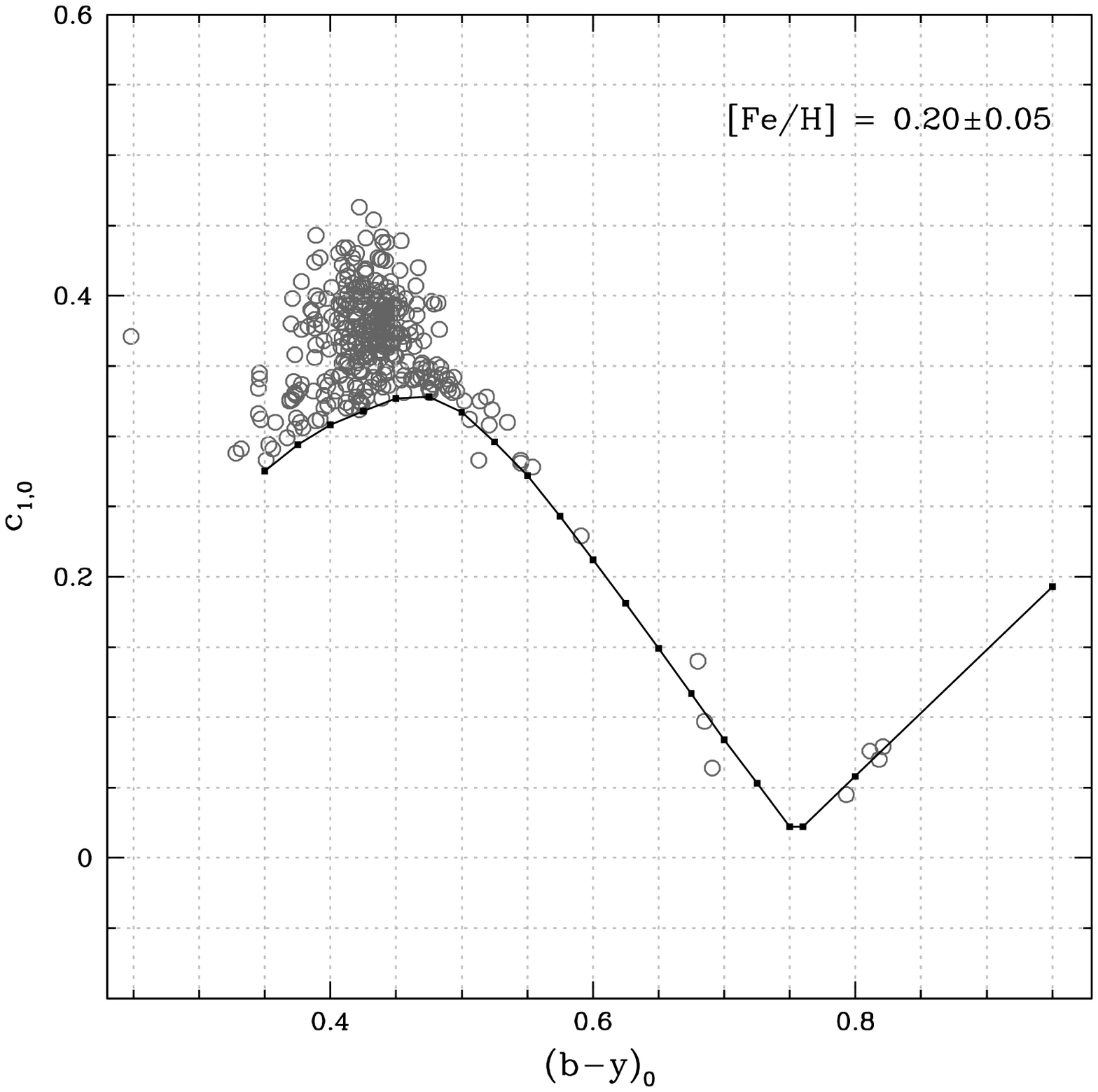}\includegraphics[width=9cm]{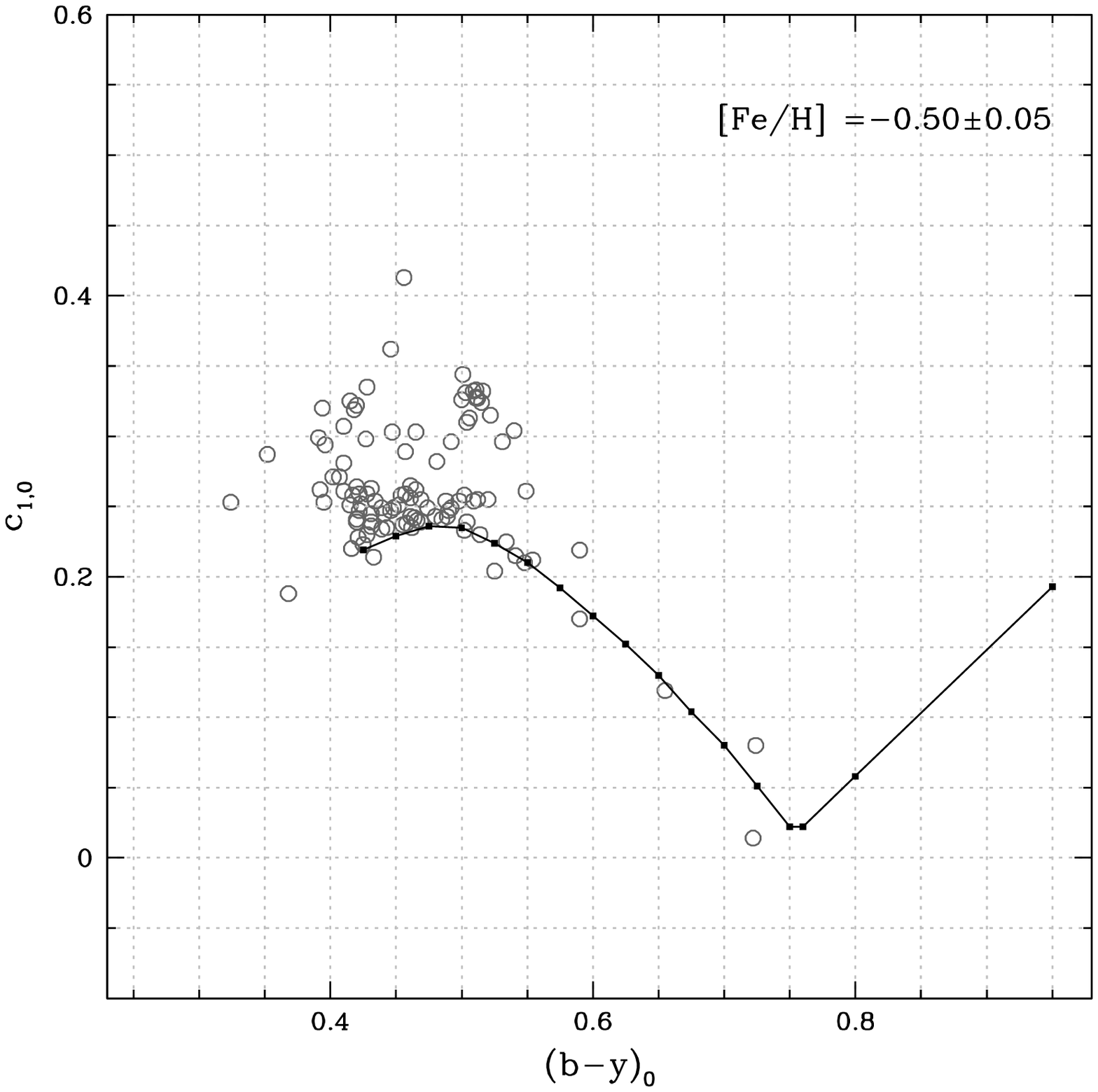}
\caption{Two examples of how the dwarf sequences in the $c_1$
  v.s. $(b-y)$ diagram, discussed in Sect.\,\ref{Sect:dwarfseq}, were
  established. The left hand panel shows dwarf stars with $0.15
  <$[M/H]$< 0.25$ and the right hand panel dwarf stars with $-0.55
  <$[M/H]$< -0.45$.  A complete set of similar plots for all
  metallicities can be found in Appendix\,\ref{ApxA} (available
  online). The standard relations are listed in
  Tables\,\ref{dwarfby:tab} and \ref{vyDwarfSeq}.}
\label{Fig:dwarfseq}
\end{figure*}

\begin{table*}
\caption{New metallicity-dependent sequences  for dwarf stars (see
  Sect.\,\ref{Sect:dwarfseq} and Figs.\,\ref{Fig:dwarfseq}, and
  \ref{Dby05} to \ref{Dbym10}).  For each range of metallicity (as
  indicated in the top two rows), we list the $c_{1,0}$ value for each
  $(b-y)_0$, as listed in the first column.}
\label{dwarfby:tab}
\center
\begin{tabular}{l r r r r r r r r r r r r r r}
\hline\hline
\noalign{\smallskip}
  [M/H]       &$0.50$    &$0.40$    &$0.30$    &$0.20$    &$0.10$    &$0.00$    &$-0.10$   &$-0.20$   &$-0.30$   &$-0.40$   &$-0.50$   &$-0.60$   &$-0.80$   &$-1.00$  \\
 $\pm$       &$0.05$    &$0.05$    &$0.05$    &$0.05$    &$0.05$    &$0.05$    &$0.05$   &$0.05$   &$0.05$   &$0.05$   &$0.05$   &$0.10$   &$0.15$   &$0.20$  \\
\noalign{\smallskip}    
 \hline
\noalign{\smallskip}
$(b-y)_0$&  $c_{1,0}$&  $c_{1,0}$&  $c_{1,0}$&  $c_{1,0}$&  $c_{1,0}$&  $c_{1,0}$&  $c_{1,0}$&  $c_{1,0}$&  $c_{1,0}$&  $c_{1,0}$&  $c_{1,0}$&  $c_{1,0}$&  $c_{1,0}$&  $c_{1,0}$ \\
\noalign{\smallskip}
\hline
\noalign{\smallskip}
0.400   &0.380  &0.345  &0.326  &0.308  &0.288  &0.271  &0.252  &0.244  &0.242  &0.228  &  -    &0.217  &  -    &  -     \\
0.425   &0.398  &0.363  &0.338  &0.318  &0.300  &0.284  &0.265  &0.252  &0.242  &0.229  &0.219  &0.211  &0.192  &0.140   \\
0.450   &0.395  &0.378  &0.347  &0.327  &0.310  &0.293  &0.278  &0.267  &0.249  &0.238  &0.229  &0.217  &0.194  &0.146   \\
0.475   &0.371  &0.371  &0.347  &0.328  &0.310  &0.298  &0.285  &0.274  &0.256  &0.247  &0.236  &0.227  &0.196  &0.154   \\
0.500   &0.341  &0.341  &0.336  &0.317  &0.305  &0.292  &0.280  &0.272  &0.259  &0.249  &0.235  &0.228  &0.196  &0.161   \\
0.525   &0.309  &0.309  &0.307  &0.296  &0.290  &0.276  &0.266  &0.262  &0.254  &0.243  &0.224  &0.220  &0.193  &0.165   \\
0.550   &0.277  &0.277  &0.276  &0.272  &0.267  &0.258  &0.245  &0.243  &0.238  &0.230  &0.210  &0.207  &0.186  &0.165   \\
0.575   &0.245  &0.245  &0.245  &0.243  &0.240  &0.235  &0.223  &0.221  &0.219  &0.211  &0.192  &0.190  &0.175  &0.163   \\
0.600   &0.213  &0.213  &0.213  &0.212  &0.210  &0.209  &0.198  &0.196  &0.195  &0.186  &0.172  &0.172  &0.162  &0.153   \\
0.625   &0.181  &0.181  &0.181  &0.181  &0.180  &0.180  &0.173  &0.171  &0.171  &0.162  &0.152  &0.152  &0.145  &0.139   \\
0.650   &0.149  &0.149  &0.149  &0.149  &0.149  &0.149  &0.145  &0.144  &0.144  &0.136  &0.130  &0.130  &0.126  &0.120   \\
0.675   &0.117  &0.117  &0.117  &0.117  &0.117  &0.117  &0.115  &0.115  &0.115  &0.109  &0.104  &0.104  &0.103  &0.097   \\
0.700   &0.084  &0.084  &0.084  &0.084  &0.084  &0.084  &0.084  &0.084  &0.084  &0.082  &0.080  &0.080  &0.077  &0.072   \\
0.725   &0.053  &0.053  &0.053  &0.053  &0.053  &0.053  &0.053  &0.053  &0.053  &0.052  &0.051  &0.051  &0.049  &0.047   \\
0.750   &0.022  &0.022  &0.022  &0.022  &0.022  &0.022  &0.022  &0.022  &0.022  &0.022  &0.022  &0.022  &0.022  &0.022   \\
0.760   &0.022  &0.022  &0.022  &0.022  &0.022  &0.022  &0.022  &0.022  &0.022  &0.022  &0.022  &0.022  &0.022  &0.022   \\
0.800   &0.058  &0.058  &0.058  &0.058  &0.058  &0.058  &0.058  &0.058  &0.058  &0.058  &0.058  &0.058  &0.058  &0.058   \\
0.950   &0.193  &0.193  &0.193  &0.193  &0.193  &0.193  &0.193  &0.193  &0.193  &0.193  &0.193  &0.193  &0.193  &0.193   \\
\noalign{\smallskip}
\hline
\end{tabular}
\end{table*}

\begin{table*}
\caption{New metallicity-dependent sequences for dwarf stars (see
  Sect.\,\ref{Sect:dwarfseq}).  For each range of metallicity (as
  indicated in the top two rows), we list the $c_{1,0}$ value for each
  $(v-y)_0$, as listed in the first column.}
\label{vyDwarfSeq}
\center
\begin{tabular}{l r r r r r r r r r r r r r r}
\hline\hline
 [M/H]       &$0.50$    &$0.40$    &$0.30$    &$0.20$    &$0.10$    &$0.00$    &$-0.10$   &$-0.20$   &$-0.30$   &$-0.40$   &$-0.50$   &$-0.60$   &$-0.80$   &$-1.00$  \\
 $\pm$       &$0.05$    &$0.05$    &$0.05$    &$0.05$    &$0.05$    &$0.05$    &$0.05$   &$0.05$   &$0.05$   &$0.05$   &$0.05$   &$0.10$   &$0.15$   &$0.20$  \\
\hline
$(v-y)_0$&  $c_{1,0}$&  $c_{1,0}$&  $c_{1,0}$&  $c_{1,0}$&  $c_{1,0}$&  $c_{1,0}$&  $c_{1,0}$&  $c_{1,0}$&  $c_{1,0}$&  $c_{1,0}$&  $c_{1,0}$&  $c_{1,0}$&  $c_{1,0}$&  $c_{1,0}$ \\
\hline
1.100   &0.381  &0.336  &0.321  &0.308  &0.292  &0.273  &0.260  &0.254  &0.245  &0.231  &0.226  &0.211  &0.190  &0.140\\
1.150   &0.393  &0.349  &0.332  &0.315  &0.299  &0.281  &0.271  &0.260  &0.247  &0.237  &0.230  &0.215  &0.195  &0.147\\
1.200   &0.400  &0.359  &0.339  &0.322  &0.306  &0.289  &0.278  &0.265  &0.253  &0.241  &0.233  &0.219  &0.198  &0.152\\
1.250   &0.395  &0.365  &0.342  &0.327  &0.309  &0.294  &0.283  &0.271  &0.257  &0.245  &0.236  &0.227  &0.201  &0.156\\
1.300   &0.385  &0.367  &0.345  &0.328  &0.310  &0.297  &0.285  &0.274  &0.260  &0.247  &0.237  &0.233  &0.202  &0.159\\
1.350   &0.375  &0.364  &0.345  &0.326  &0.308  &0.296  &0.284  &0.272  &0.261  &0.248  &0.237  &0.234  &0.203  &0.162\\
1.400   &0.363  &0.358  &0.340  &0.322  &0.305  &0.294  &0.282  &0.271  &0.260  &0.247  &0.236  &0.232  &0.202  &0.164\\
1.450   &0.350  &0.348  &0.332  &0.314  &0.298  &0.288  &0.276  &0.266  &0.255  &0.246  &0.232  &0.228  &0.200  &0.166\\
1.500   &0.337  &0.335  &0.323  &0.304  &0.288  &0.281  &0.268  &0.259  &0.249  &0.243  &0.225  &0.222  &0.196  &0.166\\
1.550   &0.324  &0.322  &0.312  &0.293  &0.279  &0.271  &0.260  &0.251  &0.242  &0.238  &0.219  &0.215  &0.192  &0.165\\
1.600   &0.309  &0.308  &0.301  &0.282  &0.268  &0.261  &0.250  &0.242  &0.232  &0.229  &0.211  &0.207  &0.186  &0.163\\
1.650   &0.296  &0.294  &0.289  &0.270  &0.257  &0.250  &0.239  &0.232  &0.221  &0.219  &0.202  &0.198  &0.177  &0.161\\
1.700   &0.279  &0.278  &0.275  &0.258  &0.245  &0.239  &0.227  &0.221  &0.208  &0.207  &0.192  &0.188  &0.168  &0.155\\
1.750   &0.264  &0.263  &0.261  &0.245  &0.232  &0.227  &0.215  &0.210  &0.195  &0.194  &0.181  &0.177  &0.159  &0.148\\
1.800   &0.246  &0.246  &0.245  &0.231  &0.218  &0.214  &0.202  &0.198  &0.183  &0.181  &0.170  &0.166  &0.150  &0.140\\
1.850   &0.229  &0.229  &0.228  &0.215  &0.203  &0.200  &0.188  &0.185  &0.170  &0.168  &0.158  &0.154  &0.140  &0.130\\
1.900   &0.211  &0.211  &0.210  &0.198  &0.187  &0.185  &0.174  &0.172  &0.157  &0.154  &0.146  &0.141  &0.129  &0.120\\
1.950   &0.191  &0.191  &0.190  &0.181  &0.171  &0.169  &0.159  &0.158  &0.143  &0.140  &0.133  &0.128  &0.118  &0.110\\
2.000   &0.169  &0.169  &0.169  &0.162  &0.154  &0.152  &0.143  &0.142  &0.129  &0.126  &0.120  &0.115  &0.107  &0.099\\
2.050   &0.146  &0.146  &0.146  &0.142  &0.136  &0.135  &0.126  &0.125  &0.116  &0.111  &0.106  &0.102  &0.095  &0.088\\
2.100   &0.122  &0.122  &0.122  &0.120  &0.116  &0.115  &0.108  &0.107  &0.101  &0.095  &0.091  &0.087  &0.083  &0.076\\
2.150   &0.097  &0.098  &0.098  &0.097  &0.097  &0.094  &0.090  &0.089  &0.085  &0.080  &0.077  &0.074  &0.070  &0.064\\
2.200   &0.074  &0.074  &0.074  &0.073  &0.073  &0.072  &0.070  &0.069  &0.067  &0.063  &0.061  &0.059  &0.056  &0.051\\
2.250   &0.050  &0.050  &0.050  &0.050  &0.050  &0.050  &0.049  &0.049  &0.048  &0.044  &0.042  &0.041  &0.040  &0.037\\
2.310   &0.022  &0.022  &0.022  &0.022  &0.022  &0.022  &0.022  &0.022  &0.022  &0.022  &0.022  &0.022  &0.022  &0.022\\
2.350   &0.120  &0.120  &0.120  &0.120  &0.120  &0.120  &0.120  &0.120  &0.120  &0.120  &0.120  &0.120  &0.120  &0.120\\
2.380   &0.200  &0.200  &0.200  &0.200  &0.200  &0.200  &0.200  &0.200  &0.200  &0.200  &0.200  &0.200  &0.200  &0.200\\
\hline
\end{tabular}
\end{table*}

\begin{table}
\caption{The upper envelope for dwarf stars in the solar neighbourhood.}
\label{upper:tab}
\center
\begin{tabular}{l r r r r r r r r r r r r r r}
\hline\hline
\noalign{\smallskip}
$c_{1,0}$ & $(b-y)_0$ & $(v-y)_0$ \\
\hline
\noalign{\smallskip}
  0.396 &0.350 &  0.895 \\
  0.423 &0.375 &  0.947 \\
  0.448 &0.400 &  1.010 \\
  0.458 &0.410 &  1.044 \\
  0.461 &0.430 &  1.085 \\
  0.450 &0.450 &  1.131 \\
  0.424 &0.470 &  1.205 \\
  0.385 &0.490 &  1.305 \\
\hline
\noalign{\smallskip}
\end{tabular}
\end{table}

Dwarf star sequences in the Str\"{o}mgren $c_{1,0}$ -- $(b-y_{0})$
plane were introduced for F-type dwarf stars by
\citet{1975AJ.....80..955C} and later extended to $(b-y)=1.0$ by
\citet{1984A&AS...57..443O}.  These sequences were drawn by hand
tracing the lower envelope of field stars in the relevant diagram.  No
attempts were made to investigate if the stellar sequences were
metallicity dependent, although this possibility was discussed already by
\citet{1964ApNr....9..333S}. It is clear, in the
$c_{1,0}$ vs. $(b-y_{0})$ diagram, when we compare the dwarf star sequence of
\citet{1984A&AS...57..443O} to the dwarf region for metal-poor stars,
given by \citet{2004A&A...422..527S}, that the metal-poor dwarf stars
have lower $c_{1,0}$ indices than the, mainly, solar metallicity stars
used to define the sequence in \citet{1984A&AS...57..443O}. This can
be seen, e.g., in Fig.\,\ref{fig:scheme}.

We are now in a position to extend the study of
\citet{1984A&AS...57..443O} and investigate the metallicity dependence
of dwarf star sequences in both the $c_{1,0}$ vs. $(b-y)_0$ diagram
and the $c_{1,0}$ vs $(v-y)_0$ diagram. For stars in our photometric
catalogue [M/H] were calculated (see Sect.\,\ref{photcat}) using the
metallicity calibrations by \citet{2005ApJ...626..446R} for dwarf and
subgiant stars with $(b-y)_0<0.80$ and the calibration by
\citet{1984A&AS...57..443O} for dwarf stars with $0.80<(b-y)_0<1.00$.

To trace the stellar (standard) sequences, we plotted $c_{1,0}$ vs.
$(b-y)_0$ and $c_{1,0}$ vs. $(v-y)_0$ for the dwarf stars, but  each time only
for a narrow range in metallicity. Following the procedure
in \cite{1984A&AS...57..443O}, we trace the lower envelope of the
stellar distribution in both the $c_{1,0}$ vs. $(b-y)_0$ and $c_{1,0}$
vs. $(v-y)_0$ diagrams.  This lower envelope is sensitive to
metallicity.  For $(b-y)_0 > 0.7$, all dwarf stars fall on a tight
relation without any dependence on metallicity. We used all stars
redder than $(b-y)_0 \sim 0.7$ to define the sequence up to $(b-y)_0
=1.0$. Our data set has no stars redder than
1.0. Figure\,\ref{Fig:dwarfseq} shows two examples of how these
tracings were done. Figures\,\ref{Dby05} to \ref{Dbym10} in
Appendix\,\ref{ApxA} show all tracings. The sequences are tabulated in
Tables\,\ref{dwarfby:tab} and \ref{vyDwarfSeq}.

Although we have extended the tracings to as blue colours as possible
in Figs.\,\ref{Dby05} to \ref{Dbym10}, it is clear that for colours
bluer than $(b-y)_0=0.4$ the data are not substantial enough in
quantity at any metallicity to provide a secure tracing. Moreover, we
use only stars classified as GKV in \citet{1993A&AS..102...89O},
\citet{1994A&AS..104..429O}, and \citet{1994A&AS..106..257O}, and
therefore exclude bluer main sequence stars. This exclusion is also
colour dependent because it depends on the metallicity of the
stars. Because of these limitations we refrain from showing the
tracings bluer than $(b-y)_0=0.4$ and $(v-y)_0 = 1.1$.

We also traced a global upper envelope for all dwarf stars.  This
upper envelope  is listed in
Table\,\ref{upper:tab}.

\subsection{The ability of the $ugriz$ photometric system to identify giant stars}

\begin{figure*}   
\centering
\includegraphics[height=6.2cm]{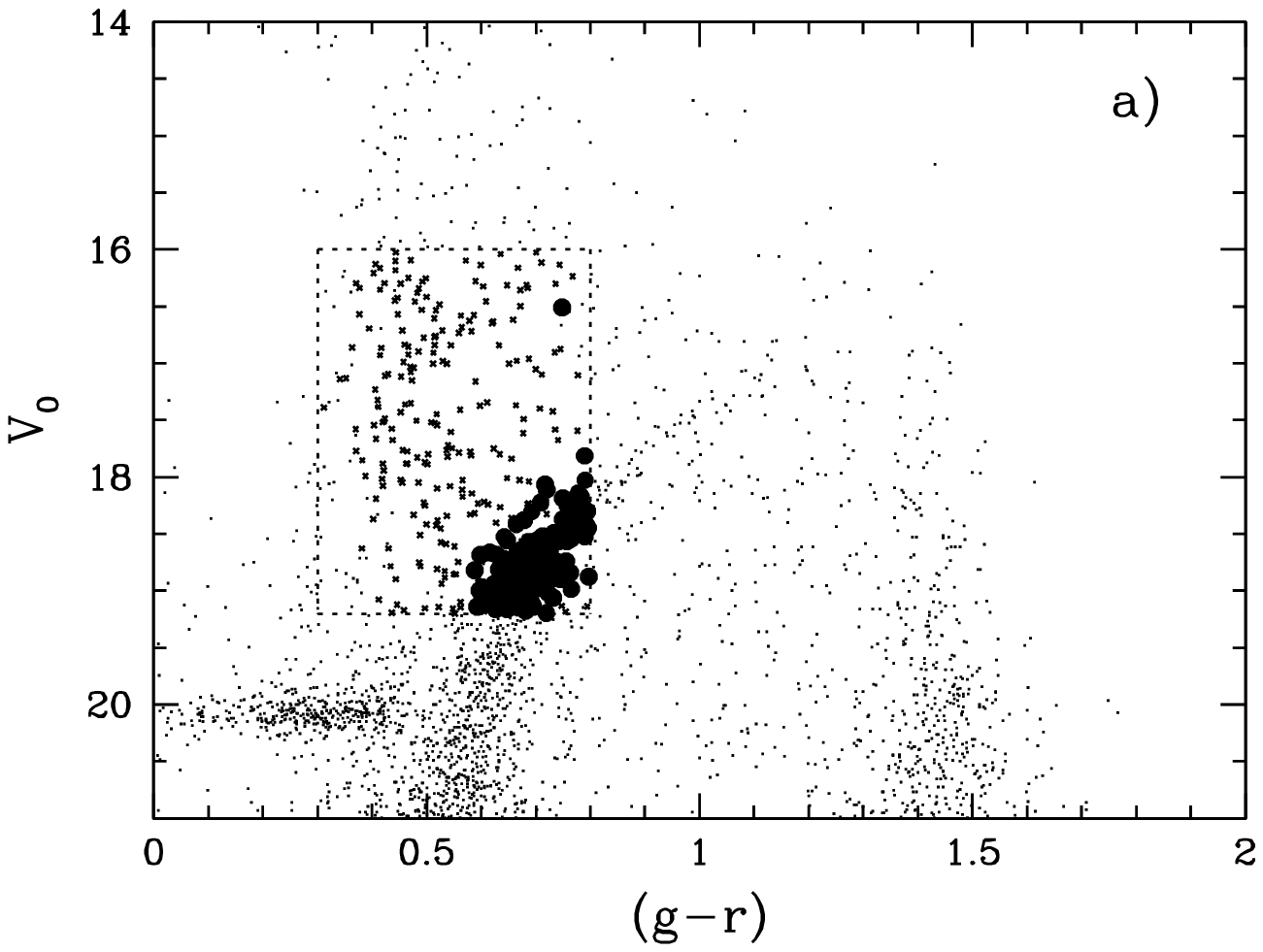}
\includegraphics[height=6.2cm]{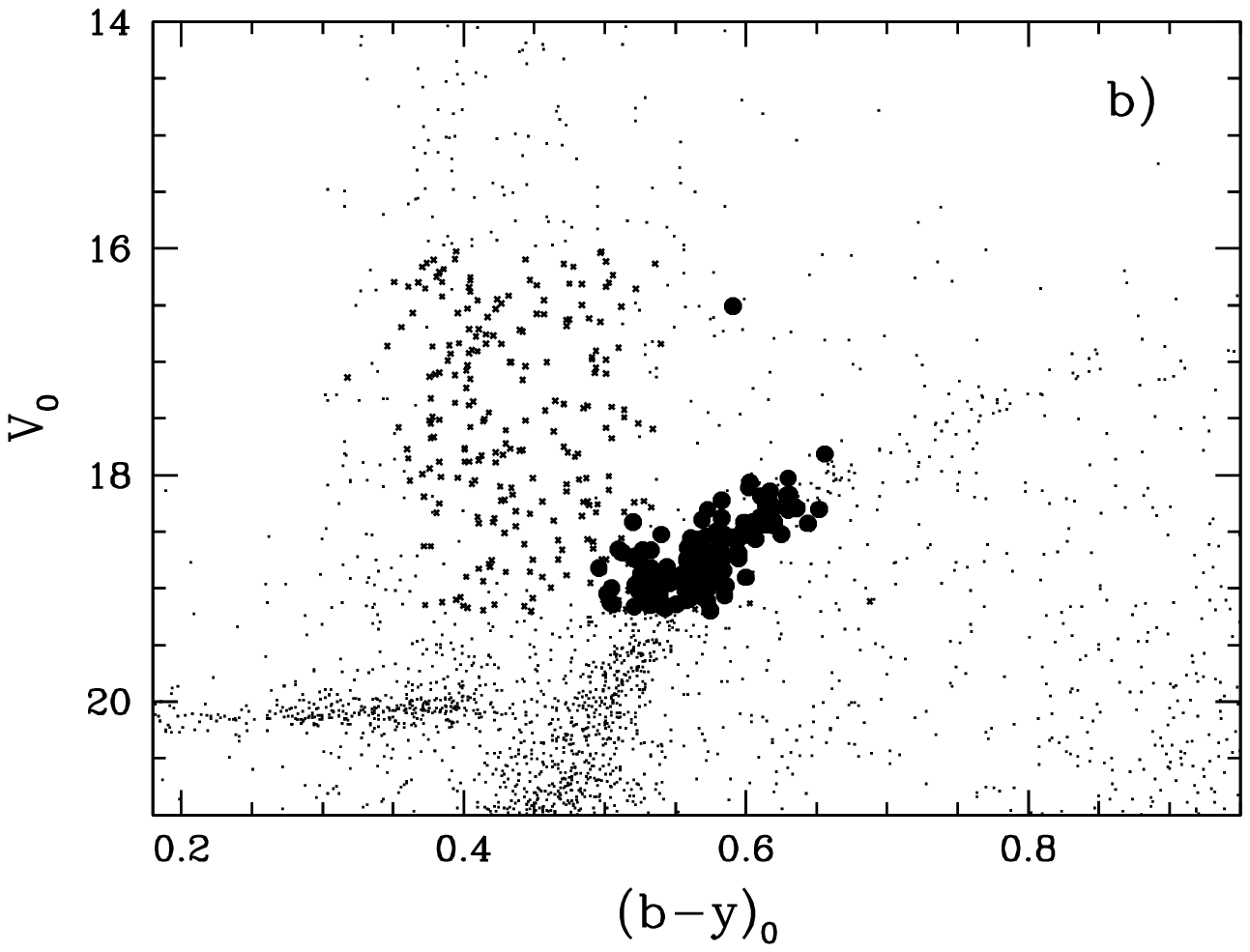}
\includegraphics[height=6.2cm]{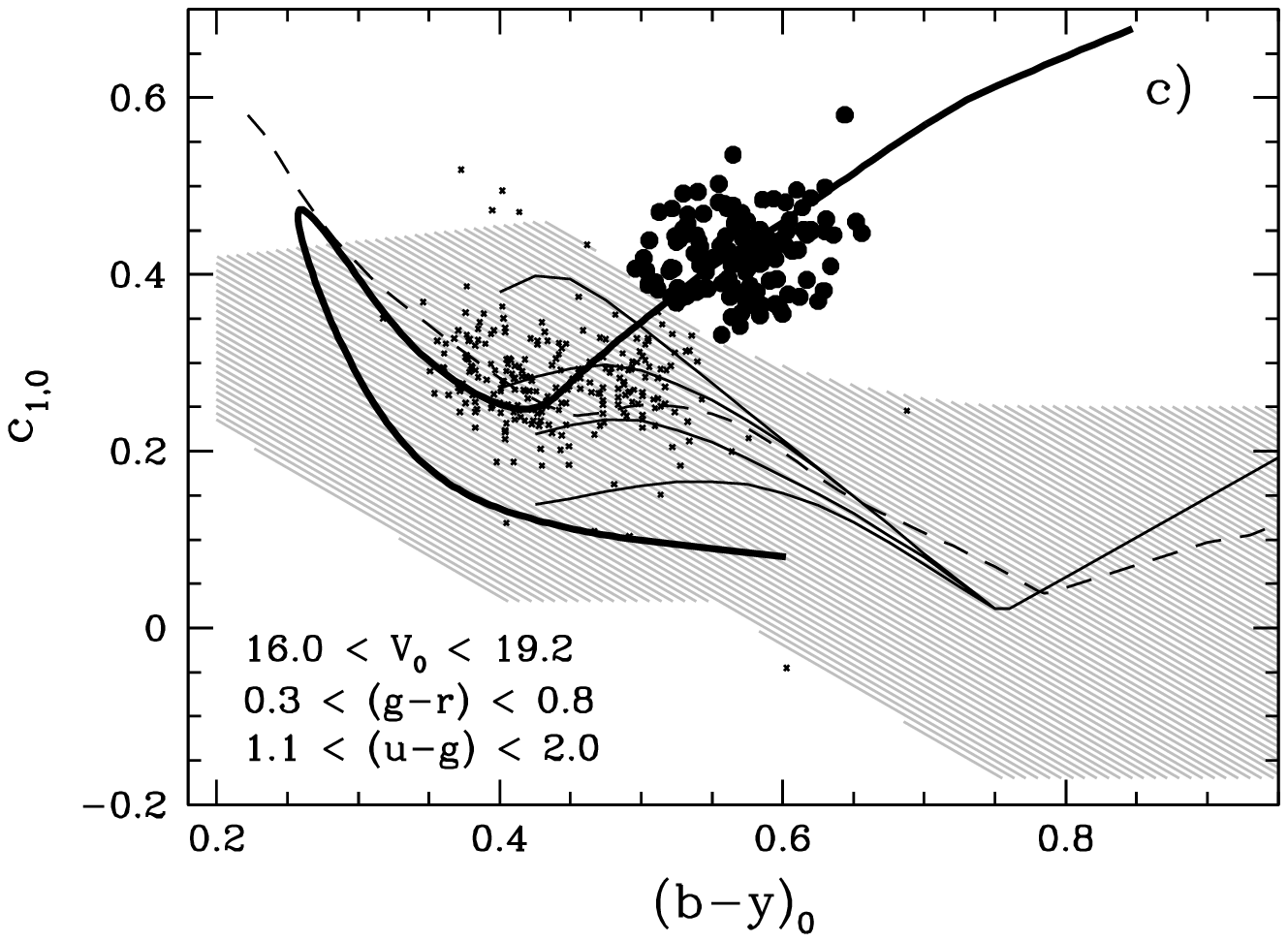}
\includegraphics[height=6.2cm]{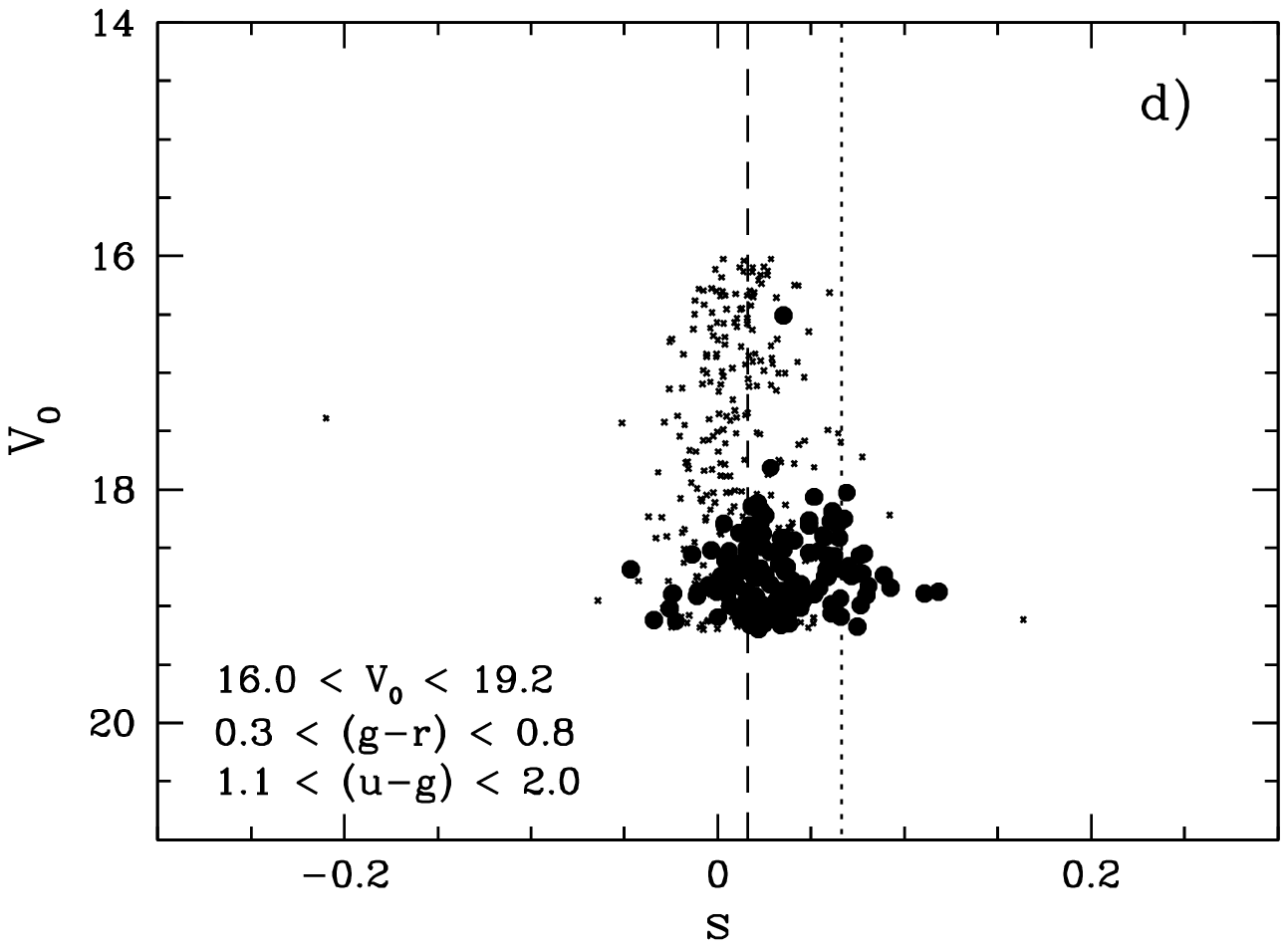}
\caption{{\bf a)} Colour--magnitude diagram showing the selection of
  stars along the line of sight towards the Draco dSph galaxy used for
  testing the giant star identification of
  \citet{2003ApJ...586..195H}.  These have $16.0<V_0<19.2$,
  $1.1<(u-g)<2.0$ and $0.3<(g-r)<0.8$ (marked with a box).  Stars
  identified as giant stars in the $c_{1,0}$ vs. $(b-y)_0$ plane are
  shown as filled dots. {\bf b)} The same stars but in a
  colour-magnitude diagram based on Str\"omgren photometry.  Same
  symbols as in panel a. The box indicated by a dotted line in a. is
  not included as it is a non-square area once mapped into this
  colour-magnitude plane. {\bf c)} Identification of giant stars
  (filled dots) in the $c_{1,0}$ vs. $(b-y)_0$ plane.  Grey hashed
  area shows the dwarf region used in \'Arnad\'ottir et al. (in prep).
  Our new dwarf star sequences (solid lines) are shown along with the
  preliminary relations by \citet{1984A&AS...57..443O} and
  \citet{1975AJ.....80..955C} (dashed line), and an isochrone with an
  age of 12 Gyr and [Fe/H]$=-2.3$ \citep[thick solid
  line,][]{1985ApJS...58..561V,2004AJ....127.1227C}.  {\bf d)} The
  distribution of identified giant stars (filled dots) in the
  $s$-index of \citet{2003ApJ...586..195H}.  Dashed line indicates the
  median $s$ of the selected stars ($0.016$) and the dotted line
  indicates the limit above which metal-poor giant stars are
  identified according to \citet{2003ApJ...586..195H}.  }
\label{Helmi03.fig}
\end{figure*}

\citet{2003ApJ...586..195H}used $ugriz$ photometry to identify
metal-poor giant stars.  We test this method using stars in the
direction of the Draco dwarf spheroidal galaxy.  The field contains
both foreground dwarf stars in the Milky Way as well as metal-poor
giant stars in the dwarf spheroidal galaxy \citep[][;\'Arnad\'ottir et
al. in prep.; Ad\'en et al. in prep.]{fariathesis,2007AA...465..357F}.

\citet{2003ApJ...586..195H} define a new colour index, $s=-0.249u +
0.794g - 0.555r + 0.24$ which is used to identify the metal-poor giant
stars. They find that metal-poor giant stars in general have larger
$s$-indices than the dwarf stars and define a giant star as a star
with an $s$-index more than 0.05 magnitudes above the median $s$-index for
the field.

We use metal-poor giant stars in the Draco dwarf spheroidal galaxy and
foreground stars along the same line-of-sight to test the ability of
the $s$-index to distinguish dwarf from giant stars. The $ugriz$
colour-magnitude diagram for the field used is shown in
Fig.\,\ref{Helmi03.fig}{\bf a}. For the comparison, we only use stars
in the colour range $1.1<(u-g)<2.0$ and $0.3<(g-r)<0.8$, where the
$s$-index is defined \citep{2003ApJ...586..195H}.  We identify
metal-poor giant stars in the direction of the Draco dwarf spheroidal
galaxy with $16.0<V_0<19.2$ using the Str\"omgren $c_{1,0}-(b-y)_0$
diagram (see Fig.\,\ref{Helmi03.fig}{\bf c}). In
Fig.\,\ref{Helmi03.fig}\,c, the dwarf region is indicated as a shaded
region (\'Arnad\'ottir et al., in prep.).

Figure \ref{Helmi03.fig}d shows the $V_0 -s$ diagram for stars
selected as  dwarf and giant stars using the Str\"omgren
$c_{1,0}-(b-y)_0$ diagram.  The metal-poor giant stars that we
identify in the $c_{1,0}$ vs. $(b-y)_0$ plane have a wide range of
$s$-index values. The dotted line indicates the $s-$value above which
metal-poor giant stars should be found. Figure \ref{Helmi03.fig}d\,
shows that metal-poor giant stars can not be distinguished from the
dwarf stars using the $s$-index.  Although the stars identified using
the $s$-index are pre-dominantly metal-poor giant stars, the $s$-index
is unable to reliably differentiate between metal-poor giant stars and
the foreground dwarf stars to good accuracy. More importantly, the
majority of the giant stars can not be identified by the $s$-index.

\section{A comparison of stellar sequences and model predictions}
\label{sect:model}

\begin{figure}   
\centering
\includegraphics[width=9cm]{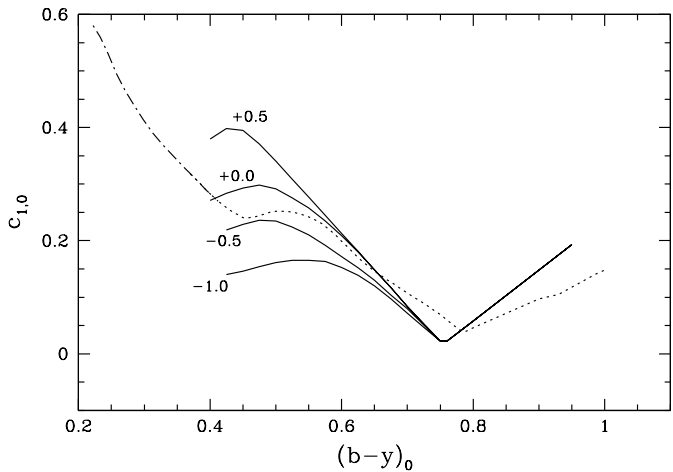}
\caption{Comparison of the new dwarf star sequences (solid lines),
  metallicities as indicated, to the preliminary relations by
  \citet{1984A&AS...57..443O} (dotted line) and by
  \citet{1975AJ.....80..955C} (dot-dashed line).  }
\label{DseqOlsCra}
\end{figure}

The stellar sequences for dwarf stars constructed in
Sect.\,\ref{Sect:dwarfseq} can be compared with model predictions
based on stellar evolutionary tracks and stellar model
atmospheres. Such comparisons are important for two reasons, they help
us to understand the physical processes occurring inside stars
(stellar evolution) and the processes in the stellar photospheres
(e.g., how well we can model the lines in the resulting stellar
spectra). Additionally, after ensuring that we understand these
processes (to a certain level), we may utilise the resulting stellar
isochrones and theoretically calculate indices to infer, e.g., the age
of a globular cluster.

In Fig.\,\ref{DseqOlsCra}, we compare our new stellar sequences for
dwarf stars with the preliminary relations of
\citet{1984A&AS...57..443O} and \citet{1975AJ.....80..955C}. As can be
seen, the metallicity dependence is significant and the lower envelope
changes by about 0.1 in $c_1$ as we change the metallicity with
0.5\,dex. For the reddest part, we agree with the preliminary
sequences in that there is only a single relation (see discussion in
Sect.\,\ref{Sect:dwarfseq}), although the slopes of the sequences
differ.

\subsection{A comparison with stellar isochrones}

\begin{figure}   
\centering
\includegraphics[width=9cm]{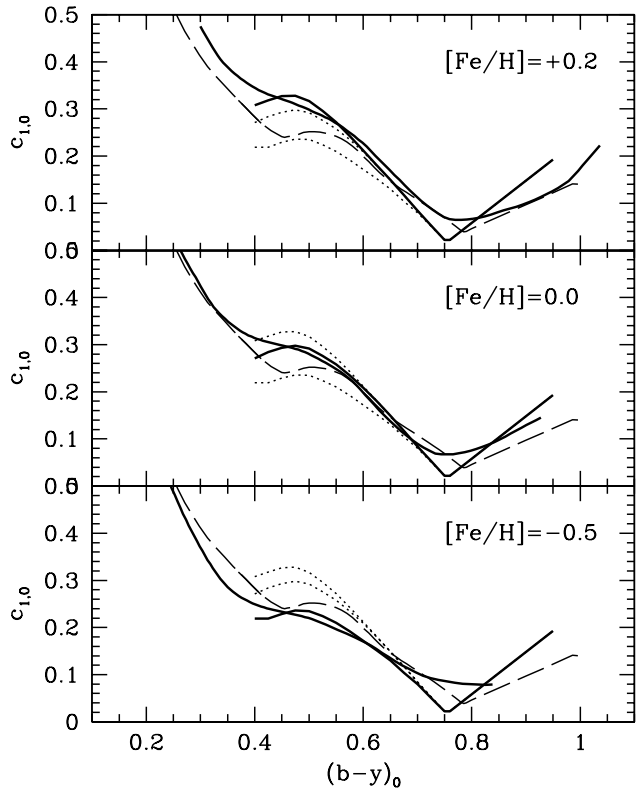}
\caption{A comparison of dwarf star sequences, as derived in this
  paper with stellar isochrones. In each panel, we show three of our
  sequences for dwarf stars for [Fe/H] = +0.20, 0.0, and --0.5. In
  each panel, the sequence with the metallicity indicated in the panel
  is shown with a thick solid line, the other two sequences are shown
  with dotted lines. The preliminary sequences by
  \citet{1984A&AS...57..443O} and by \citet{1975AJ.....80..955C} are
  shown with long dashes. An isochrone with the correct metallicity is
  also shown in each panel (thin, solid line). These isochrones are
  indicated with thick lines and all have an age of 1\,Gyr
  \citep[][]{1985ApJS...58..561V,2004AJ....127.1227C}.}
\label{compiso.fig}
\end{figure}

Few isochrones have been calculated for the Str\"omgren photometric
system, the most important set is probably that provided by
\citet{1985ApJS...58..561V} and derivations from that work.  To
convert theoretical stellar evolutionary sequences into stellar
isochrones, a colour-temperature relation is required
\citep[e.g.,][]{1986ApJS...61..509L,2004AJ....127.1227C}.  The
empirical calibration of \citet{2004AJ....127.1227C} is the most
recent and is used to convert, e.g., the isochrones of
\citet{1985ApJS...58..561V} and their derivatives onto the observed
plane. \citet{2004AJ....127.1227C} performed a detailed comparison
between stellar isochrones produced using their colour-temperature
relation and sequences of, e.g., red giant branches for globular
clusters with different metallicities, finding a good agreement.

\citet{2007AA...465..357F} preformed an additional comparison of the
stellar isochrones produced using the colour-temperature relation by
\citet{2004AJ....127.1227C} with $uvby$ photometry for field stars for
which [Fe/H] had been determined by high-resolution spectroscopy.
Their dataset is essentially identical to that  used by
\citet{2004AJ....127.1227C} to obtain the, interpolated,
colour-temperature relation for metallicities between --2\,dex and
super-solar metallicities. The comparison found some (still)
unexplained discrepancies between the data and the
isochrones. However, it was confirmed that the isochrones for about
--2\,dex and solar metallicity fit the field stars, with those
metallicities, very well. Hence, there might be some problems with the
empirical calibration needed for the colour transformation at
intermediate metallicities. Here, we therefore repeat the comparison,
this time as a comparison between our stellar sequences for dwarf
stars and the isochrones derived using the colour-temperature relation
by \citet{2004AJ....127.1227C}.

The comparison is shown in Fig.\,\ref{compiso.fig}. The stellar
sequences and the isochrones in general agree well with our sequences
for dwarf stars at $0.45<(b-y)_0<0.7$. We note, however, that the
stellar sequences trace the {\em lower} envelope of all stars that
have a narrow range of metallicities (see Table\,\ref{dwarfby:tab})
and the isochrones should reproduce the mean metallicity. Hence,
there might be some offset with respect to the $c_{1,0}$ index, but
otherwise the agreement is good for this fairly narrow magnitude range
of dwarf stars.  This comparison spans the main sequence from the
turn-off, late F--type dwarf stars to three magnitudes down the main
sequence to $M_V \sim 8$ (compare with Fig. \ref{fig:hr}\,b).

We performed a comparison between our sequences for dwarf stars, the
stellar isochrones, and the calculated indices in the $c_{\rm 1,0}$
vs. $(b-y)_{\rm 0}$ diagram. This makes for an easy comparison with
earlier works that often used $(b-y)_{\rm 0}$ as the colour along the
$x$-axis. However, the $(v-y)_{\rm 0}$ colour is more sensitive to
metallicity, as shown, e.g., by \citet{2007ApJ...670..400C}. This is
true for both giant and dwarf stars. Although the $(v-y)_{\rm 0}$ is
more sensitive to metallicity than $(b-y)_{\rm 0}$, it has the
disadvantage that is is also sensitive to the presence of CH and CN
molecules in the stellar atmosphere.
 
\subsection{A note about calculated indices}
\label{sect:ind}

\begin{figure}   
\centering
\includegraphics[width=9cm]{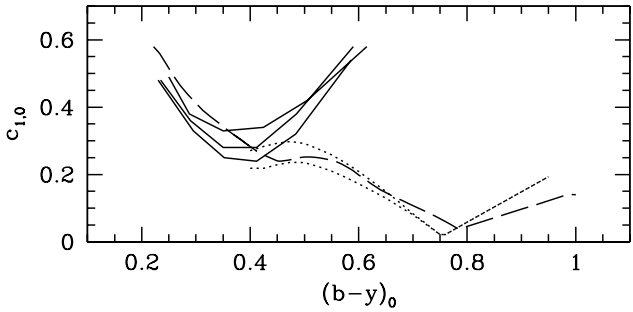}
\caption{A comparison of dwarf star sequences, as derived in this
  paper, for [Fe/H] of 0, and --0.5\,dex (dotted lines) with stellar
  indices, for stars with $\log g = 4.5$ and [Fe/H] of 0, --0.5, and
  --1.0\,dex, from \citet{2009A&A...498..527O} (solid lines).  The
  preliminary sequences by \citet{1984A&AS...57..443O} and by
  \citet{1975AJ.....80..955C} are also shown (long dashed line).}
\label{comp_onehag.fig}
\end{figure}

Theoretical indices in the Str\"omgren system have been studied in
several articles, including \citet{1986ApJS...61..509L},
\citet{1979A&A....74..313G}, and \citet{2009A&A...498..527O}.  In
Fig.\,\ref{comp_onehag.fig}, we perform a non-exhaustive comparison
between our stellar sequences for dwarf stars and indices calculated
by \citet{2009A&A...498..527O} for stars with $\log g = 4.5$. We show
stellar sequences for 0 and $-0.5$\,dex because we believe that the sequence
for $-1$\,dex is less robust (compare Fig.\,\ref{Dbym10}). It is clear
from this comparison that the calculated indices do not
reproduce the colours found for field dwarf stars in the solar
neighbourhood for $(b-y)_0>0.45$.

Based on the calculated indices, \citet{2009A&A...498..527O} derive a
metallicity calibration that is nominally valid for stars with $0.22 <
(b-y)_0 < 0.59$.  In Table\,\ref{Cal-mean-sigma}, we compare this
calibration with the spectroscopic catalogue, in the same way as for
the empirically derived metallicity calibrations.  We find an offset
of 0.33\,dex with a scatter of 0.3\,dex. This calibration clearly
reproduces the spectroscopically derived iron abundances more poorly
than the empirical calibrations available in the literature. This
shortcoming of the theoretical calibrations was already noted and
discussed by \citet{2009A&A...498..527O}.

\subsection{$\log g$ from $uvby$ photometry - a critical
evaluation \label{SectLogg}}

Although the Str\"omgren system is clearly capable of distinguishing
between dwarf and giant stars for colours redder than
$(b-y)_0\sim0.55$, the situation is far less clear when we consider
the turn-off and sub-giant region. To separate, e.g., field dwarf
stars from field sub-giants, we need a measure of their surface
gravity for which any metallicity dependence has been taken into
account, before being able to distinguish between the dwarf,
sub-giant, and giant stars in this narrow colour space (compare
Fig. \ref{fig:scheme}).

Hence, it would be desirable to derive $\log g$ directly from the
photometry itself. To our knowledge, the only $\log g$ calibration
based only on $uvby$ photometry is that of
\citet{1984A&AS...57..443O}.  If $\beta$ were to be included,
additional calibrations would be available \citep[including][where the
calibration is only shown
graphically]{2009A&A...497..209V,1993A&A...275..101E}.

\begin{figure}   
\centering
\includegraphics[width=9cm]{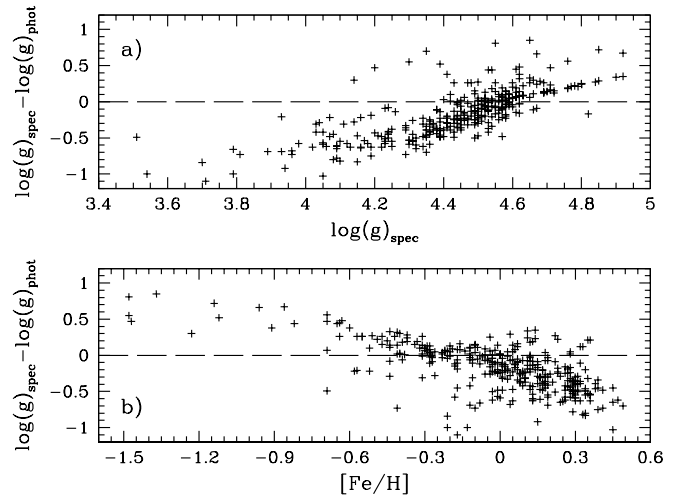}
\caption{ A comparison of $\log g$ determined using the photometric
  calibration by \citet{1984A&AS...57..443O} and $\log g$ determined
  from an abundance analysis based on high resolution spectroscopy
  \citep{2005ApJS..159..141V}.  {\bf a)} The difference as a function
  of $\log g$ determined in the spectroscopic analysis. {\bf b)} The
  difference as a function of [Fe/H].}
\label{Ols84logg.fig}
\end{figure}

\begin{figure}   
\centering
\includegraphics[width=9cm]{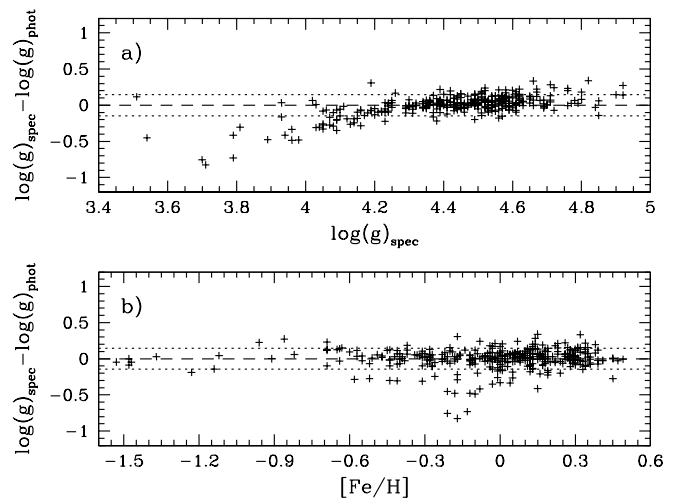}
\caption{We show our best attempt at deriving a new $\log g$
  calibration from $uvby$ photometry. {\bf a)} A comparison of $\log
  g$ determined using Eq.\,(\ref{EqLogG}) and $\log g$ determined from
  abundance analysis based on high resolution spectroscopy
  \citep{2005ApJS..159..141V}, plotted as a function of
  spectroscopically determined $\log g$.  The mean difference (dashed
  line) is $0.00$ with a $\sigma=0.15$ (dotted line).  {\bf b)} A
  comparison of $\log g$ determined using Eq.\,(\ref{EqLogG}) and
  $\log g$ determined from high resolution spectroscopy
  \citep{2005ApJS..159..141V} as a function of [Fe/H].  The mean
  difference (dashed line) is $0.00$ with a $\sigma=0.15$ (dotted
  line).}
\label{ourlogg.fig}
\end{figure}

Using the stars in Table\,B.\ref{FCC} with $\log g$ determinations from
\citet{2005ApJS..159..141V}, we test the calibration of
\citet{1984A&AS...57..443O}. Figure \ref{Ols84logg.fig} shows the
$\log g$ derived in the spectroscopic study of
\citet{2005ApJS..159..141V} ($\log g_{spec}$) minus the $\log g$
derived from the photometry ($\log g_{phot}$). As can be seen, the
calibration has a strong dependence on [Fe/H].

We now attempt the construction of a new calibration to derive $\log
g$ directly from dereddened $uvby$ photometry, using the spectroscopic
catalogue in Table\,B.\ref{FCC}. We start with a third order polynomial
in $(b-y)_0$, $m_{1,0}$, and $c_{1,0}$. We note that some calibrations
include terms in [Fe/H], which we do not because we derive [M/H] from
the same photometry and hence adding [M/H] terms would only mean
adding yet more terms to the equation without gaining any further
knowledge.

After removing terms that do not contribute significantly, we obtain the
fifteenth order polynomial

\begin{eqnarray}     
\label{EqLogG}
\log g &=& 
   -178.0420  (b-y)_0         + 109.7056 m_{1,0} \nonumber \\
&& + 47.4263  c_{1,0}          + 615.0911 (b-y)_0^2 \nonumber \\
&& + 47.0152  m_{1,0}^2        - 114.8399 c_{1,0}^2 \nonumber \\
&& - 525.0138 (b-y)_0 m_{1,0}  - 112.5602 m_{1,0} c_{1,0} \nonumber \\
&& - 598.8569 (b-y)_0^3       + 674.8341 (b-y)_0^2 m_{1,0} \nonumber \\
&& - 267.7717 (b-y)_0^2 c_{1,0}- 147.5764 m_{1,0}^2 (b-y)_0 \nonumber \\
&& + 265.3608 c_{1,0}^2 (b-y)_0+ 266.5860 (b-y)_0 m_{1,0} c_{1,0} \nonumber \\
&& + 14.3503.
\end{eqnarray}

\noindent
If we were to include [Fe/H] terms the result was a ninth order
polynomial. However, as we want to derive both metallicity and surface
gravity from the photometry itself, the 15th order polynomial
presented above is a better choice. 

Figure\,\ref{ourlogg.fig} shows a comparison with $\log g$ from
Table\,B.\ref{FCC}.  The comparison is good for stars with $\log g
\gtrsim 4.0$ but is progressively poorer towards more evolved star,
including the regime where we would most need a good calibration to
separate dwarf and sub-giant stars with similar colours! Hence, the
use of our new calibration is limited to $\log g>4.0$.
Equation\,\ref{EqLogG} is calibrated using dwarf stars in the
parameter ranges $0.236<(b-y)_0<0.616$, $0.122<c_{\rm 1,0}<0.441$,
$0.075<m_{\rm 1,0}<0.679$, and $-1.64<$[Fe/H]$<0.49$.

We also considered restricting ourselves to the region of the $c_{\rm
  1,0}-(b-y)_0$ plane where we most need a calibration
($(b-y)_0<0.55$, $0.24<c_{\rm 1,0}<0.44$, and $c_{\rm
  1,0}<-1.504*(b-y)_0+1.147$).  This also failed in the same way as
described for the wider parameter ranges, i.e. we were not able to
reliably determine the $\log g$s for subgiant stars. We also attempted
to make a calibration that would retrieve the original $\log g$s in a
synthetic stellar population, this also failed. Hence, there does not
appear to be an easy, straightforward way to derive $\log g$ directly
from the Str\"omgren $uvby$ photometry for turn-off and subgiant stars.

Based on their theoretical investigation, \citet{2009A&A...498..527O}
find that for dwarf stars cooler than the Sun $c_{\rm 1,0}$ is not a
good measure of stellar gravity. However, from our empirical
comparison of $\log g$ derived using the calibration by
\citet{1984A&AS...57..443O} and from spectroscopy we find that for
stars redder than $(b-y)_0 \sim 0.5$ the spectroscopic $\log g$
compares very well with the photometric $\log g$. For stars with
$\log g \sim 4.5$, the comparison is  also good. It thus seems that for main
sequence, cool dwarf stars the Str\"omgren system {\em is} able to
predict the surface gravity of the star.

\section{Summary}
\label{sect:sum}

As part of our studies of the properties of the Milky Way disk system
we have undertaken a critical evaluation of the Str\"omgren system's
ability to provide accurate stellar parameters and to distinguish
between dwarf, sub-giant, and giant stars.

We have found that the metallicity calibration for dwarf stars by
\citet{2005ApJ...626..446R} is the most widely applicable calibration for
determining metallicities for dwarf and subgiant stars. The
calibration of \citet{1984A&AS...57..443O} provides an extension from
$(b-y)_0=0.8$ to $(b-y)_0=1.0$.  We also note that the older
calibration of \citet{1989A&A...221...65S} performs almost
equally well, but it does not extend to as red colours as the
calibration of \citet{2005ApJ...626..446R}.

Although we have found that $uvby$ photometry can readily distinguish
between giant and dwarf stars for redder colours, it is disconcerting
that no calibration of $\log g$, for dwarf and subgiant stars, is
able to reproduce $\log g$ derived from either spectra or Hipparcos
parallaxes. In his provisional
calibration  \citet{2009A&A...497..209V} also notes the same.

Using the catalogues of \citet{1993A&AS..102...89O},
\citet{1994A&AS..104..429O}, and \citet{1994A&AS..106..257O} and the
metallicity calibration of \citet{2005ApJ...626..446R}, we have traced
new, improved standard sequences for dwarf stars. These new sequences
are metallicity dependent and provide crucial calibrations for,
e.g., stellar isochrones.

Even though we have found that stellar isochrones in the $uvby$ system
reasonably well reproduce empirical stellar sequences it is clear that
the disagreement between theoretically calculated Str\"omgren indices
and observed ones can be large.  This appears somewhat surprising as
stellar isochrones employ the same type of model atmospheres to get
the predicted colours as is often used for the elemental abundance
studies.  This state of affairs is unsatisfactory and we encourage
future theoretical studies to resolve these problems.

 As part of this work, we
have compiled a catalogue of dwarf stars with $uvby$ photometry as
well as [Fe/H] derived from high-resolution, high S/N
spectroscopy. The iron abundances have been homogenised to the scale
provided by \citet{2005ApJS..159..141V}. This catalogue is provided in
full (in electronic form) with this paper.


\begin{acknowledgements}
  We would like to thank the anonymous referee for pointing out the
  work done by \citet{2004A&A...417..301R} on the metallicity
  calibration of Str\"omgren photometry for giant stars. Bengt
  Gustafsson is thanked for a careful reading of the penultimate
  manuscript and the provision of numerous detailed comments and
  discussions that improved both the content and the style of the
  paper.  SF is a Royal Swedish Academy of Sciences Research Fellow
  supported by a grant from the Knut and Alice Wallenberg Foundation.
  This research has made use of the SIMBAD database, operated at CDS,
  Strasbourg, France.
\end{acknowledgements}

\bibliographystyle{aa} 
\bibliography{13544}

\begin{thebibliography}{107}
\expandafter\ifx\csname natexlab\endcsname\relax\def\natexlab#1{#1}\fi

\bibitem[{{Abazajian} {et~al.}(2009){Abazajian}, {Adelman-McCarthy},
  {Ag{\"u}eros}, {Allam}, {Allende Prieto}, {An}, {Anderson}, {Anderson},
  {Annis}, {Bahcall}, {Bailer-Jones}, {Barentine}, {Bassett}, {Becker},
  {Beers}, {Bell}, {Belokurov}, {Berlind}, {Berman}, {Bernardi}, {Bickerton},
  {Bizyaev}, {Blakeslee}, {Blanton}, {Bochanski}, {Boroski}, {Brewington},
  {Brinchmann}, {Brinkmann}, {Brunner}, {Budav{\'a}ri}, {Carey}, {Carliles},
  {Carr}, {Castander}, {Cinabro}, {Connolly}, {Csabai}, {Cunha}, {Czarapata},
  {Davenport}, {de Haas}, {Dilday}, {Doi}, {Eisenstein}, {Evans}, {Evans},
  {Fan}, {Friedman}, {Frieman}, {Fukugita}, {G{\"a}nsicke}, {Gates},
  {Gillespie}, {Gilmore}, {Gonzalez}, {Gonzalez}, {Grebel}, {Gunn},
  {Gy{\"o}ry}, {Hall}, {Harding}, {Harris}, {Harvanek}, {Hawley}, {Hayes},
  {Heckman}, {Hendry}, {Hennessy}, {Hindsley}, {Hoblitt}, {Hogan}, {Hogg},
  {Holtzman}, {Hyde}, {Ichikawa}, {Ichikawa}, {Im}, {Ivezi{\'c}}, {Jester},
  {Jiang}, {Johnson}, {Jorgensen}, {Juri{\'c}}, {Kent}, {Kessler}, {Kleinman},
  {Knapp}, {Konishi}, {Kron}, {Krzesinski}, {Kuropatkin}, {Lampeitl},
  {Lebedeva}, {Lee}, {Lee}, {Leger}, {L{\'e}pine}, {Li}, {Lima}, {Lin}, {Long},
  {Loomis}, {Loveday}, {Lupton}, {Magnier}, {Malanushenko}, {Malanushenko},
  {Mandelbaum}, {Margon}, {Marriner}, {Mart{\'{\i}}nez-Delgado}, {Matsubara},
  {McGehee}, {McKay}, {Meiksin}, {Morrison}, {Mullally}, {Munn}, {Murphy},
  {Nash}, {Nebot}, {Neilsen}, {Newberg}, {Newman}, {Nichol}, {Nicinski},
  {Nieto-Santisteban}, {Nitta}, {Okamura}, {Oravetz}, {Ostriker}, {Owen},
  {Padmanabhan}, {Pan}, {Park}, {Pauls}, {Peoples}, {Percival}, {Pier}, {Pope},
  {Pourbaix}, {Price}, {Purger}, {Quinn}, {Raddick}, {Fiorentin}, {Richards},
  {Richmond}, {Riess}, {Rix}, {Rockosi}, {Sako}, {Schlegel}, {Schneider},
  {Scholz}, {Schreiber}, {Schwope}, {Seljak}, {Sesar}, {Sheldon}, {Shimasaku},
  {Sibley}, {Simmons}, {Sivarani}, {Smith}, {Smith}, {Smol{\v c}i{\'c}},
  {Snedden}, {Stebbins}, {Steinmetz}, {Stoughton}, {Strauss}, {Subba Rao},
  {Suto}, {Szalay}, {Szapudi}, {Szkody}, {Tanaka}, {Tegmark}, {Teodoro},
  {Thakar}, {Tremonti}, {Tucker}, {Uomoto}, {Vanden Berk}, {Vandenberg},
  {Vidrih}, {Vogeley}, {Voges}, {Vogt}, {Wadadekar}, {Watters}, {Weinberg},
  {West}, {White}, {Wilhite}, {Wonders}, {Yanny}, {Yocum}, {York}, {Zehavi},
  {Zibetti}, \& {Zucker}}]{2009ApJS..182..543A}
{Abazajian}, K.~N., {Adelman-McCarthy}, J.~K., {Ag{\"u}eros}, M.~A., {et~al.}
  2009, \apjs, 182, 543

\bibitem[{{Ad{\'e}n} {et~al.}(2009{\natexlab{a}}){Ad{\'e}n}, {Feltzing},
  {Koch}, {Wilkinson}, {Grebel}, {Lundstr{\"o}m}, {Gilmore}, {Zucker},
  {Belokurov}, {Evans}, \& {Faria}}]{2009A&A...506.1147A}
{Ad{\'e}n}, D., {Feltzing}, S., {Koch}, A., {et~al.} 2009{\natexlab{a}}, \aap,
  506, 1147

\bibitem[{{Ad{\'e}n} {et~al.}(2009{\natexlab{b}}){Ad{\'e}n}, {Wilkinson},
  {Read}, {Feltzing}, {Koch}, {Gilmore}, {Grebel}, \&
  {Lundstr{\"o}m}}]{2009ApJ...706L.150A}
{Ad{\'e}n}, D., {Wilkinson}, M.~I., {Read}, J.~I., {et~al.} 2009{\natexlab{b}},
  \apjl, 706, L150

\bibitem[{{Alonso} {et~al.}(1996){Alonso}, {Arribas}, \&
  {Martinez-Roger}}]{1996A&AS..117..227A}
{Alonso}, A., {Arribas}, S., \& {Martinez-Roger}, C. 1996, \aaps, 117, 227

\bibitem[{{Alonso} {et~al.}(1999){Alonso}, {Arribas}, \&
  {Mart{\'{\i}}nez-Roger}}]{1999A&AS..140..261A}
{Alonso}, A., {Arribas}, S., \& {Mart{\'{\i}}nez-Roger}, C. 1999, \aaps, 140,
  261

\bibitem[{{Anthony-Twarog} \& {Twarog}(1994)}]{1994AJ....107.1577A}
{Anthony-Twarog}, B.~J. \& {Twarog}, B.~A. 1994, \aj, 107, 1577

\bibitem[{{Anthony-Twarog} \& {Twarog}(1998)}]{1998AJ....116.1922A}
{Anthony-Twarog}, B.~J. \& {Twarog}, B.~A. 1998, \aj, 116, 1922

\bibitem[{{Arce} \& {Goodman}(1999)}]{1999ApJ...512L.135A}
{Arce}, H.~G. \& {Goodman}, A.~A. 1999, \apjl, 512, L135

\bibitem[{{Arellano Ferro} \& {Mendoza V.}(1993)}]{1993AJ....106.2516A}
{Arellano Ferro}, A. \& {Mendoza V.}, E.~E. 1993, \aj, 106, 2516

\bibitem[{{Beers} {et~al.}(2002){Beers}, {Drilling}, {Rossi}, {Chiba}, {Rhee},
  {F{\"u}hrmeister}, {Norris}, \& {von Hippel}}]{2002AJ....124..931B}
{Beers}, T.~C., {Drilling}, J.~S., {Rossi}, S., {et~al.} 2002, \aj, 124, 931

\bibitem[{{Bell} \& {Gustafsson}(1978)}]{1978A&AS...34..229B}
{Bell}, R.~A. \& {Gustafsson}, B. 1978, \aaps, 34, 229

\bibitem[{{Bensby} {et~al.}(2005){Bensby}, {Feltzing}, {Lundstr{\"o}m}, \&
  {Ilyin}}]{2005A&A...433..185B}
{Bensby}, T., {Feltzing}, S., {Lundstr{\"o}m}, I., \& {Ilyin}, I. 2005, \aap,
  433, 185

\bibitem[{{Bond}(1970)}]{1970ApJS...22..117B}
{Bond}, H.~E. 1970, \apjs, 22, 117

\bibitem[{{Bond}(1980)}]{1980ApJS...44..517B}
{Bond}, H.~E. 1980, \apjs, 44, 517

\bibitem[{{Bonfils} {et~al.}(2005){Bonfils}, {Delfosse}, {Udry}, {Santos},
  {Forveille}, \& {S{\'e}gransan}}]{2005A&A...442..635B}
{Bonfils}, X., {Delfosse}, X., {Udry}, S., {et~al.} 2005, \aap, 442, 635

\bibitem[{{Bonifacio} {et~al.}(2000){Bonifacio}, {Caffau}, \&
  {Molaro}}]{2000A&AS..145..473B}
{Bonifacio}, P., {Caffau}, E., \& {Molaro}, P. 2000, \aaps, 145, 473

\bibitem[{{Calamida} {et~al.}(2007){Calamida}, {Bono}, {Stetson}, {Freyhammer},
  {Cassisi}, {Grundahl}, {Pietrinferni}, {Hilker}, {Primas}, {Richtler},
  {Romaniello}, {Buonanno}, {Caputo}, {Castellani}, {Corsi}, {Ferraro},
  {Iannicola}, \& {Pulone}}]{2007ApJ...670..400C}
{Calamida}, A., {Bono}, G., {Stetson}, P.~B., {et~al.} 2007, \apj, 670, 400

\bibitem[{{Calamida} {et~al.}(2009){Calamida}, {Bono}, {Stetson}, {Freyhammer},
  {Piersimoni}, {Buonanno}, {Caputo}, {Cassisi}, {Castellani}, {Corsi},
  {Dall'Ora}, {Degl'Innocenti}, {Ferraro}, {Grundahl}, {Hilker}, {Iannicola},
  {Monelli}, {Nonino}, {Patat}, {Pietrinferni}, {Prada Moroni}, {Primas},
  {Pulone}, {Richtler}, {Romaniello}, {Storm}, \&
  {Walker}}]{2009ApJ...706.1277C}
{Calamida}, A., {Bono}, G., {Stetson}, P.~B., {et~al.} 2009, \apj, 706, 1277

\bibitem[{{Carney}(1983)}]{1983AJ.....88..623C}
{Carney}, B.~W. 1983, \aj, 88, 623

\bibitem[{{Carollo} {et~al.}(2008){Carollo}, {Beers}, {Lee}, {Chiba}, {Norris},
  {Wilhelm}, {Sivarani}, {Marsteller}, {Munn}, {Bailer-Jones}, {Fiorentin}, \&
  {York}}]{2008Natur.451..216C}
{Carollo}, D., {Beers}, T.~C., {Lee}, Y.~S., {et~al.} 2008, \nat, 451, 216

\bibitem[{{Chen} {et~al.}(2000){Chen}, {Nissen}, {Zhao}, {Zhang}, \&
  {Benoni}}]{2000A&AS..141..491C}
{Chen}, Y.~Q., {Nissen}, P.~E., {Zhao}, G., {Zhang}, H.~W., \& {Benoni}, T.
  2000, \aaps, 141, 491

\bibitem[{{Clausen} {et~al.}(1997){Clausen}, {Larsen}, {Garcia}, {Gimenez}, \&
  {Storm}}]{1997A&AS..122..559C}
{Clausen}, J.~V., {Larsen}, S.~S., {Garcia}, J.~M., {Gimenez}, A., \& {Storm},
  J. 1997, \aaps, 122, 559

\bibitem[{{Clem} {et~al.}(2004){Clem}, {VandenBerg}, {Grundahl}, \&
  {Bell}}]{2004AJ....127.1227C}
{Clem}, J.~L., {VandenBerg}, D.~A., {Grundahl}, F., \& {Bell}, R.~A. 2004, \aj,
  127, 1227

\bibitem[{{Cohen} \& {Huang}(2009)}]{2009ApJ...701.1053C}
{Cohen}, J.~G. \& {Huang}, W. 2009, \apj, 701, 1053

\bibitem[{{Cousins}(1987)}]{1987SAAOC..11...93C}
{Cousins}, A.~W.~J. 1987, South African Astronomical Observatory Circular, 11,
  93

\bibitem[{{Crawford}(1975)}]{1975AJ.....80..955C}
{Crawford}, D.~L. 1975, \aj, 80, 955

\bibitem[{{Crawford} \& {Barnes}(1970)}]{1970AJ.....75..978C}
{Crawford}, D.~L. \& {Barnes}, J.~V. 1970, \aj, 75, 978

\bibitem[{{Edvardsson} {et~al.}(1993){Edvardsson}, {Andersen}, {Gustafsson},
  {Lambert}, {Nissen}, \& {Tomkin}}]{1993A&A...275..101E}
{Edvardsson}, B., {Andersen}, J., {Gustafsson}, B., {et~al.} 1993, \aap, 275,
  101

\bibitem[{{ESA}(1997)}]{1997yCat.1239....0E}
{ESA}. 1997, VizieR Online Data Catalog, 1239, 0

\bibitem[{{Faria}(2006)}]{fariathesis}
{Faria}, D. 2006, {Photometry of resolved stellar populations in local group
  galaxies~(PhD thesis), {\it Lund Observatory}, ISBN: 978-91-628-6905-2.}

\bibitem[{{Faria} {et~al.}(2007){Faria}, {Feltzing}, {Lundstr{\"o}m},
  {Gilmore}, {Wahlgren}, {Ardeberg}, \& {Linde}}]{2007AA...465..357F}
{Faria}, D., {Feltzing}, S., {Lundstr{\"o}m}, I., {et~al.} 2007, \aap, 465, 357

\bibitem[{{Favata} {et~al.}(1997){Favata}, {Micela}, \&
  {Sciortino}}]{1997A&A...323..809F}
{Favata}, F., {Micela}, G., \& {Sciortino}, S. 1997, \aap, 323, 809

\bibitem[{{Feltzing} \& {Gustafsson}(1998)}]{1998A&AS..129..237F}
{Feltzing}, S. \& {Gustafsson}, B. 1998, \aaps, 129, 237

\bibitem[{{Ferguson} {et~al.}(2005){Ferguson}, {Johnson}, {Faria}, {Irwin},
  {Ibata}, {Johnston}, {Lewis}, \& {Tanvir}}]{2005ApJ...622L.109F}
{Ferguson}, A.~M.~N., {Johnson}, R.~A., {Faria}, D.~C., {et~al.} 2005, \apjl,
  622, L109

\bibitem[{{Flynn} \& {Morell}(1997)}]{1997MNRAS.286..617F}
{Flynn}, C. \& {Morell}, O. 1997, \mnras, 286, 617

\bibitem[{{Gilmore} \& {Reid}(1983)}]{1983MNRAS.202.1025G}
{Gilmore}, G. \& {Reid}, N. 1983, \mnras, 202, 1025

\bibitem[{{Golay}(1974)}]{1974ASSL...41.....G}
{Golay}, M., ed. 1974, Astrophysics and Space Science Library, Vol.~41,
  {Introduction to astronomical photometry}

\bibitem[{{Grebel} \& {Richtler}(1992)}]{1992A&A...253..359G}
{Grebel}, E.~K. \& {Richtler}, T. 1992, \aap, 253, 359

\bibitem[{{Gr{\o}nbech} {et~al.}(1976){Gr{\o}nbech}, {Olsen}, \&
  {Str{\"o}mgren}}]{1976A&AS...26..155G}
{Gr{\o}nbech}, B., {Olsen}, E.~H., \& {Str{\"o}mgren}, B. 1976, \aaps, 26, 155

\bibitem[{{Grundahl} {et~al.}(2002){Grundahl}, {Stetson}, \&
  {Andersen}}]{2002A&A...395..481G}
{Grundahl}, F., {Stetson}, P.~B., \& {Andersen}, M.~I. 2002, \aap, 395, 481

\bibitem[{{Gustafsson} \& {Ardeberg}(1978)}]{1978bs...symp..145G}
{Gustafsson}, B. \& {Ardeberg}, A. 1978, in Astronomical Papers Dedicated to
  Bengt Stromgren, ed. {A.~Reiz \& T.~Andersen}, 145--152

\bibitem[{{Gustafsson} \& {Bell}(1979)}]{1979A&A....74..313G}
{Gustafsson}, B. \& {Bell}, R.~A. 1979, \aap, 74, 313

\bibitem[{{Gustafsson} \& {Nissen}(1972)}]{1972A&A....19..261G}
{Gustafsson}, B. \& {Nissen}, P.~E. 1972, \aap, 19, 261

\bibitem[{{Haywood}(2002)}]{2002MNRAS.337..151H}
{Haywood}, M. 2002, \mnras, 337, 151

\bibitem[{{Heiter} \& {Luck}(2003)}]{2003AJ....126.2015H}
{Heiter}, U. \& {Luck}, R.~E. 2003, \aj, 126, 2015

\bibitem[{{Helmi} {et~al.}(2003){Helmi}, {Ivezi{\'c}}, {Prada}, {Pentericci},
  {Rockosi}, {Schneider}, {Grebel}, {Harbeck}, {Lupton}, {Gunn}, {Knapp},
  {Strauss}, \& {Brinkmann}}]{2003ApJ...586..195H}
{Helmi}, A., {Ivezi{\'c}}, {\v Z}., {Prada}, F., {et~al.} 2003, \apj, 586, 195

\bibitem[{{Hilker}(2000)}]{2000A&A...355..994H}
{Hilker}, M. 2000, \aap, 355, 994

\bibitem[{{Holmberg} {et~al.}(2007){Holmberg}, {Nordstr{\"o}m}, \&
  {Andersen}}]{2007A&A...475..519H}
{Holmberg}, J., {Nordstr{\"o}m}, B., \& {Andersen}, J. 2007, \aap, 475, 519

\bibitem[{{Ibata} {et~al.}(2001){Ibata}, {Irwin}, {Lewis}, {Ferguson}, \&
  {Tanvir}}]{2001Natur.412...49I}
{Ibata}, R., {Irwin}, M., {Lewis}, G., {Ferguson}, A.~M.~N., \& {Tanvir}, N.
  2001, \nat, 412, 49

\bibitem[{{Ivezi{\'c}} {et~al.}(2008){Ivezi{\'c}}, {Sesar}, {Juri{\'c}},
  {Bond}, {Dalcanton}, {Rockosi}, {Yanny}, {Newberg}, {Beers}, {Allende
  Prieto}, {Wilhelm}, {Lee}, {Sivarani}, {Norris}, {Bailer-Jones}, {Re
  Fiorentin}, {Schlegel}, {Uomoto}, {Lupton}, {Knapp}, {Gunn}, {Covey},
  {Smith}, {Miknaitis}, {Doi}, {Tanaka}, {Fukugita}, {Kent}, {Finkbeiner},
  {Munn}, {Pier}, {Quinn}, {Hawley}, {Anderson}, {Kiuchi}, {Chen}, {Bushong},
  {Sohi}, {Haggard}, {Kimball}, {Barentine}, {Brewington}, {Harvanek},
  {Kleinman}, {Krzesinski}, {Long}, {Nitta}, {Snedden}, {Lee}, {Harris},
  {Brinkmann}, {Schneider}, \& {York}}]{ivezic2008}
{Ivezi{\'c}}, {\v Z}., {Sesar}, B., {Juri{\'c}}, M., {et~al.} 2008, \apj, 684,
  287

\bibitem[{{Johnson} \& {Morgan}(1953)}]{1953ApJ...117..313J}
{Johnson}, H.~L. \& {Morgan}, W.~W. 1953, \apj, 117, 313

\bibitem[{{J{\o}nch-S{\o}rensen}(1995)}]{1995A&A...298..799J}
{J{\o}nch-S{\o}rensen}, H. 1995, \aap, 298, 799

\bibitem[{{Kirby} {et~al.}(2008){Kirby}, {Simon}, {Geha}, {Guhathakurta}, \&
  {Frebel}}]{2008ApJ...685L..43K}
{Kirby}, E.~N., {Simon}, J.~D., {Geha}, M., {Guhathakurta}, P., \& {Frebel}, A.
  2008, \apjl, 685, L43

\bibitem[{{Kotoneva} {et~al.}(2002){Kotoneva}, {Flynn}, {Chiappini}, \&
  {Matteucci}}]{2002MNRAS.336..879K}
{Kotoneva}, E., {Flynn}, C., {Chiappini}, C., \& {Matteucci}, F. 2002, \mnras,
  336, 879

\bibitem[{{Lagerholm}(2008)}]{Carina08}
{Lagerholm}, C. 2008, {An investigation of metallicity and memberships of the
  Sextans dSph galaxy}

\bibitem[{{Landolt}(1992)}]{1992AJ....104..340L}
{Landolt}, A.~U. 1992, \aj, 104, 340

\bibitem[{{Lee} {et~al.}(2008){Lee}, {Beers}, {Sivarani}, {Johnson}, {An},
  {Wilhelm}, {Allende Prieto}, {Koesterke}, {Re Fiorentin}, {Bailer-Jones},
  {Norris}, {Yanny}, {Rockosi}, {Newberg}, {Cudworth}, \&
  {Pan}}]{2008AJ....136.2050L}
{Lee}, Y.~S., {Beers}, T.~C., {Sivarani}, T., {et~al.} 2008, \aj, 136, 2050

\bibitem[{{Lester} {et~al.}(1986){Lester}, {Gray}, \&
  {Kurucz}}]{1986ApJS...61..509L}
{Lester}, J.~B., {Gray}, R.~O., \& {Kurucz}, R.~L. 1986, \apjs, 61, 509

\bibitem[{{Luck} \& {Heiter}(2005)}]{2005AJ....129.1063L}
{Luck}, R.~E. \& {Heiter}, U. 2005, \aj, 129, 1063

\bibitem[{{Malyuto}(1994)}]{1994A&AS..108..441M}
{Malyuto}, V. 1994, \aaps, 108, 441

\bibitem[{{Martell} \& {Laughlin}(2002)}]{2002ApJ...577L..45M}
{Martell}, S. \& {Laughlin}, G. 2002, \apjl, 577, L45

\bibitem[{{Martell} \& {Smith}(2004)}]{2004PASP..116..920M}
{Martell}, S.~L. \& {Smith}, G.~H. 2004, \pasp, 116, 920

\bibitem[{{Mermilliod} {et~al.}(1997){Mermilliod}, {Mermilliod}, \&
  {Hauck}}]{1997A&AS..124..349M}
{Mermilliod}, J.-C., {Mermilliod}, M., \& {Hauck}, B. 1997, \aaps, 124, 349

\bibitem[{{Mishenina} {et~al.}(2004){Mishenina}, {Soubiran}, {Kovtyukh}, \&
  {Korotin}}]{2004A&A...418..551M}
{Mishenina}, T.~V., {Soubiran}, C., {Kovtyukh}, V.~V., \& {Korotin}, S.~A.
  2004, \aap, 418, 551

\bibitem[{{Nissen}(1981)}]{1981A&A....97..145N}
{Nissen}, P.~E. 1981, \aap, 97, 145

\bibitem[{{Nordstr{\"o}m} {et~al.}(2004){Nordstr{\"o}m}, {Mayor}, {Andersen},
  {Holmberg}, {Pont}, {J{\o}rgensen}, {Olsen}, {Udry}, \&
  {Mowlavi}}]{2004A&A...418..989N}
{Nordstr{\"o}m}, B., {Mayor}, M., {Andersen}, J., {et~al.} 2004, \aap, 418, 989

\bibitem[{{Ochsenbein} {et~al.}(2000){Ochsenbein}, {Bauer}, \&
  {Marcout}}]{2000A&AS..143...23O}
{Ochsenbein}, F., {Bauer}, P., \& {Marcout}, J. 2000, \aaps, 143, 23

\bibitem[{{Olsen}(1983)}]{1983A&AS...54...55O}
{Olsen}, E.~H. 1983, \aaps, 54, 55

\bibitem[{{Olsen}(1984)}]{1984A&AS...57..443O}
{Olsen}, E.~H. 1984, \aaps, 57, 443

\bibitem[{{Olsen}(1993)}]{1993A&AS..102...89O}
{Olsen}, E.~H. 1993, \aaps, 102, 89

\bibitem[{{Olsen}(1994{\natexlab{a}})}]{1994A&AS..104..429O}
{Olsen}, E.~H. 1994{\natexlab{a}}, \aaps, 104, 429

\bibitem[{{Olsen}(1994{\natexlab{b}})}]{1994A&AS..106..257O}
{Olsen}, E.~H. 1994{\natexlab{b}}, \aaps, 106, 257

\bibitem[{{Olsen}(1995)}]{1995A&A...295..710O}
{Olsen}, E.~H. 1995, \aap, 295, 710

\bibitem[{{{\"O}nehag} {et~al.}(2009){{\"O}nehag}, {Gustafsson}, {Eriksson}, \&
  {Edvardsson}}]{2009A&A...498..527O}
{{\"O}nehag}, A., {Gustafsson}, B., {Eriksson}, K., \& {Edvardsson}, B. 2009,
  \aap, 498, 527

\bibitem[{{Perry} {et~al.}(1987){Perry}, {Olsen}, \&
  {Crawford}}]{1987PASP...99.1184P}
{Perry}, C.~L., {Olsen}, E.~H., \& {Crawford}, D.~L. 1987, \pasp, 99, 1184

\bibitem[{{Perryman} {et~al.}(1997){Perryman}, {Lindegren}, {Kovalevsky},
  {Hoeg}, {Bastian}, {Bernacca}, {Cr{\'e}z{\'e}}, {Donati}, {Grenon}, {van
  Leeuwen}, {van der Marel}, {Mignard}, {Murray}, {Le Poole}, {Schrijver},
  {Turon}, {Arenou}, {Froeschl{\'e}}, \& {Petersen}}]{1997A&A...323L..49P}
{Perryman}, M.~A.~C., {Lindegren}, L., {Kovalevsky}, J., {et~al.} 1997, \aap,
  323, L49

\bibitem[{{Ram{\'{\i}}rez} \& {Mel{\'e}ndez}(2004)}]{2004A&A...417..301R}
{Ram{\'{\i}}rez}, I. \& {Mel{\'e}ndez}, J. 2004, \aap, 417, 301

\bibitem[{{Ram{\'{\i}}rez} \&
  {Mel{\'e}ndez}(2005{\natexlab{a}})}]{2005ApJ...626..446R}
{Ram{\'{\i}}rez}, I. \& {Mel{\'e}ndez}, J. 2005{\natexlab{a}}, \apj, 626, 446

\bibitem[{{Ram{\'{\i}}rez} \&
  {Mel{\'e}ndez}(2005{\natexlab{b}})}]{2005ApJ...626..465R}
{Ram{\'{\i}}rez}, I. \& {Mel{\'e}ndez}, J. 2005{\natexlab{b}}, \apj, 626, 465

\bibitem[{{Santos} {et~al.}(2001){Santos}, {Israelian}, \&
  {Mayor}}]{2001A&A...373.1019S}
{Santos}, N.~C., {Israelian}, G., \& {Mayor}, M. 2001, \aap, 373, 1019

\bibitem[{{Santos} {et~al.}(2004){Santos}, {Israelian}, \&
  {Mayor}}]{2004A&A...415.1153S}
{Santos}, N.~C., {Israelian}, G., \& {Mayor}, M. 2004, \aap, 415, 1153

\bibitem[{{Santos} {et~al.}(2005){Santos}, {Israelian}, {Mayor}, {Bento},
  {Almeida}, {Sousa}, \& {Ecuvillon}}]{2005A&A...437.1127S}
{Santos}, N.~C., {Israelian}, G., {Mayor}, M., {et~al.} 2005, \aap, 437, 1127

\bibitem[{{Schlegel} {et~al.}(1998){Schlegel}, {Finkbeiner}, \&
  {Davis}}]{1998ApJ...500..525S}
{Schlegel}, D.~J., {Finkbeiner}, D.~P., \& {Davis}, M. 1998, \apj, 500, 525

\bibitem[{{Schuster} {et~al.}(2004){Schuster}, {Beers}, {Michel}, {Nissen}, \&
  {Garc{\'{\i}}a}}]{2004A&A...422..527S}
{Schuster}, W.~J., {Beers}, T.~C., {Michel}, R., {Nissen}, P.~E., \&
  {Garc{\'{\i}}a}, G. 2004, \aap, 422, 527

\bibitem[{{Schuster} {et~al.}(2006){Schuster}, {Moitinho}, {M{\'a}rquez},
  {Parrao}, \& {Covarrubias}}]{2006A&A...445..939S}
{Schuster}, W.~J., {Moitinho}, A., {M{\'a}rquez}, A., {Parrao}, L., \&
  {Covarrubias}, E. 2006, \aap, 445, 939

\bibitem[{{Schuster} \& {Nissen}(1988)}]{1988A&AS...73..225S}
{Schuster}, W.~J. \& {Nissen}, P.~E. 1988, \aaps, 73, 225

\bibitem[{{Schuster} \& {Nissen}(1989{\natexlab{a}})}]{1989A&A...222...69S}
{Schuster}, W.~J. \& {Nissen}, P.~E. 1989{\natexlab{a}}, \aap, 222, 69

\bibitem[{{Schuster} \& {Nissen}(1989{\natexlab{b}})}]{1989A&A...221...65S}
{Schuster}, W.~J. \& {Nissen}, P.~E. 1989{\natexlab{b}}, \aap, 221, 65

\bibitem[{{Schuster} {et~al.}(1996){Schuster}, {Nissen}, {Parrao}, {Beers}, \&
  {Overgaard}}]{1996A&AS..117..317S}
{Schuster}, W.~J., {Nissen}, P.~E., {Parrao}, L., {Beers}, T.~C., \&
  {Overgaard}, L.~P. 1996, \aaps, 117, 317

\bibitem[{{Sousa} {et~al.}(2006){Sousa}, {Santos}, {Israelian}, {Mayor}, \&
  {Monteiro}}]{2006A&A...458..873S}
{Sousa}, S.~G., {Santos}, N.~C., {Israelian}, G., {Mayor}, M., \& {Monteiro},
  M.~J.~P.~F.~G. 2006, \aap, 458, 873

\bibitem[{{Strigari} {et~al.}(2008){Strigari}, {Bullock}, {Kaplinghat},
  {Simon}, {Geha}, {Willman}, \& {Walker}}]{2008Natur.454.1096S}
{Strigari}, L.~E., {Bullock}, J.~S., {Kaplinghat}, M., {et~al.} 2008, \nat,
  454, 1096

\bibitem[{{Str{\"o}mgren}(1963)}]{1963QJRAS...4....8S}
{Str{\"o}mgren}, B. 1963, \qjras, 4, 8

\bibitem[{{Str{\"o}mgren}(1964)}]{1964ApNr....9..333S}
{Str{\"o}mgren}, B. 1964, Astrophysica Norvegica, 9, 333

\bibitem[{{Thor{\'e}n} \& {Feltzing}(2000)}]{2000A&A...363..692T}
{Thor{\'e}n}, P. \& {Feltzing}, S. 2000, \aap, 363, 692

\bibitem[{{Twarog} {et~al.}(2003){Twarog}, {Anthony-Twarog}, \& {De
  Lee}}]{2003AJ....125.1383T}
{Twarog}, B.~A., {Anthony-Twarog}, B.~J., \& {De Lee}, N. 2003, \aj, 125, 1383

\bibitem[{{Twarog} {et~al.}(2007){Twarog}, {Vargas}, \&
  {Anthony-Twarog}}]{2007AJ....134.1777T}
{Twarog}, B.~A., {Vargas}, L.~C., \& {Anthony-Twarog}, B.~J. 2007, \aj, 134,
  1777

\bibitem[{{Valenti} \& {Fischer}(2005)}]{2005ApJS..159..141V}
{Valenti}, J.~A. \& {Fischer}, D.~A. 2005, \apjs, 159, 141

\bibitem[{{van Leeuwen}(2007)}]{2007hnrr.book.....V}
{van Leeuwen}, F. 2007, {Hipparcos, the New Reduction of the Raw Data}
  (Hipparcos, the New Reduction of the Raw Data. By Floor van Leeuwen,
  Institute of Astronomy, Cambridge University, Cambridge, UK Series:
  Astrophysics and Space Science Library, Vol. 350 20 Springer Dordrecht)

\bibitem[{{van Leeuwen}(2009)}]{2009A&A...497..209V}
{van Leeuwen}, F. 2009, \aap, 497, 209

\bibitem[{{Vandenberg} \& {Bell}(1985)}]{1985ApJS...58..561V}
{Vandenberg}, D.~A. \& {Bell}, R.~A. 1985, \apjs, 58, 561

\bibitem[{{VandenBerg} {et~al.}(2006){VandenBerg}, {Bergbusch}, \&
  {Dowler}}]{2006ApJS..162..375V}
{VandenBerg}, D.~A., {Bergbusch}, P.~A., \& {Dowler}, P.~D. 2006, \apjs, 162,
  375

\bibitem[{{von Hippel} \& {Bothun}(1993)}]{1993ApJ...407..115V}
{von Hippel}, T. \& {Bothun}, G.~D. 1993, \apj, 407, 115

\bibitem[{{von Hippel}(1992)}]{1992AJ....104.1765V}
{von Hippel}, T.~A. 1992, \aj, 104, 1765

\bibitem[{{Woolf} \& {Wallerstein}(2005)}]{2005MNRAS.356..963W}
{Woolf}, V.~M. \& {Wallerstein}, G. 2005, \mnras, 356, 963

\bibitem[{{Yasuda} {et~al.}(2007){Yasuda}, {Fukugita}, \&
  {Schneider}}]{2007AJ....134..698Y}
{Yasuda}, N., {Fukugita}, M., \& {Schneider}, D.~P. 2007, \aj, 134, 698

\bibitem[{{Yong} \& {Lambert}(2003)}]{2003PASP..115...22Y}
{Yong}, D. \& {Lambert}, D.~L. 2003, \pasp, 115, 22

\bibitem[{{York} {et~al.}(2000){York}, {Adelman}, {Anderson}, {Anderson},
  {Annis}, {Bahcall}, {Bakken}, {Barkhouser}, {Bastian}, {Berman}, {Boroski},
  {Bracker}, {Briegel}, {Briggs}, {Brinkmann}, {Brunner}, {Burles}, {Carey},
  {Carr}, {Castander}, {Chen}, {Colestock}, {Connolly}, {Crocker}, {Csabai},
  {Czarapata}, {Davis}, {Doi}, {Dombeck}, {Eisenstein}, {Ellman}, {Elms},
  {Evans}, {Fan}, {Federwitz}, {Fiscelli}, {Friedman}, {Frieman}, {Fukugita},
  {Gillespie}, {Gunn}, {Gurbani}, {de Haas}, {Haldeman}, {Harris}, {Hayes},
  {Heckman}, {Hennessy}, {Hindsley}, {Holm}, {Holmgren}, {Huang}, {Hull},
  {Husby}, {Ichikawa}, {Ichikawa}, {Ivezi{\'c}}, {Kent}, {Kim}, {Kinney},
  {Klaene}, {Kleinman}, {Kleinman}, {Knapp}, {Korienek}, {Kron}, {Kunszt},
  {Lamb}, {Lee}, {Leger}, {Limmongkol}, {Lindenmeyer}, {Long}, {Loomis},
  {Loveday}, {Lucinio}, {Lupton}, {MacKinnon}, {Mannery}, {Mantsch}, {Margon},
  {McGehee}, {McKay}, {Meiksin}, {Merelli}, {Monet}, {Munn}, {Narayanan},
  {Nash}, {Neilsen}, {Neswold}, {Newberg}, {Nichol}, {Nicinski}, {Nonino},
  {Okada}, {Okamura}, {Ostriker}, {Owen}, {Pauls}, {Peoples}, {Peterson},
  {Petravick}, {Pier}, {Pope}, {Pordes}, {Prosapio}, {Rechenmacher}, {Quinn},
  {Richards}, {Richmond}, {Rivetta}, {Rockosi}, {Ruthmansdorfer}, {Sandford},
  {Schlegel}, {Schneider}, {Sekiguchi}, {Sergey}, {Shimasaku}, {Siegmund},
  {Smee}, {Smith}, {Snedden}, {Stone}, {Stoughton}, {Strauss}, {Stubbs},
  {SubbaRao}, {Szalay}, {Szapudi}, {Szokoly}, {Thakar}, {Tremonti}, {Tucker},
  {Uomoto}, {Vanden Berk}, {Vogeley}, {Waddell}, {Wang}, {Watanabe},
  {Weinberg}, {Yanny}, \& {Yasuda}}]{2000AJ....120.1579Y}
{York}, D.~G., {Adelman}, J., {Anderson}, Jr., J.~E., {et~al.} 2000, \aj, 120,
  1579

\end{thebibliography}

\Online   

\begin{appendix}
\onecolumn
\section{Stellar sequences \label{ApxA}}


\begin{figure*}
\sidecaption
  \includegraphics[width=12cm]{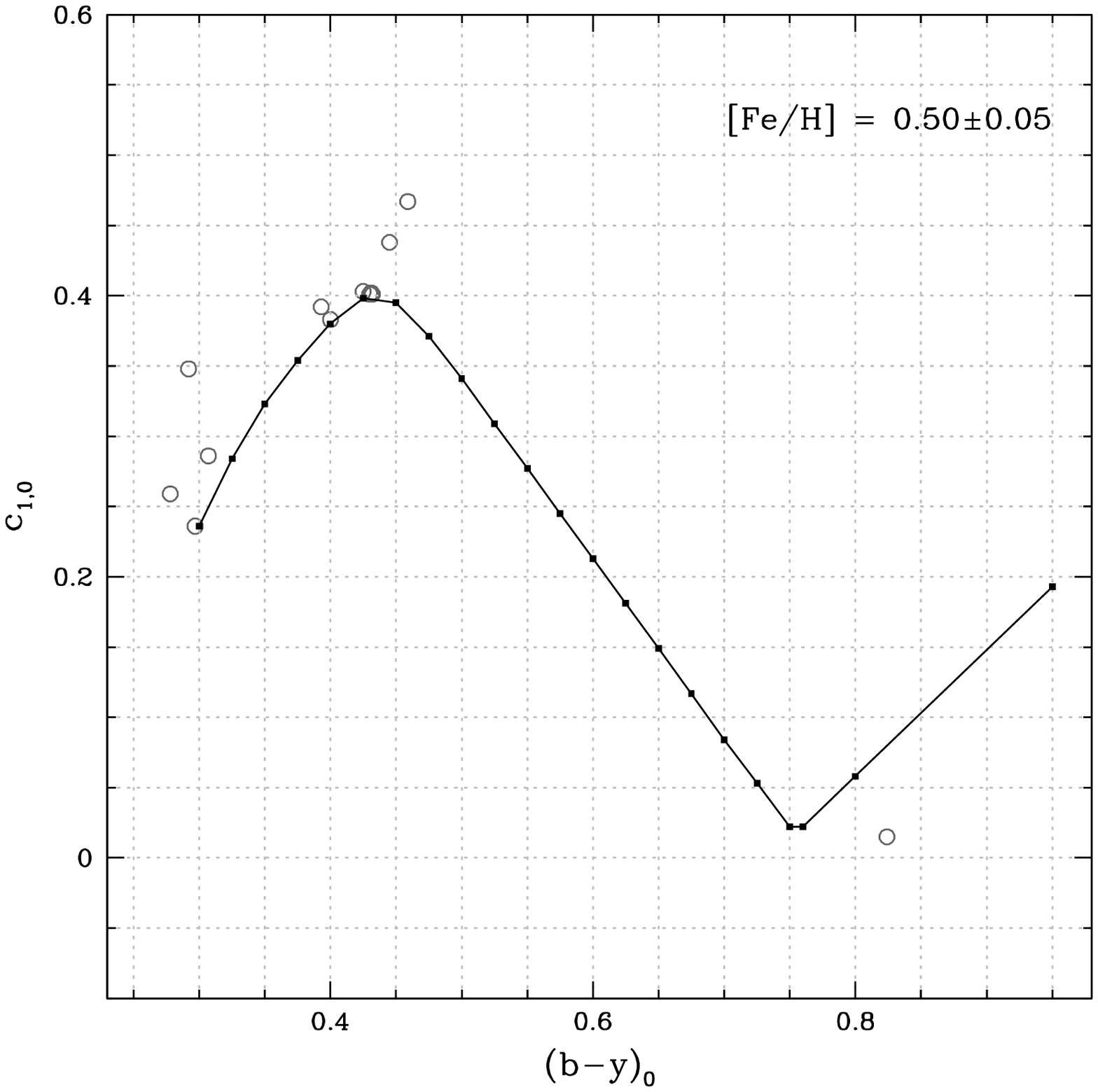}
   \caption{The figure shows how the dwarf star sequence was traced from nearby 
dwarf stars with  [Fe/H]$=0.50 \pm 0.05$ plotted in the $c_{1,0}$ vs $(b-y)_0$ diagram.}
     \label{Dby05}
\end{figure*}

\begin{figure*}
\sidecaption
  \includegraphics[width=12cm]{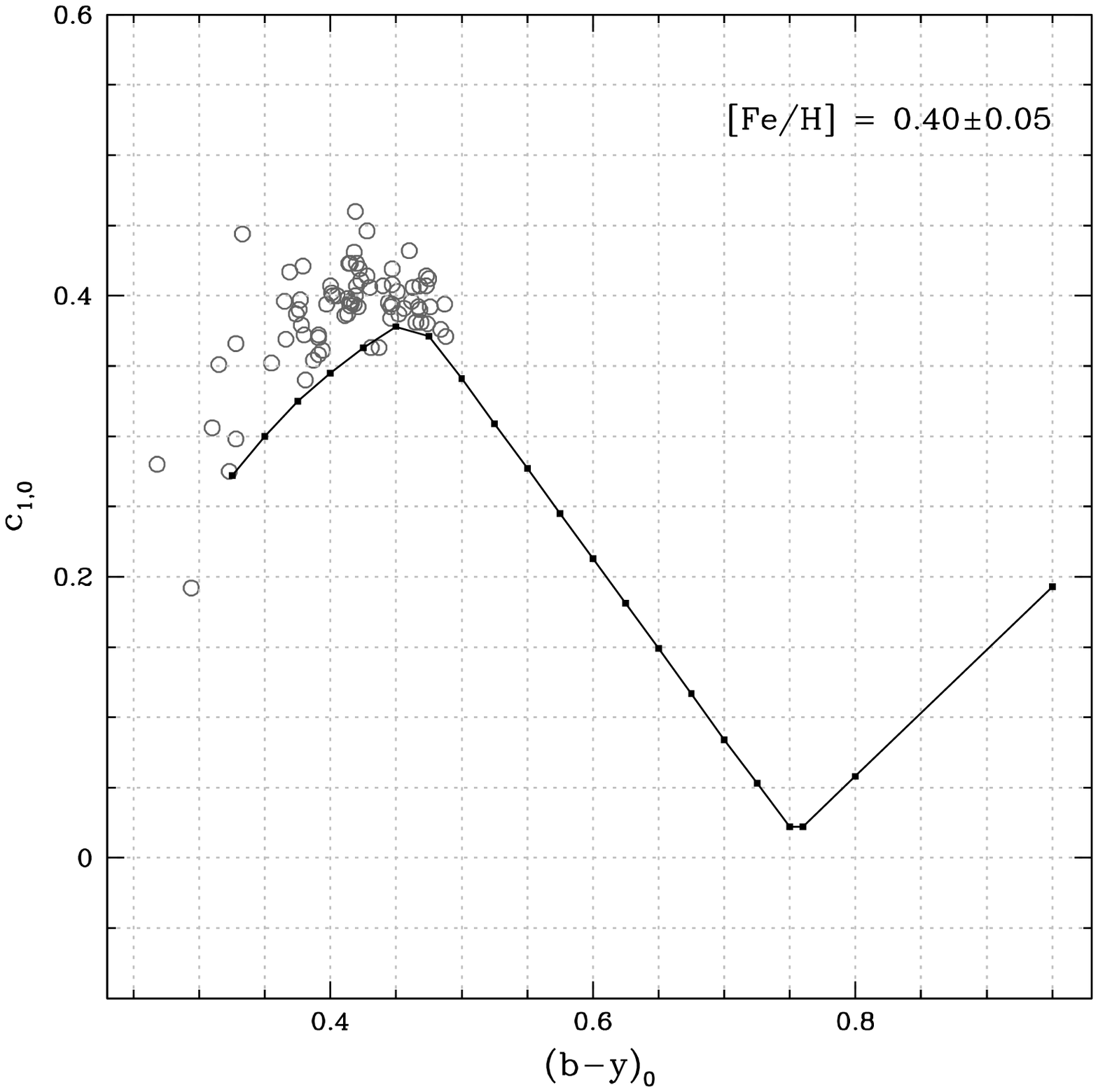}
   \caption{The figure shows how the dwarf star sequence was traced from nearby 
dwarf stars with  [Fe/H]$=0.40 \pm 0.05$ plotted in the $c_{1,0}$ vs $(b-y)_0$ diagram.}
     \label{Dby04}
\end{figure*}

\begin{figure*}
\sidecaption
  \includegraphics[width=12cm]{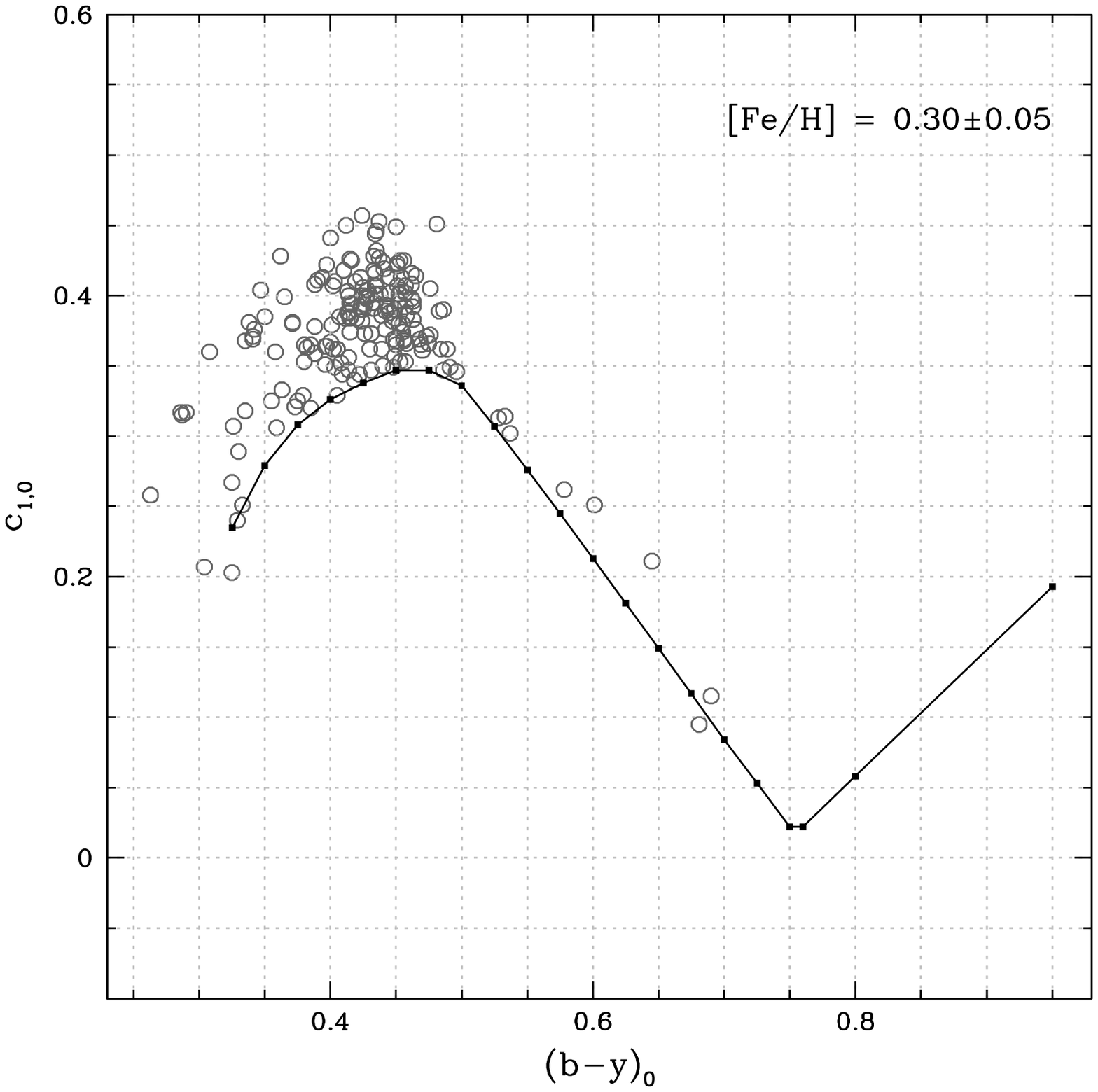}
   \caption{The figure shows how the dwarf star sequence was traced from nearby 
dwarf stars with  [Fe/H]$=0.30 \pm 0.05$ plotted in the $c_{1,0}$ vs $(b-y)_0$ diagram.}
     \label{Dby03}
\end{figure*}

\begin{figure*}
\sidecaption
  \includegraphics[width=12cm]{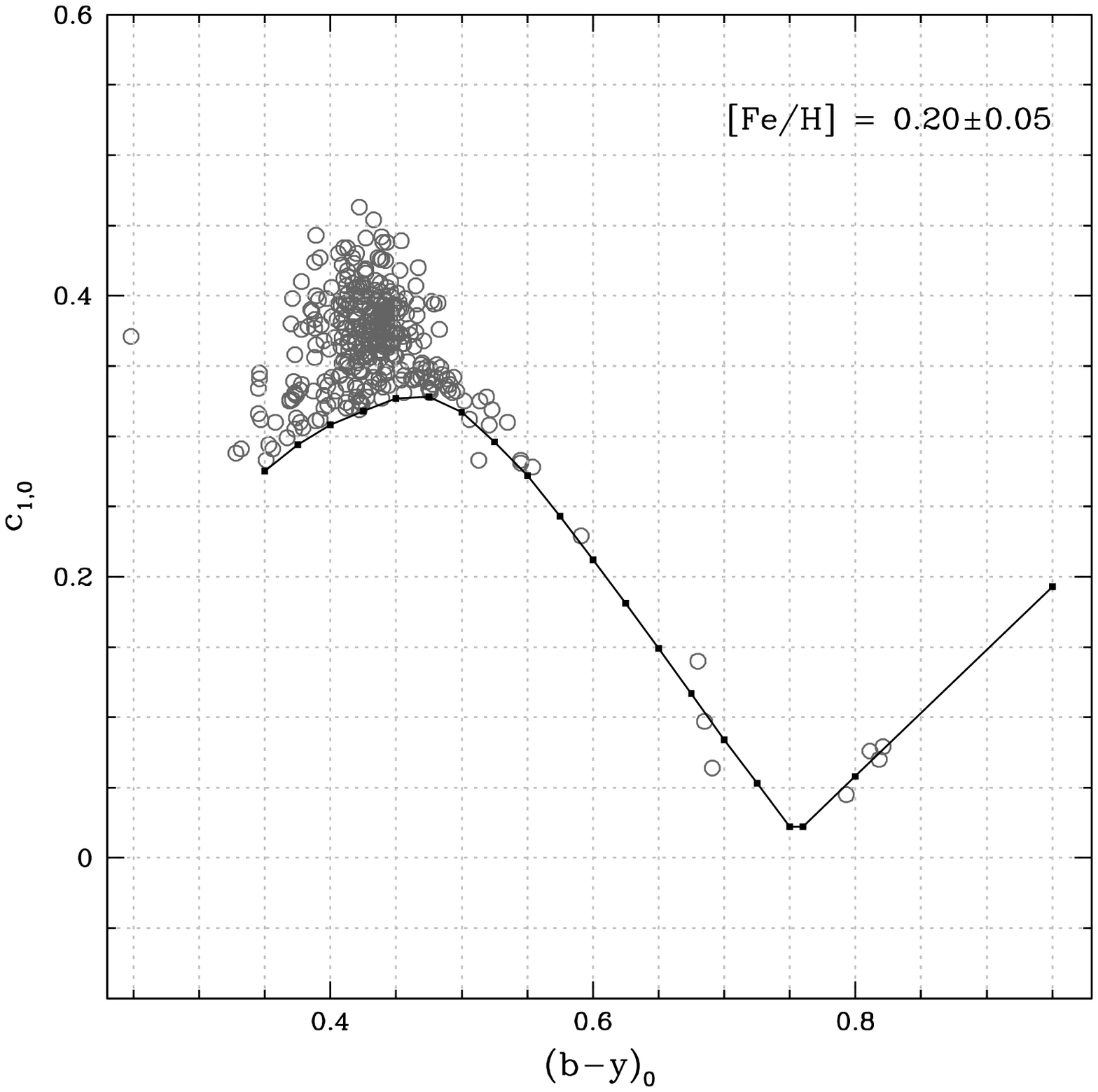}
   \caption{The figure shows how the dwarf star sequence was traced from nearby 
dwarf stars with  [Fe/H]$=0.20 \pm 0.05$ plotted in the $c_{1,0}$ vs $(b-y)_0$ diagram.}
     \label{Dby02}
\end{figure*}

\begin{figure*}
\sidecaption
  \includegraphics[width=12cm]{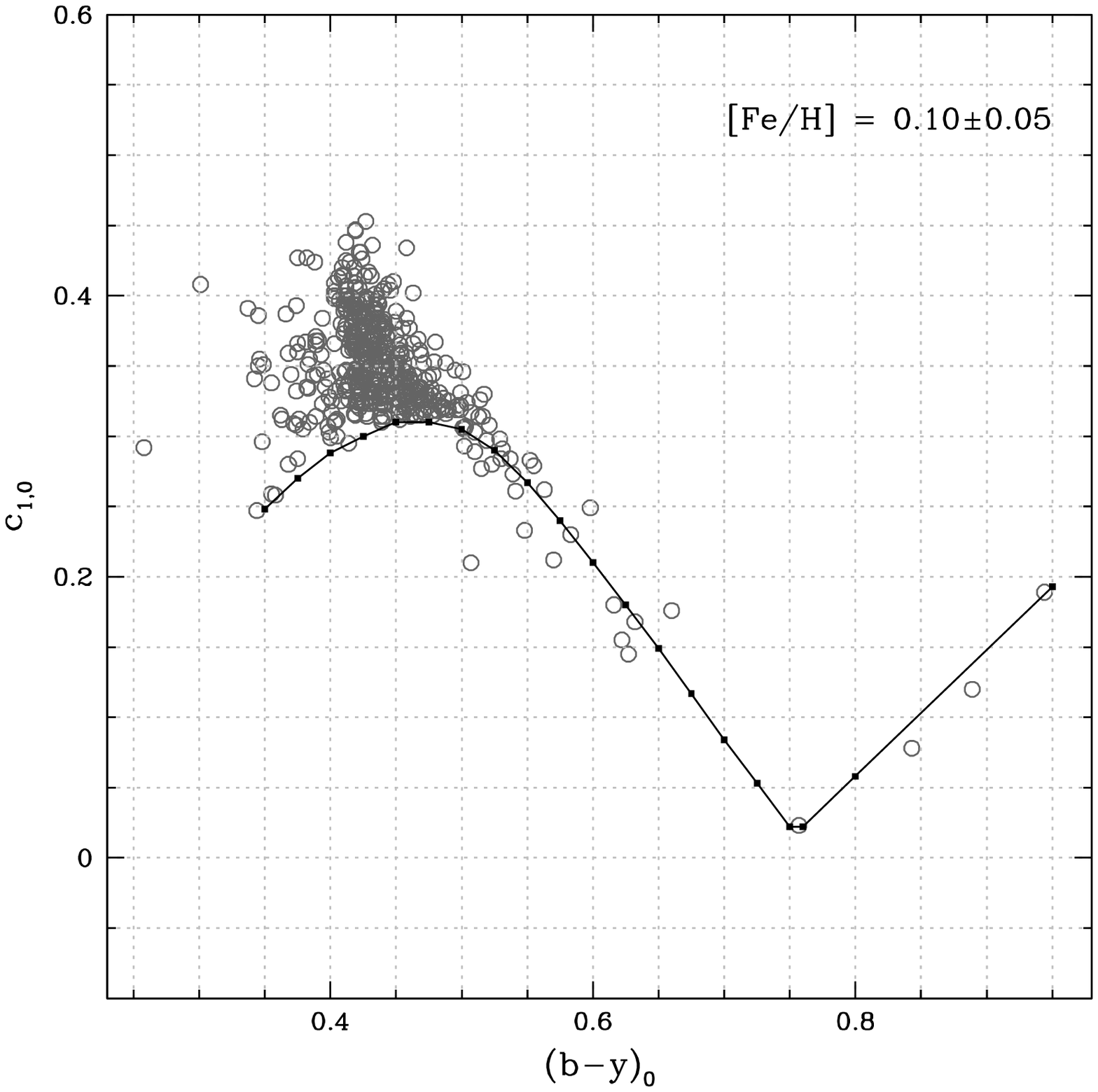}
   \caption{The figure shows how the dwarf star sequence was traced from nearby 
dwarf stars with  [Fe/H]$=0.10 \pm 0.05$ plotted in the $c_{1,0}$ vs $(b-y)_0$ diagram.}
     \label{Dby01}
\end{figure*}

\begin{figure*}
\sidecaption
  \includegraphics[width=12cm]{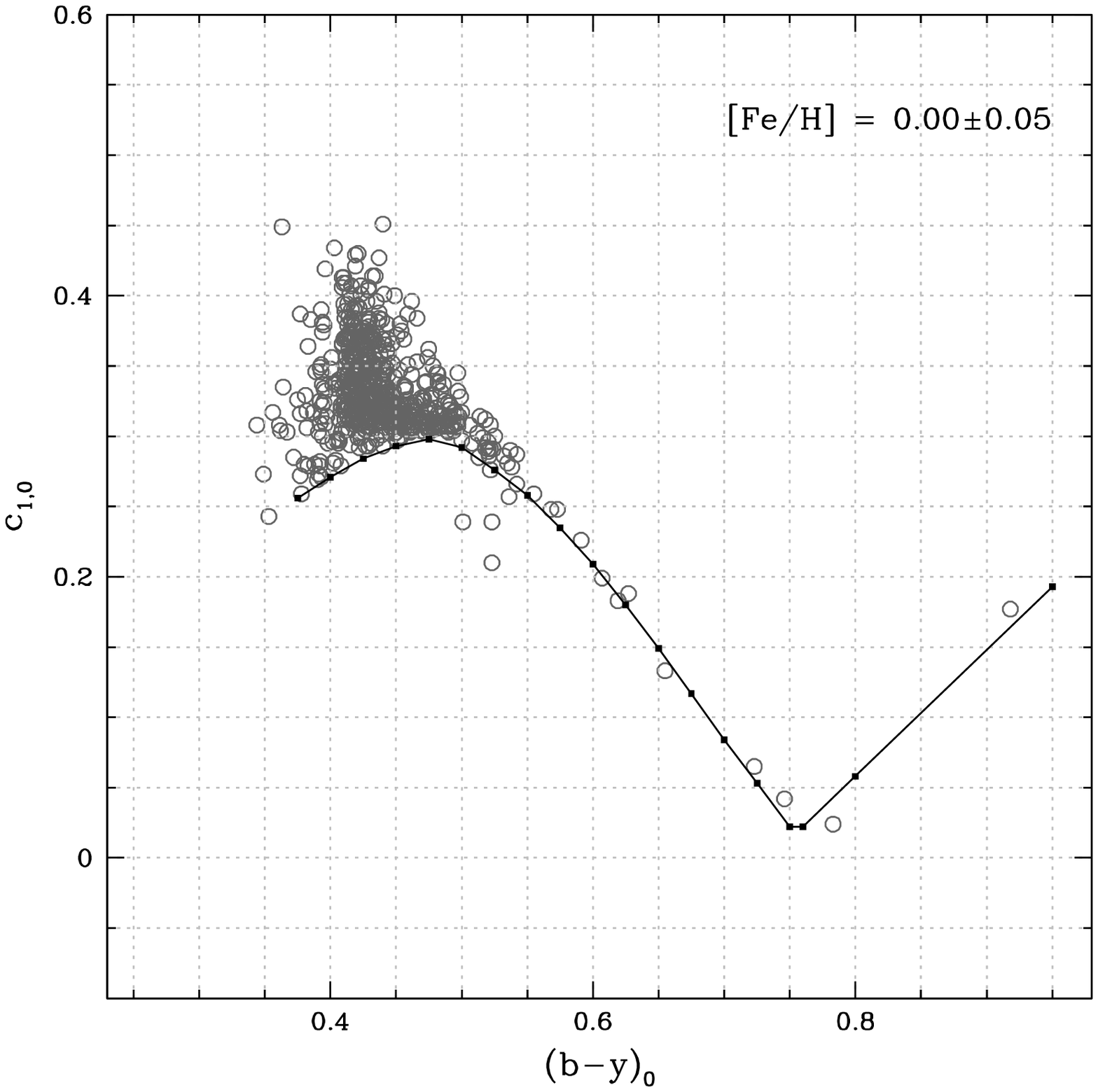}
   \caption{The figure shows how the dwarf star sequence was traced from nearby 
dwarf stars with  [Fe/H]$=0.00 \pm 0.05$ plotted in the $c_{1,0}$ vs $(b-y)_0$ diagram.}
     \label{Dby00}
\end{figure*}

\begin{figure*}
\sidecaption
  \includegraphics[width=12cm]{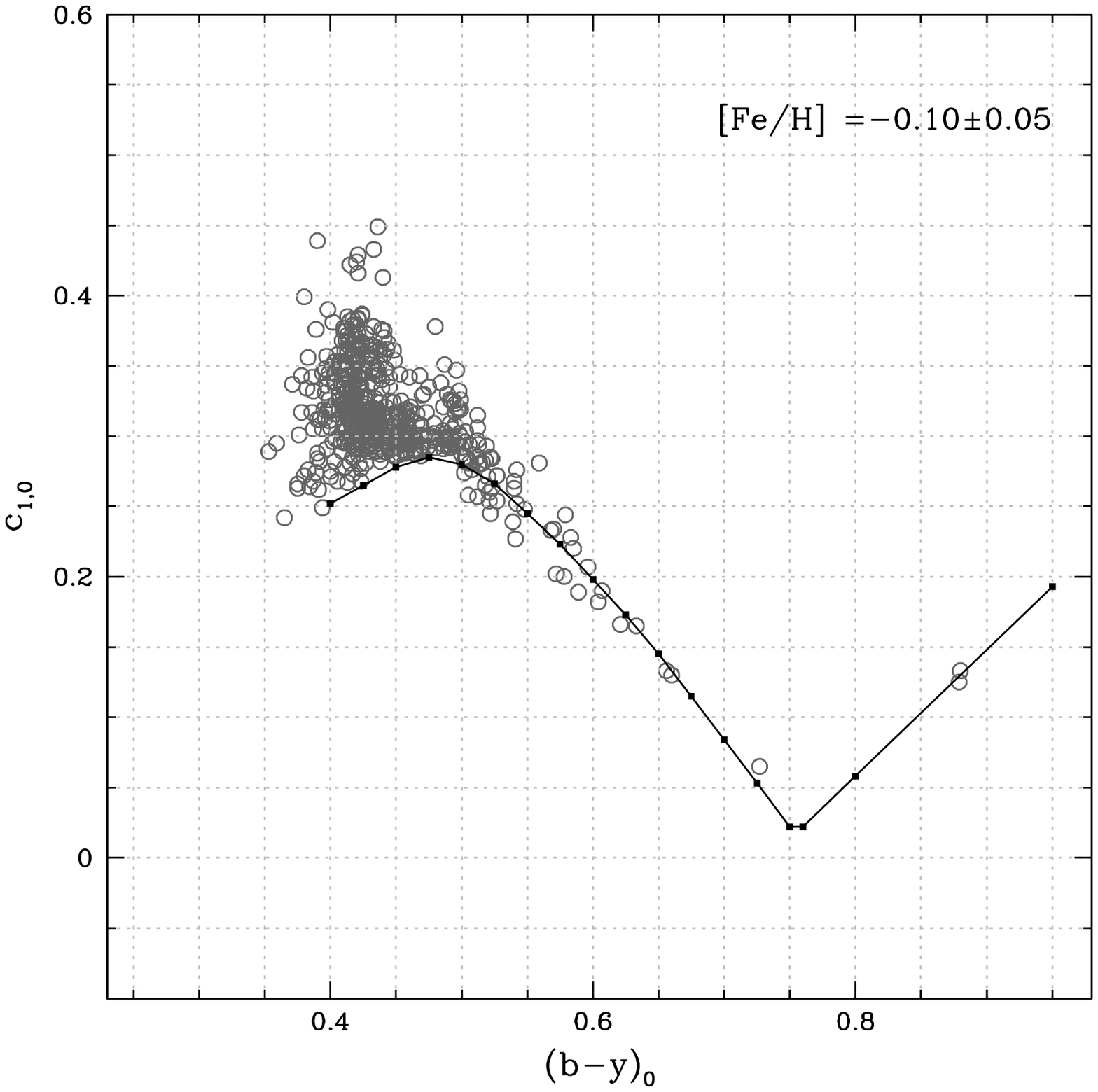}
   \caption{The figure shows how the dwarf star sequence was traced from nearby 
dwarf stars with  [Fe/H]$=-0.10 \pm 0.05$ plotted in the $c_{1,0}$ vs $(b-y)_0$ diagram.}
     \label{Dbym01}
\end{figure*}

\begin{figure*}
\sidecaption
  \includegraphics[width=12cm]{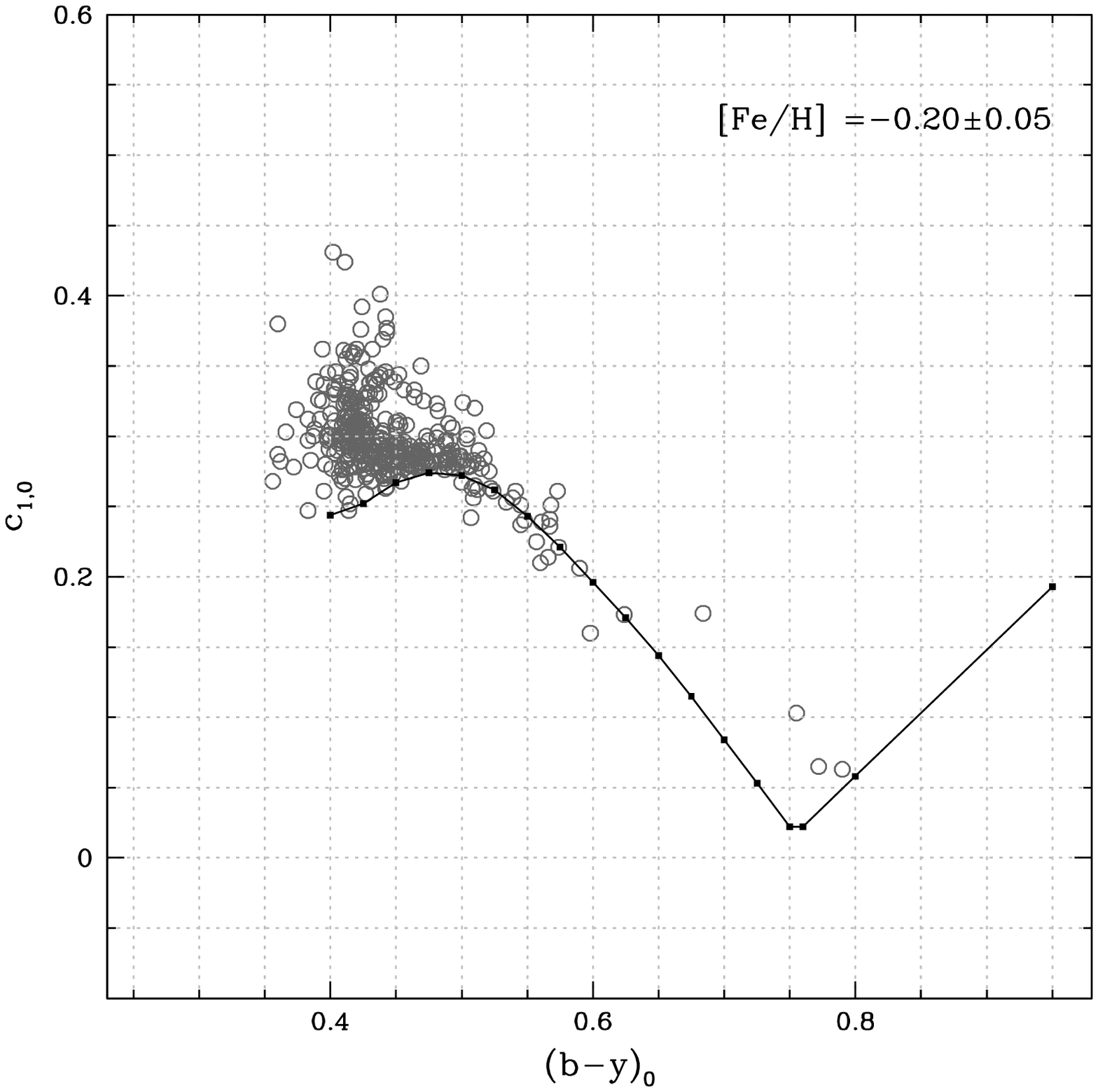}
   \caption{The figure shows how the dwarf star sequence was traced from nearby 
dwarf stars with  [Fe/H]$=-0.20 \pm 0.05$ plotted in the $c_{1,0}$ vs $(b-y)_0$ diagram.}
     \label{Dbym02}
\end{figure*}

\begin{figure*}
\sidecaption
  \includegraphics[width=12cm]{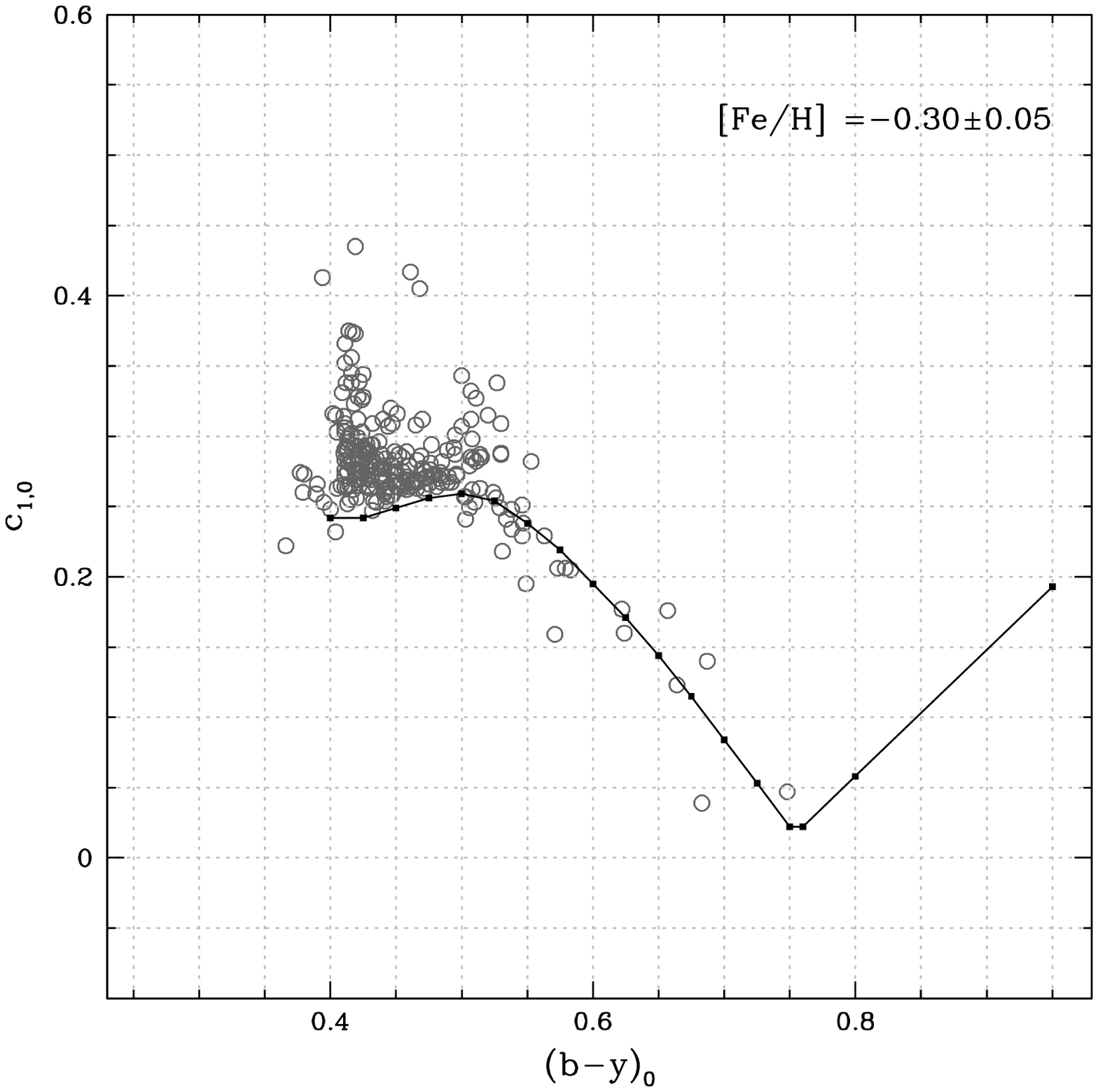}
   \caption{The figure shows how the dwarf star sequence was traced from nearby 
dwarf stars with  [Fe/H]$=-0.30 \pm 0.05$ plotted in the $c_{1,0}$ vs $(b-y)_0$ diagram.}
     \label{Dbym03}
\end{figure*}

\begin{figure*}
\sidecaption
  \includegraphics[width=12cm]{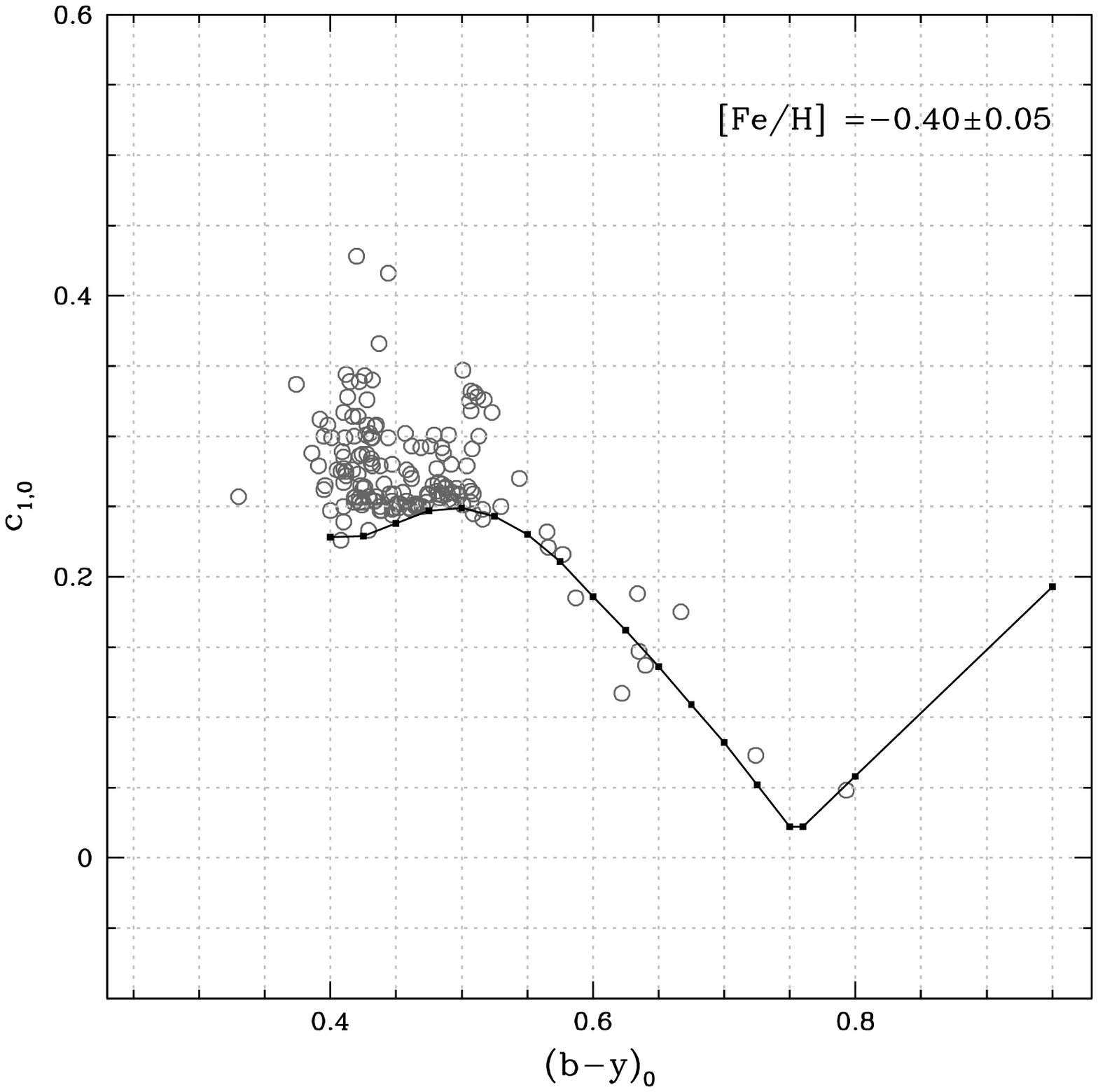}
   \caption{The figure shows how the dwarf star sequence was traced from nearby 
dwarf stars with  [Fe/H]$=-0.40 \pm 0.05$ plotted in the $c_{1,0}$ vs $(b-y)_0$ diagram.}
     \label{Dbym04}
\end{figure*}

\begin{figure*}
\sidecaption
  \includegraphics[width=12cm]{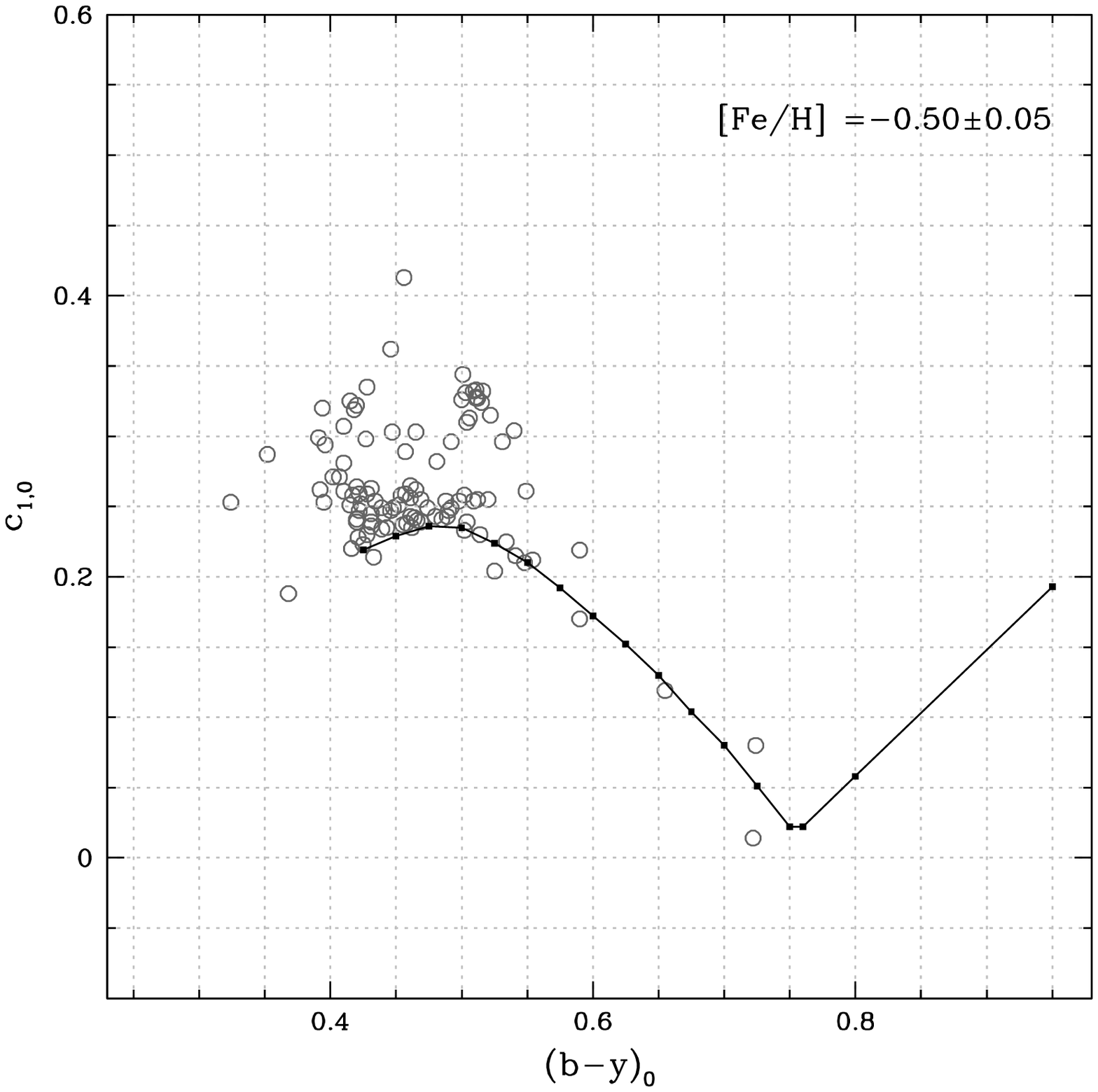}
   \caption{The figure shows how the dwarf star sequence was traced from nearby 
dwarf stars with  [Fe/H]$=-0.50 \pm 0.05$ plotted in the $c_{1,0}$ vs $(b-y)_0$ diagram.}
     \label{Dbym05}
\end{figure*}

\begin{figure*}
\sidecaption
  \includegraphics[width=12cm]{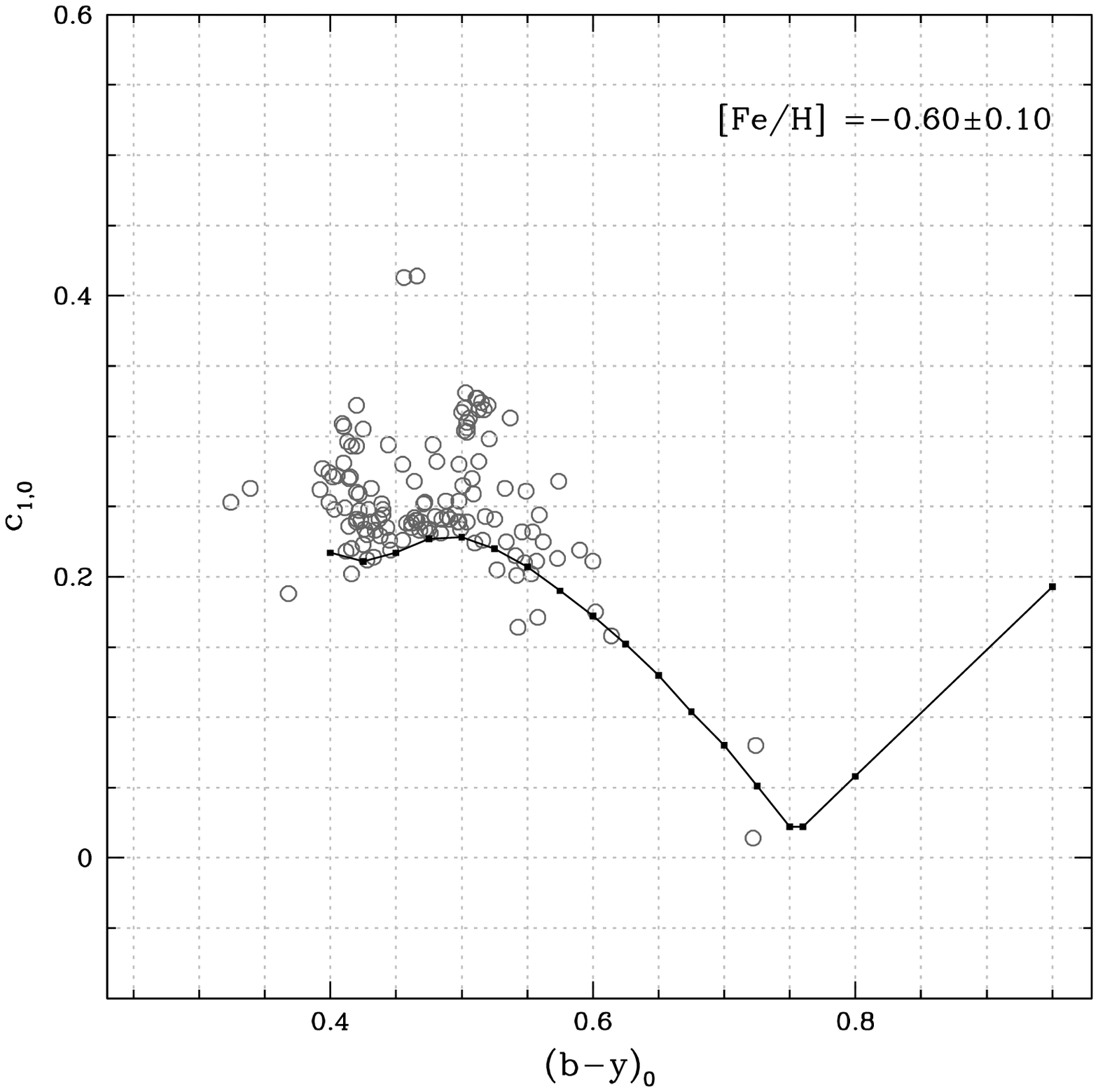}
   \caption{The figure shows how the dwarf star sequence was traced from nearby 
dwarf stars with  [Fe/H]$=-0.60 \pm 0.10$ plotted in the $c_{1,0}$ vs $(b-y)_0$ diagram.}
     \label{Dbym06}
\end{figure*}

\begin{figure*}
\sidecaption
  \includegraphics[width=12cm]{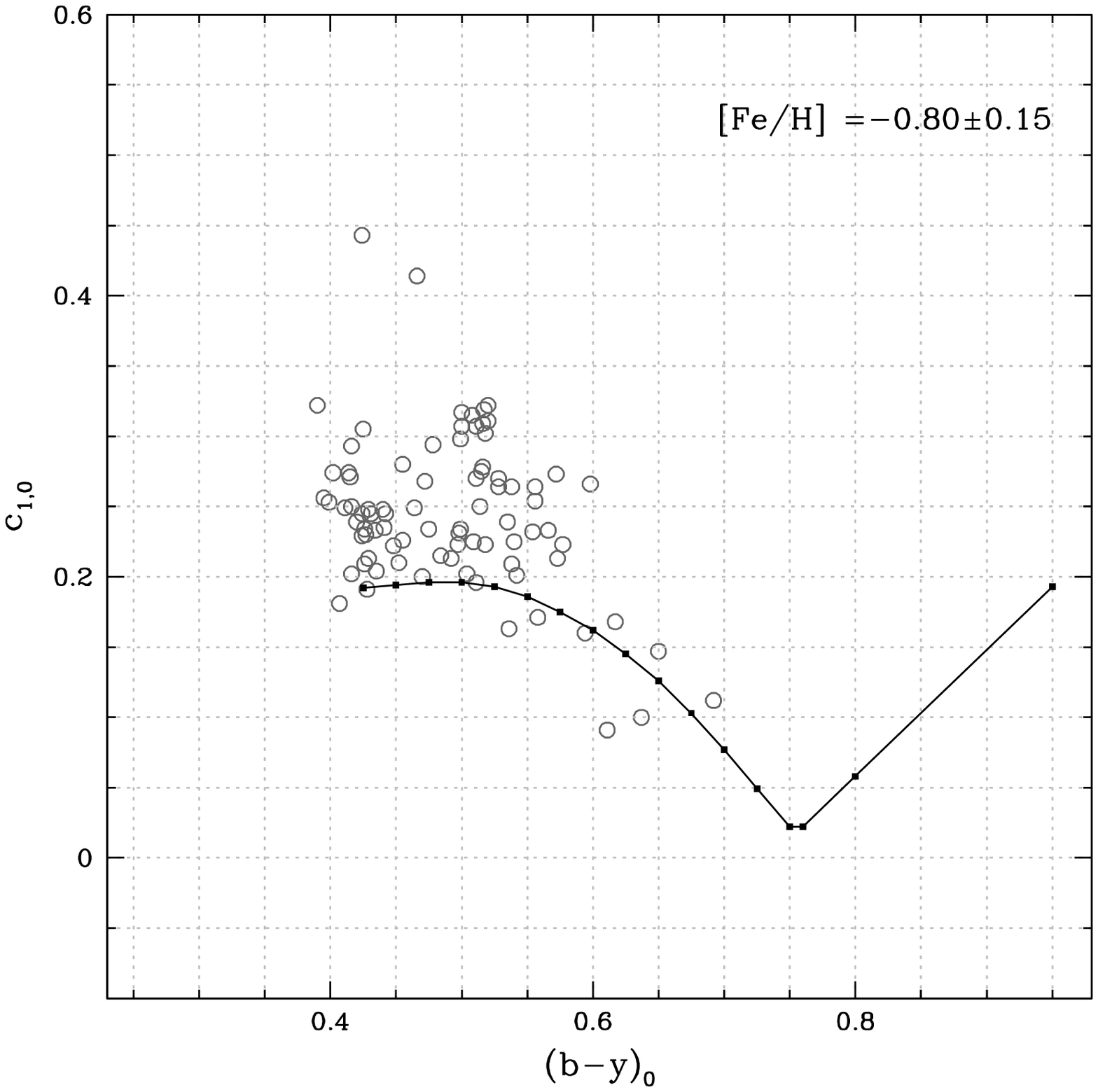}
   \caption{The figure shows how the dwarf star sequence was traced from nearby 
dwarf stars with  [Fe/H]$=-0.80 \pm 0.10$ plotted in the $c_{1,0}$ vs $(b-y)_0$ diagram.}
     \label{Dbym08}
\end{figure*}

\begin{figure*}
\sidecaption
  \includegraphics[width=12cm]{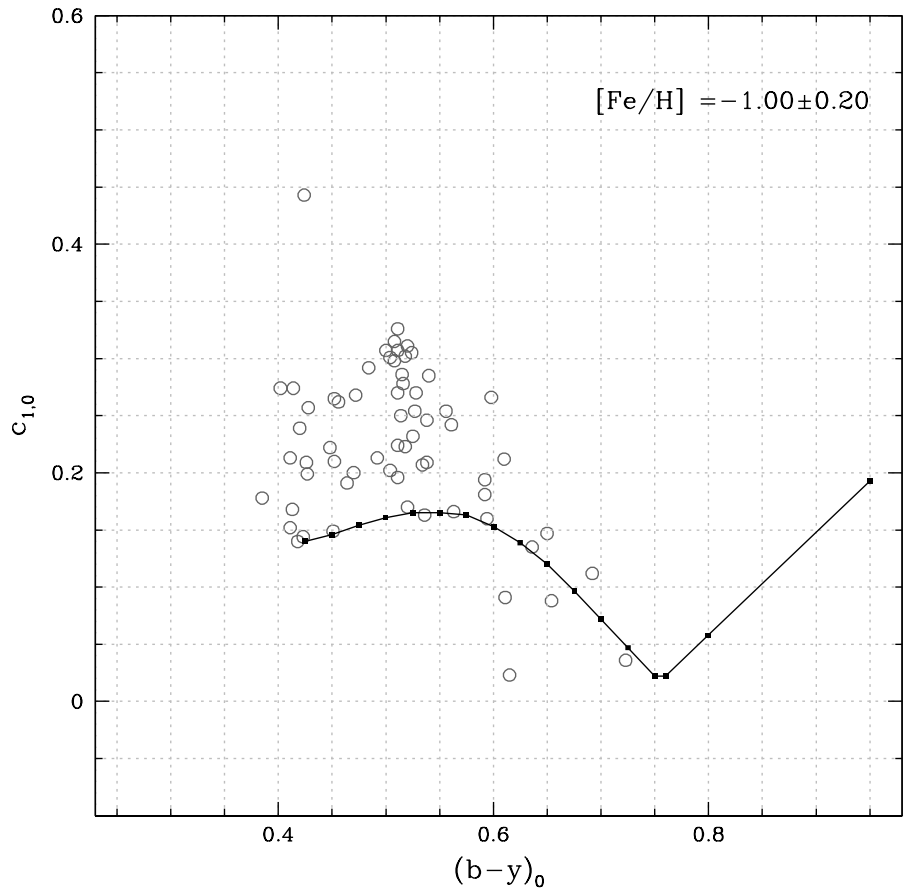}
   \caption{The figure shows how the dwarf star sequence was traced from nearby 
dwarf stars with  [Fe/H]$=-1.00 \pm 0.20$ plotted in the $c_{1,0}$ vs $(b-y)_0$ diagram.}
     \label{Dbym10}
\end{figure*}

\end{appendix}

\begin{appendix}
\onecolumn
\section{Table containing the data collected to test calibrations of [Fe/H] in Sect.\, \label{ApxB}}

How the catalogue is constructed is explained in detail in
Sect.\,\ref{Sect:cats}.

Column 1 lists the Hipparcos number of the star and Col. 2 gives an
alternative stellar name.  Column 3 gives the photometry reference
(SN88 for \citet{1988A&AS...73..225S}, O84 for
\citet{1984A&AS...57..443O}, O93 for \citet{1993A&AS..102...89O}, and
O94 for \citet{1994A&AS..104..429O}) and Columns 4 to 7 give the
$uvby$ photometry. Column 8 gives the colour excess of the star.
Columns 9 to 12 give the dereddened $uvby$ photometry.  Column 13 and
14 give the average [Fe/H] \citep[on the][scale]{2005ApJS..159..141V}
and the full range of [Fe/H] (on the same scale as in column 13) if
the star was found in more than one study.  Columns 15 and 16 give the
number of references for the [Fe/H] and lists them (1:
\citet{2005ApJS..159..141V} , 2: \citet{1997A&A...323..809F}, 3:
\citet{1998A&AS..129..237F}, 4: \citet{2000A&AS..141..491C}, 5:
\citet{2000A&A...363..692T}, 6: \citet{2001A&A...373.1019S}, 7:
\citet{2003AJ....126.2015H}, \citet{2003PASP..115...22Y}, 9:
\citet{2004A&A...418..551M}, 10: \citet{2004A&A...415.1153S}, 11:
\citet{2005A&A...442..635B}, 12: \citet{2005AJ....129.1063L}, 13:
\citet{2005A&A...437.1127S}, 14: \citet{2005MNRAS.356..963W}, and 15:
\citet{2006A&A...458..873S}).
The data will be made publicly available through CDS.

\onllongtabL{1}{
\scriptsize{
\begin{longtable}{r l l r r r r r r r r r r r l l}
\caption{Dwarf stars with both Str\"{o}mgren photometry and [Fe/H] based on high resolution, high S/N spectroscopy.\label{FCC} }\\
  \hline \hline
\multirow{2}{*}{HIP} &\multirow{2}{*}{Name} & Ref. &  \multicolumn{1}{c}{\multirow{2}{*}{$V$}} &  \multicolumn{1}{c}{\multirow{2}{*}{$(b-y)$}} &  \multicolumn{1}{c}{\multirow{2}{*}{$m_{1}$}} &  \multicolumn{1}{c}{\multirow{2}{*}{$c_{1}$}} & \multirow{2}{*}{E$(B-V)$} &  \multicolumn{1}{c}{\multirow{2}{*}{$V_{0}$}} &  \multicolumn{1}{c}{\multirow{2}{*}{$(b-y)_0$}} &  \multicolumn{1}{c}{\multirow{2}{*}{$m_{1,0}$}} &  \multicolumn{1}{c}{\multirow{2}{*}{$c_{1,0}$}} & \multirow{2}{*}{$<$[Fe/H]$>$} & [Fe/H] &   \multirow{2}{*}{N}&  Ref.\\
  & & $uvby$ &  & & & & & & & & & & range & & [Fe/H] \\
\hline
    400   & HD 225261   & O93   &  7.824&  0.453&  0.271&  0.252 & $<0.02$&  7.824&  0.453&  0.271&  0.252& $   -0.44$&     ... &   1&  1   \\
    544   & HD    166   & O93   &  6.089&  0.459&  0.286&  0.311 & $<0.02$&  6.089&  0.459&  0.286&  0.311& $    0.14$&     0.08&   3&  1,7,12 \\
   1499   & HD 1461     & O93   &  6.471&  0.421&  0.244&  0.360 & $<0.02$&  6.471&  0.421&  0.244&  0.360& $    0.18$&     ... &   1&  1   \\ 
   1931   & HD   2039   & O93   &  9.008&  0.412&  0.218&  0.394 & $<0.02$&  9.008&  0.412&  0.218&  0.394& $    0.30$&     0.04&   2&  1,10    \\
   1936   & HD 2025     & O93   &  7.884&  0.538&  0.473&  0.234 & $<0.02$&  7.884&  0.538&  0.473&  0.234& $   -0.25$&     0.07&   2&  1,2 \\
\hline
  \end{longtable}
 } }

\end{appendix}

\end{document}